\documentclass[10pt,onecolumn]{report}

\usepackage[utf8x]{inputenc}
\usepackage{color}
\usepackage{bbm} 

\usepackage[paper=a4paper,left=30mm,right=30mm,top=30mm,bottom=30mm]{geometry}

\usepackage{amsfonts,amsmath,amssymb,stmaryrd}

\usepackage{graphicx}
\usepackage{subfigure}  
\usepackage{bbm} 
\usepackage{hyperref}
\usepackage[pdftex]{epsfig}
\usepackage{mathrsfs}
\usepackage{verbatim}
\usepackage{centernot}
\usepackage{ulem}
\usepackage{wrapfig}
\usepackage{cite}

\usepackage{titlesec}

\titleformat
{\chapter} 
[display] 
{\bfseries\Large} 
{} 
{-15ex} 
{ \rule{\textwidth}{1pt}
    \vspace{1ex}
    \centering 
    \hspace{-2ex}
    \thechapter
    \hspace{1ex}
} 
[
\vspace{-1.3ex}%
\rule{\textwidth}{0.3pt}
] 

 \titleformat{\section}
  {\normalfont\large\bfseries}
  {\thesection}{1em}{}

 \titleformat{\subsection}
  {\normalfont\bfseries}
  {\thesubsection}{1em}{}

\renewcommand{\l}{\left(}
\renewcommand{\r}{\right)}

\newcommand{\gs}{\text{gs}}

\newcommand{\bra}[1]{\langle#1|}
\newcommand{\ket}[1]{|#1\rangle}

\renewcommand{\H}{\hat{\mathcal{H}}}
\renewcommand{\c}{\hat{c}}
\renewcommand{\a}{\hat{a}}
\newcommand{\cd}{\hat{c}^\dagger}
\newcommand{\ad}{\hat{a}^\dagger}

\renewcommand{\d}{\hat{d}}
\newcommand{\dd}{\hat{d}^\dagger}

\newcommand{\G}{\hat{\Gamma}}
\newcommand{\Gd}{\hat{\Gamma}^\dagger}
\newcommand{\hc}{\text{h.c.}}
\newcommand{\MF}{\text{MF}}

\newcommand{\RG}{\text{RG}}
\newcommand{\psd}{\hat{\psi}^\dagger}
\newcommand{\ps}{\hat{\psi}}
\newcommand{\I}{\text{I}}
\newcommand{\p}{\text{p}}
\newcommand{\f}{\text{F}}
\newcommand{\s}{\text{S}}
\renewcommand{\sf}{\text{MIX}}

\newcommand{\U}{\hat{U}}
\newcommand{\Ud}{\hat{U}^\dagger}
\newcommand{\F}{\hat{F}}

\newcommand{\ph}{\text{ph}}
\newcommand{\IB}{\text{IB}}
\newcommand{\B}{\text{B}}
\newcommand{\eff}{\text{eff}}
\newcommand{\BB}{{\rm BB}}
\newcommand{\rd}{{\rm red}}
\newcommand{\h}[1]{\hat{#1}}
\renewcommand{\sc}{{\rm sc}}
\newcommand{\BZ}{{\rm BZ}}

\usepackage{array}

\usepackage{cancel,ifthen}

\usepackage{bm}	
\renewcommand{\vec}[1]{\bm{#1}}

\begin{document}
\normalem	

\begin{titlepage}
\vspace*{2cm}

{\noindent
\Large \bf
\textbf{New theoretical approaches to Bose polarons}}

\vspace{0.6cm}

\noindent
{\normalsize
\textsc{$~^{1,2,3}$Fabian Grusdt}  ~ and \textsc{ $~^3$Eugene Demler} } 

\vspace{0.3cm}

\hspace{-0.25cm}
\emph{
$~^1$ Department of Physics and Research Center OPTIMAS, TU Kaiserslautern, Germany}

\emph{$~^2$ Graduate School Materials Science in Mainz, 67663 Kaiserslautern, Germany}

\emph{$~^3$ Department of Physics, Harvard University, Cambridge, Massachusetts 02138, USA}

\vspace{15cm}

Proceedings of the International School of Physics "Enrico Fermi"
\end{titlepage}

\newpage
\tableofcontents

\newpage
\chapter{Summary}

The Fr\"ohlich polaron model describes a ubiquitous class of problems concerned with understanding properties of a single mobile particle interacting with a bosonic reservoir. Originally introduced in the context of electrons interacting with phonons in crystals, this model found applications in such diverse areas as strongly correlated electron systems, quantum information, and high energy physics. In the last few years this model has been applied to describe impurity atoms immersed in Bose-Einstein condensates of ultracold atoms. The tunability of microscopic parameters in ensembles of ultracold atoms and the rich experimental toolbox of atomic physics should allow to test many theoretical predictions and give us new insights into equilibrium and dynamical properties of polarons.
In these lecture notes we provide an overview of common theoretical approaches that have been used to study BEC polarons, including Rayleigh-Schr\"odinger and Green's function perturbation theories, self-consistent Born approximation, mean-field approach, Feynman's variational path integral approach, Monte Carlo simulations, renormalization group calculations, and Gaussian variational ansatz. We focus on the renormalization group approach and provide details of analysis that have not been presented in earlier publications. We show that this method helps to resolve striking discrepancy in polaron energies obtained using mean-field approximation and Monte Carlo simulations. We also discuss applications of this method to the calculation of the effective mass of BEC polarons. As one experimentally relevant example of a non-equililbrium problem we consider Bloch oscillations of Bose polarons and demonstrate that one should find considerable deviations from the commonly accepted phenomenological Esaki-Tsu model. We review which parameter regimes of Bose polarons can be achieved in various atomic mixtures.

\newpage
\chapter{Introduction}

Properties of quantum systems can be modified dramatically when they interact with an environment. One of the first  systems in which this phenomenon has been recognized is an electron moving in a deformable crystal. As originally pointed out by Landau and Pekar \cite{Landau1946,Landau1948}, Fr\"ohlich \cite{Froehlich1954} and Holstein \cite{Holstein1959b,Holstein1959} a single
electron can cause distortion of the ionic lattice that is sufficient to provide a strong modification of the electron motion. Qualitatively this can be understood as electron  moving together with its screening cloud of phonons, which not only renormalizes the effective mass of the electron but can also make its propagation (partially) incoherent. The resulting "dressed" electron has been termed {\it polaron}. Phononic dressing of electrons is important for understanding many solid state materials including ionic crystals and polar semiconductors \cite{Mahan2000,PolaronsAdvMat2007,alexandrov2009advances,Devreese2009}, and even high temperature superconductors \cite{Mishchenko2011,Zhou2008}. It has also been discussed in the context of electrons on the surface of liquid helium \cite{Shikin1973,Jackson1981,Devreese2009}. The importance of polaronic renormalization of electron systems goes beyond academic curiosity. In several technologically relevant materials, such as organic semiconductors used in flexible displays, unusual temperature dependence of electron mobility arises from the strong coupling of electrons to phonons \cite{Gershenson2006,Ortmann2011}. By now the idea of polaronic dressing has been extended far beyond electron-phonon systems and has become an important paradigm in physics. One important example is charge carriers in systems with strong magnetic fluctuations, such as holes doped into antiferromagnetic Mott insulators \cite{Dagotto1994,Nagaev1967} or electrons in magnetic semiconductors \cite{Kaminski2002,Fiete2003}, which can be described as magnetic polarons. Even in the Standard Model of high energy physics the way the Higgs field  produces masses of other particles \cite{Higgs1,Higgs2} is closely related to the mechanism of polaronic dressing. 

Taken in a broader perspective, polarons represent an example  of quantum impurity systems, in which a single impurity introduces interactions (or at least non-trivial dynamics) in the many-body system which is hosting it. Such systems have long been a fertile ground for testing analytical, field-theoretical, and numerical methods for studying quantum many-body systems. Several fundamental paradigms in condensed matter physics have been introduced in the context of quantum impurity models and then extended to other physical problems. The importance of orthogonality catastrophe (OC) has been first realized in connection with the X-ray absorption, in which a  time dependent core hole potential leads to a non-trivial dynamics of a Fermi sea of conduction electrons \cite{Anderson1967}. Subsequent work showed that orthogonality catastrophe also plays an important role in electron transport in mesoscopic systems, such as quantum dots \cite{Geim1994,Abanin2004}. The spin-bath model, describing a two level system interacting with a bath of harmonic oscillators, provided a universal paradigm for the description of quantum systems with dissipation. This class of models not only exhibits phase transitions between quantum and classical behaviors \cite{Leggett1987} but also serves as a framework for describing qubit coupling to the environment \cite{Bar-Gill2012}. The Kondo effect, in which scattering of a localized spin on conduction electrons leads to the formation of a bound state, was the starting point of the application of the renormalization group approach to condensed matter systems. The crucial aspect of most impurity problems is that simple perturbative and mean-field approaches are not sufficient. In cases of OC, spin-bath models, and Kondo effect powerful analytical methods have been employed including bosonization (see e.g. \cite{Gogolin2004,Giamarchi2003}), renormalization group \cite{Wilson1975,Anderson1970,Hewson1993}, slave particles representation \cite{Read1983,Hewson1993}, and Bethe ansatz solution \cite{Wiegmann1983,Andrei1980}. 

By contrast, the toolbox of analytical methods that have been utilized for describing mobile impurities interacting with phonon baths in two and three spatial dimensions has been limited. Even though historically the polaron problem has been formulated first, it turned out to be much less amenable to analytical treatment. An important common feature of the spin-bath, OC, and Kondo models is that they are effectively one-dimensional (1d). This may seem surprising since in the two latter systems one considers impurities inside 3d Fermi gases. However, crucial simplification arises from the fact that impurities are fully localized and interact with conduction band electrons only via $s$-wave scattering. Thus only electrons scattering in the angular momentum $l=0$ channel need to be considered. When the problem is reduced to 1d, it is possible to employ special non-perturbative techniques not available in higher 
dimensions \cite{Tsvelik2001,Giamarchi2003}. On the other hand analysis of mobile impurities in $d>1$ requires dealing with full complexities of higher dimensional problems (the special case of mobile impurities in 1d systems has been discussed by \cite{Castella1993,Tsukamoto1998,Keil2000,Lamacraft2008,Lamacraft2009,Mathy2012,Schecter2012a,Schecter2012,Catani2012,Gamayun2014,Gamayun2014a,Schecter2014,Arzamasovs2014,Kantian2014,Burovski2014,Knap2014,Gamayun2015}). Thus most of the progress in understanding of polarons came from numerical methods such as diagrammatic Monte Carlo analysis \cite{Prokofev1998,Mishchenko2000}. Two exceptions come from recently suggested semi-analytical methods that used non-perturbative techniques of renormalization group analysis \cite{Grusdt2015RG,Grusdt2015DSPP} and Gaussian variational wavefunction \cite{Shchadilova2014} and demonstrated excellent agreement with Monte-Carlo calculations (for a review of earlier work on variational wavefunctions see Ref. \cite{Lakhno2015}). The goal of these lecture notes is to provide a pedagogical introduction into the problem of mobile impurities interacting with phonon baths in three dimensions, with an emphasis on the renormalization group (RG) approach proposed in Ref. \cite{Grusdt2015RG}. We do not attempt to review the full body of literature on the subject of polarons but provide a self-contained overview of basic ideas. The discussion will be centered on a very specific experimental realization of the polaron system: an impurity atom interacting with a Bose-Einstein Condensate (BEC) of ultracold atoms, which we will call BEC polaron. For a pedagogical review with more emphasis on solid state polarons see Ref.\cite{Devreese2013}.

Recent revival of interest in the problem of mobile impurities interacting with a bath of phonons \cite{Tempere2009,Novikov2009,Cucchietti2006,sacha2006self,kalas2006interaction,Bruderer2008,Bruderer2008a,Bruderer2007,Privitera2010,Casteels2011a,Casteels2012,Casteels2011,blinova2013single,Rath2013,Shashi2014RF,Li2014,Levinsen2015,Kain2014,Grusdt2015RG,Shchadilova2014,Christensen2015,Schmidt2015,Ardila2015} comes from the rapid progress in the field of ultracold atoms. This new experimental platform not only allows to create a large variety of polaronic systems with different impurity masses and tunable 
interactions \cite{Catani2012,schmid2010dynamics,Spethmann2012,Fukuhara2013,Scelle2013,Grusdt2015DSPP,Hohmann2015}, but also provides a new toolbox for studying equilibrium and dynamical properties of 
polarons \cite{Astrakharchik2004,Bruderer2008,Bruderer2010,Johnson2011,Schecter2012a,Knap2014,Fukuhara2013,Dasenbrook2013,Scelle2013}.  Most of our discussion will be concerned with the so-called Fr\"ohlich Hamiltonian
\begin{eqnarray}
{\cal H} = \frac{\vec{P}^2}{2M} + \int d^3 \vec{k} ~ \omega_k \a_{\vec{k}}^\dagger \a_{\vec{k}} + \int d^3 \vec{k} ~ V_k e^{ i \vec{k} \cdot \vec{R}} ( \a_{\vec{k}} + \a_{-\vec{k}}^\dagger).
\label{Frolich_first}
\end{eqnarray}
Here $\vec{R}$ and $\vec{P}$ are the impurity position and momentum operators respectively, $M$ is the impurity mass, $\a_k$ is the annihilation operator of phonons at wavevector $\vec{k}$, $\omega_k$ describes the phonon dispersion, and $V_k$ is the matrix element of the impurity-phonon interaction at momentum $k$ (see chapter \ref{Chap:DerivationFroehlich} for detailed discussion of the Hamiltonian \eqref{Frolich_first}). While this model was originally introduced in the context of electron-phonon systems, it has been shown to describe the interaction between impurity atoms and Bogoliubov modes of a BEC \cite{Cucchietti2006,Sacha2006,Tempere2009}, in the regime where the quantum depletion of the condensate around the impurity is small and scattering of phonons at finite momentum can be neglected (see also derivation below). We note that, while we focus on the specific BEC polaron problem, the theoretical methods that we review here -- and in the case of the RG, develop -- are generic and can be applied to a whole variety of polaron models. Readers interested in polaronic systems other than in the cold atoms context should be able to follow our discussion while skipping the details specific to ultracold quantum gases. We note that Fermi polarons have also been a subject of considerable theoretical \cite{prokof2008fermi,punk2009polaron,schmidt2011excitation,Mathy2011,massignan2012polarons,Massignan_review} and experimental \cite{Schirotzek2009,chevy2010ultra,Koschorreck2012,kohstall2012metastability,Zhang2012,Massignan_review} interest in the context of ultracold atoms. They will not be discussed in this review.

\begin{figure}[t!]
\centering
\epsfig{file=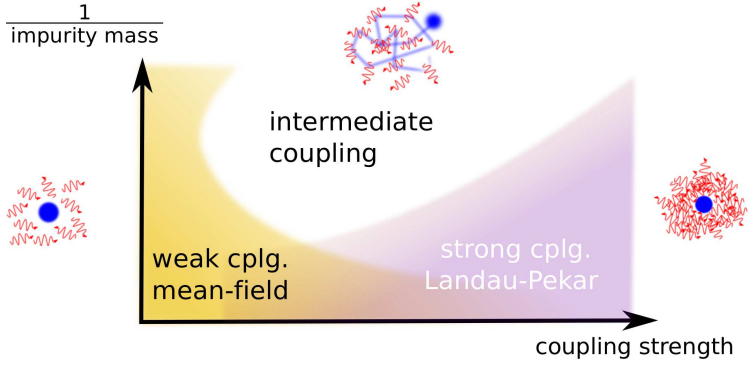, width=0.65\textwidth}
\caption{Qualitative phase diagram of the Fr\"ohlich model, as a function of the impurity-phonon coupling strength ($\alpha$) and the inverse impurity mass ($M^{-1}$). Note that there are no transitions, but rather cross-overs between different regimes. In the weak coupling regime the impurity (blue) carries along a cloud of phonons (red). In the strong coupling regime 
large number of phonons make the impurity atom effectively trapped in the induced potential. In the intermediate coupling regime the impurity can mediate interactions between phonons which results in the build up of correlations between the phonons. A quantitative analysis of this phase diagram in two dimensions was presented in Ref.\cite{Grusdt2015DSPP}.}
\label{fig:QualPhaseDiagFROH}
\end{figure}

Different regimes of the Fr\"ohlich model are illustrated in FIG.\ref{fig:QualPhaseDiagFROH}. This model is characterized by two important parameters: the  impurity mass $M$ and the impurity-phonon dimensionless coupling strength $\alpha$. In the weak coupling regime, the polaron can be thought of as a quasi free impurity carrying a loosely bound screening cloud of phonons. In the strong coupling regime, on the other hand, the screening cloud is so large that the impurity becomes effectively self-trapped in the resulting  potential. These two regimes have been previously described by a weak coupling mean-field theory \cite{Lee1953} and a strong coupling approach based on the adiabatic approximation \cite{Landau1946,Landau1948} respectively. 
The most interesting and challenging regime in FIG.\ref{fig:QualPhaseDiagFROH} comes in the intermediate coupling case which can be realized only for sufficiently light impurities. In this case Landau's picture of an impurity carrying a polarization cloud needs to be refined because the light impurity has an additional role of mediating interactions between phonons. As a result the polaron state develops strong correlations between phonons. An efficient analytical description of such highly-correlated intermediate coupling polarons is challenging and has not been available until recently. Feynman developed a variational all-coupling theory \cite{Feynman1955}, which we will review here. While it has been remarkably accurate in the case of optical phonons \cite{Peeters1985Feynman}, recent numerical quantum Monte Carlo calculations \cite{Prokofev1998,Vlietinck2015} have shown that Feynman's theory does not provide an adequate description at intermediate coupling strengths when phonons with a Bogoliubov dispersion are considered. The advantage of Feynman's approach is simultaneously its main limitation. It approximates the polaron ground state using only two variational parameters. This explains why the approach does not provide an adequate description of polarons when phonon dispersions with multiple energy scales are considered. By contrast the RG method that we review in these lecture notes has been demonstrated to be in excellent agreement with Monte Carlo calculations. In these lecture notes we will introduce the RG approach and discuss its application to the BEC polaron problem. We emphasize again that the RG method is valid for generic polaron models and is particularly powerful in systems with multiple energy scales in the phonon dispersion and/or impurity phonon coupling. In this sense, the RG approach complements Feynman's variational theory and Monte Carlo calculations. While it is not as easy for analytical calculations as Feynman's ansatz, it is far more accurate but still does not require heavy numerics involved in full Monte Carlo calculations (see e.g. Ref. \cite{Mishchenko2005} for a review). Most importantly the RG approach provides a lot of physical insight into the nature of the polaronic states at intermediate coupling, not accessible in Monte Carlo.

One of the subtle issues special to interacting systems of ultracold atoms is the question of ultra-violet (UV) divergences. These have to be regularized and the answer should be expressed in terms of the physical scattering length independently of the high momentum cutoff $\Lambda_0$. We discuss this issue explicitly in the case of Bose polarons and show that the Hamiltonian \eqref{Frolich_first} has two types of divergences. The first one is linear in $\Lambda_0$ and manifests itself already at the mean-field level \cite{Shashi2014RF,Rath2013} (this has been recognized in earlier work, see e.g. Refs. \cite{Tempere2009,BeiBing2009}). The other divergence is logarithmic in $\Lambda_0$ \cite{Grusdt2015RG} and appears only when higher order corrections (in the interactions $V_k$) are included (see also recent work \cite{Christensen2015}). 
We discuss how the power-law divergence is cancelled when one includes mean-field terms not written explicitly in \eqref{Frolich_first} and analyzes modifications to the two-particle Lippmann-Schwinger equation due to many-body physics, We also explain that the physical origin of the logarithmic divergence. As we discuss below, this divergence arises from exciting phonon modes at high energies. Traditional approaches such as mean-field, Feynman variational ansatz, and Landau-Pekar strong coupling expansion have a built-in suppression of the high energy modes, which precludes them from capturing the log-divergence. One needs more sophisticated techniques, such as renormalization group approach discussed in these lecture notes or Gaussian variational wavefunctions introduced in \cite{Shchadilova2014}, to capture this physics.\\

These lecture notes are organized as follows. In Chap.\ref{Chap:DerivationFroehlich} we begin by deriving the Bogoliubov-Fr\"ohlich polaron Hamiltonian describing an impurity in a BEC. Chap.\ref{chap:CommonApproaches} is devoted to a review of common theoretical approaches to the Fr\"ohlich model, with special emphasis on the application to the BEC polaron. In Chap.\ref{chap:RGapproach} we give a detailed pedagogical introduction to the RG approach for describing intermediate coupling polarons. The RG derivations of ground state polaron properties presented in this chapter have been presented previously in the Ph.D. thesis of one of us (F.G.).
In Chap.\ref{chap:UVregularization} we discuss the important issue of proper UV regularization of the polaron energies, specific to systems of ultracold atoms. Readers who are interested in more generic polaronic models can skip this chapter. We discuss experimentally relevant parameters for quantum gases in Chap.\ref{chap:ResultsForExperiments}, where we also present previously unpublished numerical results of the RG method. In Chap.\ref{chap:polaronBO} we present examples of non-equilibrium BEC polaron problems, and we discuss in particular the problem of polaron Bloch oscillations in a lattice potential. Finally we close with an outlook in Chap.\ref{Chap:Outlook}.

\newpage
\chapter{Derivation of the Fr\"ohlich Hamiltonian}
\label{Chap:DerivationFroehlich}

In this section we introduce a microscopic model that describes an impurity atom immersed in a BEC and discuss why, in a wide regime of parameters, it is equivalent to  
the celebrated Fr\"ohlich model. The microscopic model is presented in Sec. \ref{Sec:microHam}, with derivation of  the effective Fr\"ohlich Hamiltonian given in Secs. \ref{Sec:FroehlichDerivation}, \ref{Sec:FroehlichDerivationFull}. We examine parameters entering the Fr\"ohlich model and motivate the definition of the dimensionless coupling constant $\alpha$ in Sec. \ref{Sec:CharcteristicScales}. In Sec. \ref{Sec:LippmannSchwinger} we review the connection between the effective impurity-boson interaction strength $g_{\IB}$ and the corresponding scattering length $a_{\rm IB}$ that arises from the Lippmann-Schwinger equation. We will rely on this discussion in subsequent chapters where we discuss UV divergences of the Fr\"ohlich model.

\section{Microscopic Hamiltonian: Impurity in a BEC}
\label{Sec:microHam}

\begin{figure}[b]
\centering
\epsfig{file=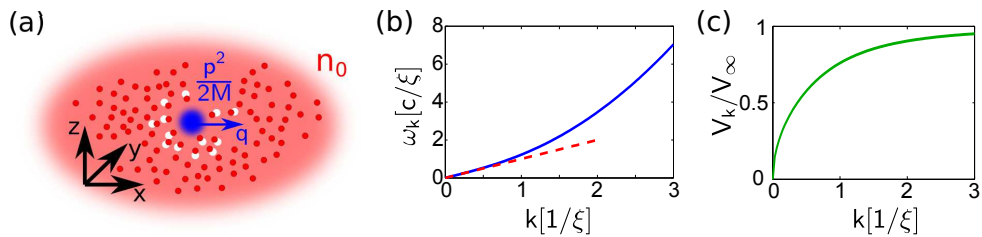, width=0.85\textwidth}
\caption{We consider a single impurity immersed in a three dimensional homogeneous BEC (a). The total momentum $\vec{q}$ of the system is conserved, and the interaction of the impurity with the Bogoliubov phonons of the BEC leads to the formation of a polaron. The dispersion relation of the Bogoliubov phonons in the BEC, with an acoustic behavior $c k$ (dashed) for small momenta, is shown in (b) and their scattering amplitude $V_k$ with the impurity in (c).}
\label{fig:SetupPolarons}
\end{figure}

The starting point for our discussion is a many-body system that consists of impurity (I) atoms immersed in a  $d$-dimensional BEC of bosons (B), see FIG. \ref{fig:SetupPolarons} (a). The density of the BEC will be denoted by $n_0$. For concreteness, in the rest of this section we will restrict ourselves to three dimensional systems, $d=3$, but we return to the general case later in these lecture notes. We describe host bosons using the second quantized field operator $\hat{\phi}(\vec{r})$ and denote their mass by $m_\text{B}$. Impurity atoms are described 
by the field operator $\hat{\psi}(\vec{r})$ and we denote their mass by $M$. 

In the following we will restrict our analysis to the case of small impurity concentration so that the average distance between impurities is much larger than the polaron size and the inter-boson distance $n_0^{-1/3}$. This allows us to consider a conceptually simpler single-impurity problem. Later on we will also find it convenient to switch to the first quantized representation of the impurity atom, but for now we stay with the second quantized representation of both BEC and impurity atoms. 

We introduce contact pseudopotentials $g_\IB$ and $g_\BB$ to describe impurity/boson and boson/boson interactions (for pedagogical introduction into pseudpotentials see Ref. \cite{Pethick2008}) and find the microscopic many-body Hamiltonian (we set $\hbar = 1$)
\begin{multline}
 \H = \int d^3\vec{r} ~ \hat{\phi}^{\dagger}(\vec{r}) \left[ -\frac{\nabla^2}{2 m_\B} 
+ \frac{g_{\text{BB}}}{2} \hat{\phi}^\dagger(\vec{r}) \hat{\phi}(\vec{r})\right] \hat{\phi}(\vec{r}) +
 \\ + \int d^3\vec{r} ~ \hat{\psi}^{\dagger}(\vec{r}) \left[ -\frac{\nabla^2}{2 M} 
+ g_{\IB} \hat{\phi}^{\dagger}(\vec{r}) \hat{\phi}(\vec{r}) \right] \hat{\psi}(\vec{r}).
 \label{eq:Hmicro}
\end{multline}
In writing the last equation we neglected external potentials for the impurity and Bose atoms such as coming from the confining parabolic potentials. We assume that these potentials are smooth and shallow enough so that one can perform analysis for a uniform system and then average over density distribution. Later we will also consider extensions of the  equation (\ref{eq:Hmicro}) to systems with an optical lattice.

\section{Fr\"ohlich Hamiltonian in a BEC}
\label{Sec:FroehlichDerivation}
We now show that Eq.\eqref{eq:Hmicro} can be reduced to an effective Fr\"ohlich polaron Hamiltonian that describes interaction of impurity atoms with the collective phonon excitations of the BEC \cite{Tempere2009}.
Before presenting a detailed derivation we give a short preview. Using Bogoliubov approximation to describe the BEC and keeping the lowest order terms for non-condensed bosonic atoms, we  arrive at
\begin{multline}
\H = g_\IB n_0 + \int d^3 \vec{k} ~ \Biggl[  \int d^3 \vec{r} ~ \psd(\vec{r}) \ps(\vec{r}) V_{\vec{k}} e^{i \vec{k}\cdot \vec{r}} \l \a_{\vec{k}} + \ad_{-\vec{k}} \r  + \omega_{k} \ad_{\vec{k}} \a_{\vec{k}}  \Biggr] - \\
- \int d^3 \vec{r}~ \psd(\vec{r})  \frac{\vec{\nabla}^2}{2 M} \ps(\vec{r}) = g_{\rm IB} n_0 + \H_{\rm FROH}.
\label{eq:HFroh}
\end{multline}
The second term in Eq.\eqref{eq:HFroh} has Bogoliubov phonons described by creation and annihilation operators $\a^{\dagger}_{\vec{k}}$ and $\a_{\vec{k}}$. They obey canonical commutation relations (CCRs), $[\a_{\vec{k}},\ad_{\vec{k}'}]=\delta(\vec{k}-\vec{k}')$. The corresponding Bogoliubov dispersion is given by
\begin{equation}
\omega_k = c k \sqrt{1+\frac{1}{2} \xi^2 k^2},
\label{eq:BogoliubovDispersion}
\end{equation}
and is shown in FIG.\ref{fig:SetupPolarons} (b). In addition there are impurity-phonon interactions, which are characterized by the scattering amplitude
\begin{equation}
V_{\vec{k}} = \sqrt{n_0} (2 \pi)^{-3/2} g_\IB \l \frac{(\xi k)^2}{2 + (\xi k)^2} \r^{1/4},
 \label{eq:Vktilde}
\end{equation}
which is plotted in FIG.\ref{fig:SetupPolarons} (c). 
We will refer to the model (\ref{eq:HFroh}) as the \emph{Bogoliubov-Fr\"ohlich} Hamiltonian. Note that we set the overall energy to be zero in the absence of impurities 
$E(g_\IB=0)=0$.

Bruderer et al. \cite{Bruderer2007} pointed out an important constraint on the validity of 
\eqref{eq:HFroh}. This approximation works only when condensate depletion in the vicinity of the impurity is small, resulting in a condition on the interaction strength,
\begin{equation}
 |g_\IB| \ll 4 c \xi^2.
 \label{eq:condBogoFroh}
\end{equation}
Here  $\xi$ is the healing length and $c$ is the speed of sound  in the BEC. They are given by
\begin{equation}
\xi = 1 / \sqrt{2 m_{\text{B}} g_{\text{BB}} n_0}, \qquad c = \sqrt{g_{\text{BB}} n_0 / m_{\text{B}}}
\label{eq:BECprop}
\end{equation} 
Additional discussion of experimental conditions required for the model (\ref{eq:HFroh}) to be applicable can be found in Refs.\cite{Grusdt2015RG,Hohmann2015}.

We note that applicability of the Fr\"ohlich model to describe Bose polarons is still a subject of debate (see Ref. \cite{Ardila2015,Christensen2015} and discussion of the importance of two phonon terms in section \ref{sec:GSenergyPolaron}). However this model is interesting for its own sake because of its connection to many other physical systems. Thus it will be the focus of our discussion in these lecture notes.

\section{Microscopic derivation of the Fr\"ohlich model}
\label{Sec:FroehlichDerivationFull}
Next we turn to the formal derivation of the Fr\"ohlich Hamiltonian \eqref{eq:HFroh} from the microscopic model \eqref{eq:Hmicro}. Note that, although this section is elementary, we present a detailed discussion here for completeness. We start by considering periodic boundary conditions for bosons (we assume period $L$ in every direction) and take the limit $L \rightarrow \infty$ in the end. Thus the boson field operator can be expressed in terms of a discrete set of bosonic modes $\d_{\vec{k}}^{(\dagger)}$,
\begin{equation}
\hat{\phi}(\vec{r}) =L^{-3/2} \sum_{\vec{k}} e^{i \vec{k} \cdot \vec{r}} \d_{\vec{k}}.
\end{equation}
The bosonic field obeys CCRs, $[\hat{\phi}(\vec{r}),\hat{\phi}^\dagger(\vec{r}')]=\delta(\vec{r} - \vec{r}')$, hence the discrete set of operators $\d_{\vec{k}}$ obey CCRs, $[\d_{\vec{k}},\dd_{\vec{k}'}] = \delta_{\vec{k},\vec{k}'}$.

To describe the BEC and its elementary excitations, we use standard Bogoliubov theory (see e.g. \cite{Pitaevskii2003,Pethick2008}) and introduce a macroscopic population of the state with zero momentum, $\d_0 = \sqrt{N_0}$. Here $N_0$ is the extensive number of atoms inside the condensate. The relation between Bogoliubov phonons $\a_{\vec{k}}^{(\dagger)}$ and original atomic operators is given by $\d_{\vec{k}} = \cosh \theta_{\vec{k}} \a_{\vec{k}} - \sinh \theta_{\vec{k}} \ad_{-\vec{k}}$. Within Bogoliubov theory the boson Hamiltonian has a simple quadratic form
\begin{equation}
\H_{\text{B}} = \sum_{\vec{k}} \omega_k \ad_{\vec{k}} \a_{\vec{k}}.
\end{equation}
Parameters of the Bogoliubov transformation are given by
\begin{equation}
\cosh \theta_{\vec{k}} = \frac{1}{\sqrt{2}} \sqrt{\frac{k^2/2 m_{\text{B}} + g_{\text{BB}} n_0 }{ \omega_k } +1}, \qquad \sinh \theta_{\vec{k}} = \frac{1}{\sqrt{2}} \sqrt{\frac{k^2/2 m_{\text{B}} + g_{\text{BB}} n_0 }{ \omega_k } - 1}.
\label{eq:parametersBogoliubov}
\end{equation}

\begin{figure}[t]
\centering
\epsfig{file=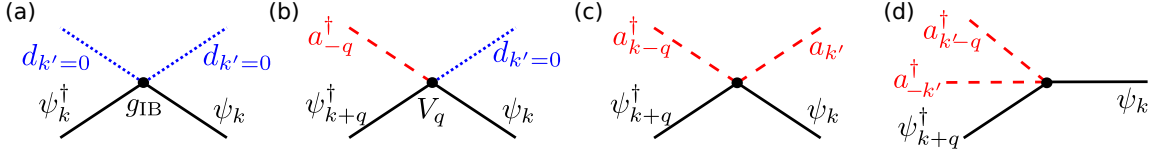, width=\textwidth}
\caption{Different vertices describing the BEC-MF shift of the impurity (a) and impurity-phonon interactions (b-d). Solid (black) lines denote the impurity field, dotted (blue) lines the macroscopic condensate (density $n_0$) and dashed (red) lines the Bogoliubov phonons. The last two vertices (c,d) are neglected to obtain the Fr\"ohlich Hamiltonian describing an impurity in a BEC.}
\label{fig:IBvortices}
\end{figure}

Now we can derive the effective boson-impurity interaction. We rewrite the microscopic density-density contact interaction in the Hamiltonian as
\begin{equation}
\H_{\text{IB}} = g_\IB \int d^3\vec{r} ~ \psd(\vec{r}) \ps(\vec{r}) \hat{\phi}^\dagger (\vec{r}) \hat{\phi}(\vec{r}) = \frac{g_\IB}{L^3} \int d^3\vec{r} ~ \psd(\vec{r})  \ps(\vec{r}) \sum_{\vec{k},\vec{k}'}  e^{i \l \vec{k} - \vec{k}' \r \cdot \vec{r} } \dd_{\vec{k}} \d_{\vec{k}'}.
\end{equation}
Taking into account the macroscopic occupation of the condensate (i.e. the state at $k=0$) gives
\begin{equation}
\H_{\text{IB}} = \frac{g_\IB}{L^3} \int d^3\vec{r} ~ \psd(\vec{r}) \ps(\vec{r}) \left[ N_0 + \sum_{\vec{k} \neq 0} \l e^{i \vec{k} \cdot \vec{r}} \sqrt{N_0} \dd_{\vec{k}} + \hc \r + \sum_{\vec{k},\vec{k}' \neq 0}  e^{i \l \vec{k} - \vec{k}' \r \cdot \vec{r} } \dd_{\vec{k}} \d_{\vec{k}'} \right].
\label{eq:HIBdiscrecte}
\end{equation}
FIG.\ref{fig:IBvortices} illustrates different processes in the Hamiltonian \eqref{eq:HIBdiscrecte} after expressing bosonic operators $\d_{\vec{k}}$ in terms of Bogoliubov phonons.

In the spirit of the Bogoliubov approximation we reduce equation (\ref{eq:HIBdiscrecte}) by keeping only the lowest order terms in $\d_{\vec{k} \neq 0}$, i.e. we disregard the last term in this equation. This follows the usual assumption that the condensate density $n_0$ is much larger than the density of excited phonons 
$n_\text{ph}$.\footnote{Note that although $N_0$ is a macroscopic number and thus 
$\sqrt{N_0} \gg 1$ in the thermodynamic limit, this by itself does not justify dropping the two-boson vertices $\sim \dd_{\vec{k}} \d_{\vec{k}'}$. Eq.\eqref{eq:condFroh} provides the relevant condition.}. We thus obtain the condition
\begin{equation}
\frac{n_\text{ph}}{n_0} \ll 1.
\label{eq:condFroh}
\end{equation}
We will return to this condition later and derive a simpler expression below. 

After expressing the original atomic operators $\d_{\vec{k}}$ using phonon operators $\a_{\vec{k}}$ and disregarding the last term in (\ref{eq:HIBdiscrecte}) we obtain the  Fr\"ohlich Hamiltonian with discreet values of momenta,
\begin{equation}
\H = \sum_{\vec{k}} \omega_k \ad_{\vec{k}} \a_{\vec{k}} + g_\IB \frac{N_0}{L^3} +
 \int d^3\vec{r} ~ \psd(\vec{r}) \l - \frac{\nabla^2}{2 M} + \sum_{\vec{k}\neq 0} V_k^{\text{disc}} e^{-i\vec{k} \cdot \vec{r}} \l \a_{\vec{k}} +  \ad_{-\vec{k}} \r \r \ps(\vec{r}). 
 \label{eq:HfrohDiscrete}
\end{equation}
The scattering amplitude $V_k^\text{disc}$ is determined by Bogoliubov parameters from Eq.\eqref{eq:parametersBogoliubov} \cite{Tempere2009}
\begin{equation}
V_{k}^{\text{disc}} = \frac{g_\IB}{ L^{3}} \sqrt{N_0} \l \frac{k^2 /2 m_{\text{B}} }{2 g_{\text{BB}} n_0 + k^2 /2 m_{\text{B}}}\r^{1/4}.
\end{equation}
Note that $V_{k}^{\text{disc}}$ by itself is not well-behaved in the limit $L \rightarrow \infty$. However summations over $\vec{k}$ will be well defined.

As the last step we  consider thermodynamic limit by taking $L \rightarrow \infty$, while the BEC density $N_0/L^3 = n_0$ remains fixed. To do so, continuous operators $\a(\vec{k})$ are defined in the usual way, which are no longer dimensionless but acquire the dimension $L^{3/2}$.
\begin{equation}
\a(\vec{k}) := \l \frac{L}{2 \pi} \r^{3/2} \a_{\vec{k}}, \qquad [\a(\vec{k}) , \ad(\vec{k}')] = \delta(\vec{k} - \vec{k}'), \qquad \sum_{\vec{k}} = \frac{L^3}{\l 2 \pi \r^3} \int d^3 \vec{k}.
\end{equation}
 As a consequence, the scattering amplitude $V_{k}^{\text{disc}}$ becomes modified and the expression \eqref{eq:Vktilde} for $V_k$ is obtained, which is well-behaved in thermodynamic limit. From the discrete Fr\"ohlich Hamiltonian \eqref{eq:HfrohDiscrete} the continuum version Eq.\eqref{eq:HFroh} is obtained. (Note that in Eq.\eqref{eq:HFroh} we denoted the continuous operators $\a(\vec{k})$ by $\a_{\vec{k}}$ again for simplicity of notations.)

Finally we use the Fr\"ohlich Hamiltonian \eqref{eq:HFroh} to simplify the condition in Eq. \eqref{eq:condFroh}. In order to estimate the phonon density in the vicinity of the impurity, we consider the effect of quantum depletion due to one-phonon vertices in Eq.\eqref{eq:HIBdiscrecte}, see also FIG.\ref{fig:IBvortices} (b). To this end we calculate the phonon number $N_\text{ph}$ due to these processes perturbatively (in the interaction strength $g_\IB$) from the Fr\"ohlich Hamiltonian,
\begin{equation}
N_\text{ph} = \int d^3 \vec{k} ~ \l \frac{V_k}{\omega_k} \r^2 = \frac{n_0 g_\IB^2}{\sqrt{2} \pi^2 c^2 \xi}.
\end{equation}
To understand the length scales of the phonon screening cloud, let us consider the asymptotic expressions for $V_k$ and $\omega_k$,
\begin{equation}
V_k \sim \begin{cases}
k^0=1  & k \gtrsim 1/\xi \\
\sqrt{k} & k \lesssim 1/\xi
\end{cases},
\qquad 
\omega_k \sim \begin{cases}
k  & k \gtrsim 1/\xi \\
k^2 & k \lesssim 1/\xi
\end{cases}.
\end{equation}
Thus for small $k\lesssim 1/\xi$ the phonon density $k^2 V_k^2 / \omega_k^2 \sim k$, whereas for large $k \gtrsim 1/\xi$ it scales like $k^2 V_k^2 / \omega_k^2 \sim k^{-2}$. Thus most of the depleted phonons have momentum $k \approx 1/ \xi$, and the phonon cloud has a spatial extend on the order of $\xi$. Consequently we can estimate $n_\text{ph} \approx N_\text{ph} \xi^{-3}$ and the condition \eqref{eq:condFroh} is equivalent to
\begin{equation}
|g_\IB| \ll  \pi 2^{1/4} c ~ \xi^2 = 3.736... ~ c ~ \xi^2,
\label{eq:condgIB}
\end{equation}
which is essentially the same condition as Eq.\eqref{eq:condBogoFroh} suggested in Ref. \cite{Bruderer2007}.

\section{Characteristic scales and the polaronic coupling constant}
\label{Sec:CharcteristicScales}
To get a qualitative understanding of the physical content of the Bogoliubov-Fr\"ohlich Hamiltonian, it is instructive to work out the relevant length and energy scales of different processes. The comparison of different energy scales will then naturally lead to the definition of the dimensionless coupling constant $\alpha$ characterizing Fr\"ohlich polarons. 

To begin with, we note that the free impurity Hamiltonian is scale-invariant and no particular length scale is preferred. This is different for the many-body BEC state, where boson-boson interactions give rise to the healing length $\xi$ as a characteristic length scale. This can be seen for instance from the Bogoliubov dispersion, which has universal low-energy behavior $\omega_k \sim c k$ and high-energy behavior $\omega_k \sim k^2 /2 m_\B$, with a cross-over around $k \approx 1/\xi$. Note that the same length scale appears in the impurity-boson interactions, where $V_k$ saturates for momenta $k \gtrsim 1/\xi$. Thus we find it convenient to measure lengths in units of $\xi$. 

Because of the scale invariance, no particular energy scale can be assigned to the free impurity. We saw that the Bogoliubov dispersion is scale invariant asymptotically for both $k \rightarrow 0$ and $\infty$, but at the characteristic energy 
\begin{equation}
E_\text{ph} \simeq c / \xi
\label{eq:defEph}
\end{equation}
its behavior is non-universal. The impurity-phonon interaction in the Fr\"ohlich model, on the other hand, is characterized by an energy scale, which can be derived as follows. Consider a localized impurity ($M \rightarrow \infty$), such that $\psd(\vec{r}) \ps(\vec{r}) \rightarrow \delta(\vec{r})$ can formally be replaced by a delta-function in the microscopic Hamiltonian \eqref{eq:Hmicro}. To estimate the magnitude of the relevant terms in $\hat{\phi}^\dagger \hat{\phi}$, we recall that the interaction terms in the Fr\"ohlich Hamiltonian \eqref{eq:HFroh} 
contain both the condensate and finite momentum atoms. We take $\d_{k=0} \rightarrow \sqrt{n_0}$ and $\d_{k \neq 0} \rightarrow \sqrt{n_{\ph}^0} \sim \xi^{-\frac{3}{2}}$, where we used $\xi$ as a natural lengthscale for phonons. Note that we used $n_\ph^0$ as an intrinsic density of the BEC quantum depletion and not $n_\ph$ as discussed above in Eq.\eqref{eq:condgIB}. The typical impurity-phonon interactions scale like $g_\IB \sqrt{n_0 n_\text{ph}^0}$ and we find 
\begin{equation}
E_\IB \sim g_\IB \sqrt{n_0 \xi^{-3}}.
\label{Eq_E_IB}
\end{equation} 
We point out that equation \eqref{Eq_E_IB} was obtained without momentum integration and thus should be understood as characteristic interaction scale per single $k$-mode, just like
\eqref{eq:defEph} defines $E_\text{ph}$ for a typical phonon with $k \sim \xi^{-1}$.

Now we can define a single dimensionless coupling constant $\alpha$ that will characterize the impurity boson interaction strength. While it may seem natural to define it as a ratio 
$E_\IB / E_\ph$, we observe that the Fr\"ohlich Hamiltonian is invariant under sign changes of $g_\IB$ (up to an overall energy shift due to the BEC-MF term $g_\IB n_0$). Hence it is more common to define the dimensionless coupling constant as $\alpha \simeq (E_\IB / E_\ph )^2$. The latter can be expressed using the impurity-boson scattering length $a_\IB$ \cite{Tempere2009}. In the limit $M \rightarrow \infty$ the reduced mass for impurity-boson pair is given by $m_\rd = (1/M + 1/m_\B)^{-1} = m_\B$. Using the simplest (Born) relation between the interaction strength $g_\IB$ and the universal scattering length, $a_\IB = m_\rd g_\IB / ( 2 \pi ) + \mathcal{O}(g_\IB^2)$ (this relation is reviewed in the next section together with higher order corrections to it) we obtain $E_\IB / E_\ph = \sqrt{8 \pi^2} a_\IB \sqrt{n_0 \xi}$.  This gives us 
\begin{equation}
\alpha := 8 \pi a_\IB^2 n_0 \xi
\label{eq:defAlpha}
\end{equation}
introduced previously in Ref. \cite{Tempere2009}. In the infinite mass case, $M \to \infty$ the last equation reduces to $\alpha = (E_\IB / E_\ph )^2 / \pi$, where an additional numerical factor $\pi^{-1}$ was introduced to recover the result \eqref{eq:defAlpha} of Ref.\cite{Tempere2009}. 

We emphasize again that in the following we use the convention that $\alpha$ is defined through the scattering length $a_\IB$ as in Eq.\eqref{eq:defAlpha} independently of the impurity mass\footnote{When $M < \infty$ the relation of $\alpha$ to $E_\IB/E_\ph$ is slightly modified because $E_\IB$ depends explicitly on $g_\IB$, which relates to $a_\IB$ via $m_\rd$, thus depending on $M$.}. We also note that the coupling constant can be formulated in a slightly different -- but equivalent -- way as \cite{Tempere2009}
\begin{equation}
\alpha = \frac{a_\IB^2}{ a_\BB \xi }.
\end{equation}
While the simplicity of this expression is appealing, Eq.\eqref{eq:defAlpha} is more clear in regards to the physics of the Fr\"ohlich Hamiltonian. Other non-equivalent definitions of the coupling constant have been given in Ref.\cite{Cucchietti2006,blinova2013single,Grusdt2014BO}.

We also note that a natural mass scale in the problem comes from BEC atoms $m_\B$. Hence in Fig. \ref{fig:QualPhaseDiagFROH} we distinguish between light, $M < m_\B$, and heavy, $M> m_\B$, impurities cases.

\section{Lippmann-Schwinger equation}
\label{Sec:LippmannSchwinger}
In this section we discuss how the effective interaction strength $g_\IB$ can be related to the experimentally accessible scattering length $a_\IB$. The following discussion is specific to ultracold atoms, and can be omitted by readers interested in polarons in other contexts. In Eq.\eqref{eq:Hmicro}, $g_\IB$ is related to $a_\IB$ through the Lippmann-Schwinger equation. Later we will see that the ground state energy of the Bogoliubov Fr\"ohlich Hamiltonian \eqref{eq:HFroh} contains terms scaling with $a_\IB^2$ (and higher orders). Thus to obtain reliable results, we need to evaluate the BEC MF shift $g_\IB n_0$ at least to order $a_\IB^2$, which requires us to consider higher order solutions of the Lippmann-Schwinger equation below. 

Experimentally, the interaction strengths $g_\IB$ and $g_\BB$ can not be measured directly because they are only convenient parameters in the simplified model \eqref{eq:Hmicro}. The actual inter-atomic potentials are far more complicated, and in many cases not even well known. Nevertheless there is a fundamental reason why these interactions can be modeled by simple contact interactions, specified by a parameter $g$ at a given momentum cutoff $\Lambda_0$. When performing two-body scattering experiments, the measured scattering amplitude $f_k$ takes a \emph{universal} form in the low-energy limit (see e.g. \cite{Bloch2008} and references therein),
\begin{equation}
f_k = - \frac{1}{1/a + i k}.
\label{eq:fkUniversal}
\end{equation}
It is determined solely by the scattering length $a$ (here we restrict ourselves to $s$-wave scattering for simplicity). Since scattering amplitudes can be directly accessed in cold atom experiments, the scattering lengths $a_\IB$ and $a_\BB$ fully characterize the interactions between low-energetic atoms. 

To connect the value of a scattering length $a$ to the contact interaction strength $g$ in the simplified model, the following argument can be used. The form of the scattering amplitude given by equation \eqref{eq:fkUniversal} is universal in the low-energy limit, regardless of the details of a particular interaction potential. The complicated microscopic potential $V_\text{mic}(x)$ may thus be replaced by \emph{any} pseudopotential $V_\text{pseudo}(x)$, as long as it reproduces the same universal scattering length $a$. In particular, we can choose a simple 3D contact interaction potential $V_\text{pseudo}(x) = g \delta (\vec{x})$, where the interaction strength $g$ is chosen such that it gives the correct scattering length
\begin{equation}
a = - \lim_{k \to 0} f_k(g).
\label{eq:Connection_ga}
\end{equation}
This equation implicitly defines $g(a)$ as a function of $a$ (for a fixed $\Lambda_0$).

To calculate the interaction strength $g_\IB$ from the scattering length $a_\IB$, which is assumed to be known from a direct measurement, we now need to calculate the scattering amplitude $f_k(g_\IB)$ by solving the two-body scattering problem with the pseudopotential $g_\IB \delta(\vec{r})$. The easiest way to accomplish this is to use the $T$-matrix formalism, see e.g. \cite{Ballentine1998}, where the scattering amplitude is given by
\begin{equation}
f_{k} = - \frac{m_\rd}{2 \pi} \bra{k,s} \hat{T} \l \omega=\frac{k^2}{2 m_\rd} \r \ket{k,s}.
\end{equation}
Here $\ket{k,s}$ denotes a spherical $s$-wave (zero angular momentum $\ell=0$ of the relative wavefunction) with wave vector $k$, and we introduced the reduced mass $m_\rd$ of the two scattering partners. In the case of the impurity-boson scattering, $m_\rd^{-1} = M^{-1} + m_\B^{-1}$, whereas for the boson-boson scattering we would have $m_\rd = m_\B/2$. 

To calculate the $T$-matrix, we need to solve the Lippmann-Schwinger equation (LSE),
\begin{equation}
\hat{T} = \l 1 + \hat{T} \hat{G}_0 \r \hat{V},
\end{equation}
where the free propagator is given by
\begin{equation}
\hat{G}_0(\omega) = \int d^3 \vec{k} ~ \frac{\ket{\vec{k}}\bra{\vec{k}}}{\omega - \frac{k^2}{2 m_\rd} + i \epsilon}.
\label{eq:G0LSE}
\end{equation}
To zeroth order the $T$-matrix is given by the scattering potential, $\hat{T}_0 = \hat{V}$ where $\hat{V} = \int d^3 \vec{r} ~ \ket{\vec{r}} g_\IB \delta(\vec{r}) \bra{\vec{r}}$, and all higher orders can be solved by a Dyson series.

From the first order LSE we can derive relations \cite{Bloch2008}
\begin{equation}
a_\IB = \frac{m_\rd}{2 \pi} g_\IB, \qquad \qquad a_\BB = \frac{m_\B}{4 \pi} g_\BB,
\label{eq:aIBgIBaBBgBB}
\end{equation}
which we will use to calculate the scattering amplitude $V_k$ in Eq.\eqref{eq:Vktilde} as well as the BEC properties in Eq.\eqref{eq:BECprop}. Later we will show that proper regularization of the polaron energies requires the solution of the LSE to second-order in $g_{\rm IB}$ \cite{Tempere2009,Rath2013,Shashi2014RF}. Hence we will now obtain the relation between $a_\IB$ and $g_\IB$ valid up to second order. Formally the second order result is given by $\hat{T} = \hat{T}_0 + \hat{T}_0 \hat{G}_0 \hat{T}_0$. Thus using expressions for $\hat{T}_0$ and $\hat{G}_0$ one easily finds
\begin{align}
a_\IB  &= \frac{g_\IB}{2 \pi} m_\rd - \frac{2 g_{\IB}^2}{(2\pi)^3} m_\rd \int_0^{\Lambda_0} dk ~ k^2  \frac{1}{k^2 / 2 m_\rd} + \mathcal{O}(a_\IB^3) \\
 &=  \frac{g_\IB}{2 \pi} m_\rd - \frac{4 g_{\IB}^2}{(2\pi)^3} m_\rd^2 \Lambda_0 + \mathcal{O}(a_\IB^3).
 \label{eq:aIBgIBscndOrder}
\end{align}
Note that the integral on the right-hand-side of this equation is UV-divergent, and in order to regularize it we introduced a sharp momentum cut-off at $\Lambda_0$. We will return to this important issue in chapter \ref{chap:UVregularization}.

\newpage
\chapter{Overview of common theoretical approaches}
\label{chap:CommonApproaches}

In this chapter, we review common theoretical approaches that have been applied to solve polaron Hamiltonians (like the one due to Fr\"ohlich) in the limits of weak and strong coupling. In Sec.\ref{sec:PertApproach} we start by giving an overview of commonly used perturbative methods. In Sec.\ref{sec:LangFirsov} we discuss the solution of the BEC polaron problem at infinite impurity mass. Next, in Sec.\ref{sec:LLP}, we review a powerful exact method developed by Lee, Low and Pines (LLP) which utilizes the conservation of the polaron momentum $\vec{q}$. It provides a useful starting point for subsequent approximate mean-field treatment, which we discuss in Sec.\ref{sec:MFtheory}. The influential treatment of the strong coupling regime, due to Landau and Pekar, will be presented in Sec.\ref{sec:LandauPekar}. In Sec.\ref{sec:Feynman} we review Feynman's variational path-integral approach, which is considered to be the first successful all-coupling polaron theory. Finally, in Sec.\ref{sec:quantMC}, we discuss numerical quantum Monte Carlo approaches to the polaron problem. A related discussion can be found in the lecture notes by J. Devreese \cite{Devreese2013}, focusing on solid state problems however.

\section{Perturbative approaches}
\label{sec:PertApproach}

\begin{figure}[b!]
\centering
\epsfig{file=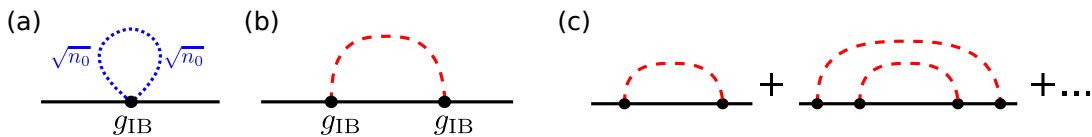, width=0.95\textwidth}
\caption{In (a) and (b) the two Feynman diagrams contributing to the self-energy of the interacting impurity in the Born-approximation are shown. (a) corresponds to the BEC mean-field shift, where $n_0$ is the BEC density. (b) corresponds to the leading-order contribution from the Fr\"ohlich Hamiltonian. In (c) the non-crossing Feynman diagrams are shown which are taken into account in the self-consistent Born approximation.}
\label{fig:FeynmanDiagsPert}
\end{figure}

In the weak coupling regime the polaron problem can be solved using many-body perturbation theory. We will now discuss two perturbative approaches, the first based on wavefunctions (see Sec.\ref{sec:RayleighSchroedinger}) and the second on Green's functions (see Sec.\ref{sec:GreensFunctions}). In both cases there are two relevant Feynman diagrams, which are shown in FIG.\ref{fig:FeynmanDiagsPert}. The BEC mean-field shift $g_{\rm IB} n_0$ in FIG.\ref{fig:FeynmanDiagsPert} (a) is already included in the Fr\"ohlich Hamiltonian \eqref{eq:HFroh}.

\subsection{Rayleigh-Schr\"odinger perturbation theory}
\label{sec:RayleighSchroedinger}
We start from a standard perturbative expansion of the Fr\"ohlich Hamiltonian \eqref{eq:HFroh} in powers of the interaction strength $\alpha$. In the context of many-body theory this is also referred to as Rayleigh-Schr\"odinger perturbation theory, see e.g. Ref.\cite{Hubac2010} for a review.

For $\alpha=0$ the non-interacting Hamiltonian is easily diagonalized in the basis of phonon number states $\ket{n_{\vec{k}}}$ and impurity momentum sates $\psi_{\vec{q}}(\vec{r})$. The corresponding eigenenergies read
\begin{equation}
E(n_{\vec{k}},\vec{q}) = g_{\rm IB}n_0 + \int d^3 \vec{k} ~ \omega_k n_{\vec{k}}+ \frac{\vec{q}^2}{2 M}.
\end{equation}
To obtain the polaron ground state energy at a given total momentum $\vec{q}$ we start from the non-interacting eigenstate $\psi_{\vec{q}}(\vec{r}) \otimes \ket{0}$. To first order in $V_k$ there is no contribution to the energy, because $V_k \bra{n_{\vec{k}}} \a_{\vec{k}} + \ad_{-\vec{k}} \ket{n_{\vec{k}}} = 0$. To first order in $\alpha$ (i.e. to second order in $V_k$) one virtual phonon can be created and annihilated, see FIG.\ref{fig:FeynmanDiagsPert} (b). The corresponding ground state energy is
\begin{equation}
E_0(\vec{q}) =g_\IB n_0 +  \frac{\vec{q}^2}{2 M}  - \int^{\Lambda_0} d^3 \vec{k} ~ \frac{V_k^2}{\omega_k + \frac{k^2}{2M}-\frac{\vec{q} \cdot \vec{k}}{M}}  + \mathcal{O}(\alpha^2).
\label{eq:E0pert}
\end{equation}

Two comments are in order concerning equation \eqref{eq:E0pert}. Firstly, the ground state energy becomes divergent in the presence of a pole in the integrand. This happens for impurity momenta $|\vec{q}| \geq M c$ above which the impurity velocity $\vec{q}/M$ exceeds the speed of sound $c$ of the phonons. In this regime the impurity becomes supersonic and Cherenkov radiation is emitted by the moving impurity. A dynamical treatment is required to describe these processes in detail and the Rayleigh-Schr\"odinger perturbation theory breaks down.

The second comment concerns the UV behavior of the result \eqref{eq:E0pert}. When the UV cut-off $\Lambda_0$ is sent to infinity, the integral scales like $\int^{\Lambda_0} dk ~ k^2 / \omega_k \sim \Lambda_0$ and diverges. This effect is specific to ultracold atoms, and Chap.\ref{chap:UVregularization} is devoted to a detailed discussion of the proper regularization procedure.

\subsection{Green's function perturbation theory and self-consistent Born}
\label{sec:GreensFunctions}
The Rayleigh-Schr\"odinger perturbation theory is a powerful method for obtaining leading-order results for the polaron energy. Unfortunately it becomes intractable rather quickly when higher order terms are included. A more powerful method relies on using Green's functions. In particular, this approach allows to use self-consistent approximation schemes, of which the self-consistent Born approximation is the most popular one. 

An imaginary-time Green's function formalism for the polaron problem can be set up, see e.g. Refs. \cite{Mahan2000,Altland2010}. It has been applied to the polaron problem by many authors, see e.g. Ref. \cite{Kain2014} for a recent analysis of polarons in dipolar quantum gases. The Green's function of the impurity reads
\begin{equation}
G(\vec{q},i\omega_n) =\l i \omega_n - \frac{\vec{q}^2}{2 M} - \Sigma(\vec{q},i\omega_n) \r^{-1},
\end{equation}
where $\omega_n$ are Matsubara frequencies. Within the Born approximation the self-energy $\Sigma(\vec{q},i\omega_n)$ is determined by the Feynman diagrams in FIG.\ref{fig:FeynmanDiagsPert} (a) and (b),
\begin{equation}
\Sigma(\vec{q},i\omega_n) = g_{\rm IB} n_0 - \beta^{-1} \sum_{i \nu_m} \int d^3 \vec{k} ~ |V_{k}|^2 D_0(\vec{k},i\nu_m) G_0(\vec{q} - \vec{k},i\omega_n - i \nu_m).
\label{eq:BornSelfEnergy}
\end{equation}
Here $\beta^{-1} \sum_{i \nu_m}$ denotes the sum over all phonon Matsubara frequencies $\nu_m = 2 \pi m / \beta$, where the inverse temperature $\beta= 1 / k_{\rm B} T$ should be sent to infinity to obtain ground state properties. The free phonon Green's function is given by
\begin{equation}
D_0(\vec{k},i\nu_m) = 2 \omega_k \l (i \nu_m)^2 - \omega_k^2 \r^{-1}
\end{equation}
and the free impurity Green's function reads $G_0(\vec{q},i\omega_n) = ( i \omega_n - \vec{q}^2 / 2 M )^{-1}$.

Analytical continuation of the self-energy \eqref{eq:BornSelfEnergy} in Born-approximation yields the retarded self-energy \cite{Kain2014},
\begin{equation}
\Sigma^{\rm R}(\vec{q},\omega) = g_{\rm IB} n_0 - \int d^3 \vec{k} ~ \frac{|V_k|^2}{\omega_k - \omega + (\vec{q} - \vec{k})^2 / 2 M + i 0^-}.
\label{eq:SelfEnergyRetarded}
\end{equation}
To derive corrections to the polaron ground state energy analytically, we evaluate the self-energy at the free impurity resonance $\omega = \vec{q}^2 / 2M$. This yields
\begin{equation}
E_0 = \frac{\vec{q}^2 }{ 2M} + {\rm Re} \Sigma^{\rm R}(\vec{q}, \vec{q}^2 / 2M) = \frac{\vec{q}^2 }{ 2M} + g_{\rm IB} n_0 - \int d^3 \vec{k} ~ \frac{|V_k|^2}{\omega_k + \frac{\vec{k}^2}{2M} - \frac{\vec{q} \cdot \vec{k} }{ M} },
\end{equation}
which coincides with the Rayleigh-Schr\"odinger result \eqref{eq:E0pert}. For a discussion of Cherenkov radiation in the Green's function formalism see e.g. Ref.\cite{Kain2014}. From Eq.\eqref{eq:SelfEnergyRetarded} the full self-energy can easily be calculated numerically within the Born-approximation.

To extend the perturbative approach beyond the Rayleigh-Schr\"odinger theory, the \emph{self-consistent} Born approximation can be employed. It relies on the self-consistent solution of Eq.\eqref{eq:BornSelfEnergy}, in which $G_0$ is replaced by the dressed Green's function $G$,
\begin{equation}
\Sigma(\vec{q},i\omega_n) = g_{\rm IB} n_0 - \beta^{-1} \sum_{i \nu_m} \int d^3 \vec{k} ~ \frac{ |V_{k}|^2 D_0(\vec{k},i\nu_m) }{ i \omega_n - i \nu_m -  \frac{(\vec{q} - \vec{k})^2}{2 M} - \Sigma(\vec{q} - \vec{k},i\omega_n - i \nu_m) }.
\label{eq:BornSelfEnergySelfCons}
\end{equation}
This approximation corresponds to summing up an infinite number of  non-crossing diagrams as shown in FIG.\ref{fig:FeynmanDiagsPert} (c), see e.g. Ref.\cite{Altland2010}. 

Eq.\eqref{eq:BornSelfEnergySelfCons} can be solved numerically for $\Sigma(\vec{q},\omega)$. In order to obtain some analytical insight, we make an additional approximation and consider the following ansatz for the self-energy,
\begin{equation}
\Sigma(\vec{q},\omega) = \varepsilon + \frac{\vec{q}^2}{2 W}.
\end{equation}
Here $\varepsilon$ denotes a correction to the ground state energy and $W$ gives a correction to the impurity mass; for simplicity we ignored the frequency dependence. By analytical continuation of Eq.\eqref{eq:BornSelfEnergySelfCons} we obtain
\begin{equation}
\varepsilon + \frac{\vec{q}^2}{2 W} = g_{\rm IB} n_0 - \int d^3 \vec{k} ~ \frac{|V_k|^2}{\omega_k - \omega + \frac{(\vec{q} - \vec{k})^2}{ 2 \mathcal{M} } + \varepsilon + i 0^-},
\label{eq:selfConsBornApproxSelfEngy}
\end{equation}
cf. Eq.\eqref{eq:SelfEnergyRetarded}. On the right hand side of the equation we introduced the renormalized mass
\begin{equation}
\mathcal{M}^{-1} = M^{-1} + W^{-1}.
\end{equation}

As above, we are interested in the self-energy at the frequency $\omega$ where the self-consistently determined impurity Green's function has a pole, $\omega = \vec{q}^2 / 2 \mathcal{M} + \varepsilon$. Putting this into Eq.\eqref{eq:selfConsBornApproxSelfEngy} and considering $\vec{q} = 0$ first we obtain
\begin{equation}
\varepsilon = g_{\rm IB} n_0 - \int d^3 \vec{k} ~ \frac{|V_k|^2}{\omega_k + \frac{\vec{k}^2}{2 \mathcal{M}}}.
\end{equation}
To derive the expression for the renormalized mass $\mathcal{M}$ we expand Eq.\eqref{eq:selfConsBornApproxSelfEngy} around $\vec{q} = 0$. Comparison of terms of order $\mathcal{O}(q^2)$ on both sides yields
\begin{equation}
\mathcal{M}^{-1} = M^{-1} - \frac{8 \pi}{3} \mathcal{M}^{-2} 
\int_0^{\Lambda_0} dk ~ \frac{|V_k|^2 k^4}{\l \omega_k + \frac{\vec{k}^2}{2 \mathcal{M}} \r^3}.
\label{eq:selfConsBornMass}
\end{equation}

The resulting ground state energy from the self-consistent Born approximation reads
\begin{equation}
E_0(\vec{q}) |_{\rm SCB} = \frac{\vec{q}^2 }{ 2M} + {\rm Re} \Sigma^{\rm R}(\vec{q}, \vec{q}^2 / 2\mathcal{M} + \varepsilon) = g_{\rm IB} n_0 + \frac{\vec{q}^2}{2 \mathcal{M}} - \int d^3 \vec{k} ~ \frac{|V_k|^2}{\omega_k + \frac{\vec{k}^2}{2 \mathcal{M}}}.
\label{eq:selfConsBornEnergyResult}
\end{equation}
Note that both (\ref{eq:E0pert}) and (\ref{eq:selfConsBornEnergyResult}) are UV divergent and
require introducing the cut-off $\Lambda_0$.  From Eq.\eqref{eq:selfConsBornEnergyResult} it is apparent that $\mathcal{M}$ corresponds to the effective mass of the polaron.

\section{Exact solution for infinite mass }
\label{sec:LangFirsov}
We start by solving the problem of a localized impurity with an infinite mass $M\to\infty$. Although this corresponds to a special case of a static potential for the bosons, the infinite mass limit illustrates how the phonon cloud in a polaron can be described quantitatively. The corresponding Fr\"ohlich Hamiltonian \eqref{eq:HFroh} now reads
\begin{equation}
\H = g_\IB n_0 + \int d^3 \vec{k} ~ \Biggl[ V_{\vec{k}} \l \a_{\vec{k}} + \ad_{-\vec{k}} \r  + \omega_{k} \ad_{\vec{k}} \a_{\vec{k}}  \Biggr],
\label{eq:HinfMass}
\end{equation}
where we assumed an impurity localized at $\vec{r}=\vec{0}$. We denote the phonon part of the wavefunction as $\ket{\Psi_\ph}_a$ so that the total wavefunction is given by $\ket{\Phi} = \ket{\Psi_\ph}_a \otimes \psd(\vec{0}) \ket{0}_\I$.

To obtain the phonon wavefunction $\ket{\Psi_\ph}_a$ of the polaron ground state, we apply a unitary transformation to the Hamiltonian \eqref{eq:HinfMass},
\begin{equation}
\hat{U} = \exp \l \int d^3 \vec{k} ~ \alpha^{\infty}_{\vec{k}} ~ \ad_{\vec{k}} - (\alpha^{\infty}_{\vec{k}})^* ~ \a_{\vec{k}}  \r.
\label{eq:LangFirsov}
\end{equation}
This transformation describes a coherent displacement of phonon operators,
\begin{equation}
\hat{U}^\dagger \a_{\vec{k}} \hat{U}  = \a_{\vec{k}} + \alpha_{\vec{k}}^\infty.
\end{equation}

Using the symmetry $V_{-\vec{k}}^* = V_{\vec{k}}$, the transformed Hamiltonian reads
\begin{equation}
\hat{U}^\dagger \H \hat{U} = g_\IB n_0 + \int d^3 \vec{k} ~ \Biggl[ \omega_{k} \ad_{\vec{k}} \a_{\vec{k}} +  \l \ad_{\vec{k}} \l V_{\vec{k}} + \omega_k \alpha_{\vec{k}}^\infty \r + \hc \r+  V_{\vec{k}} \l \alpha_{\vec{k}}^\infty + (\alpha_{\vec{k}}^\infty)^* \r + \omega_k |\alpha_{\vec{k}}^\infty|^2 \Biggr].
\end{equation}
By choosing $\alpha^\infty_{\vec{k}} = - V_{\vec{k}} / \omega_k$, the interaction terms linear in $\a_{\vec{k}}^{(\dagger)}$ can be eliminated. The ground state of the resulting Hamiltonian is given by the vacuum $\ket{0}_a$ and its energy is
\begin{equation}
E_0 = g_\IB n_0 - \int d^3\vec{k} ~ \frac{|V_{\vec{k}}|^2}{\omega_k}.
\end{equation}
Finally, the infinite-mass polaron wavefunction is
\begin{equation}
\ket{\Phi} = \exp \left[ - \int d^3 \vec{k} ~ \frac{V_{\vec{k}}}{\omega_k} \l \ad_{\vec{k}} -  \a_{\vec{k}} \r \right] \ket{0}_a \otimes \psd(\vec{0}) \ket{0}_\I.
\label{eq:gsLangFirsov}
\end{equation}

\section{Lee-Low-Pines treatment}
\label{sec:LLP}
Now we turn our attention to mobile impurities with a finite mass $M < \infty$, and consider a translationally invariant polaron model (it may even be a \emph{discrete} translational invariance). In this case the momentum of the polaron is a conserved quantity (quasimomentum in the lattice case). In a seminal paper Lee, Low and Pines (LLP) \cite{Lee1953} demonstrated how the conservation of polaron momentum can be made explicit. Using this as a starting point they developed a weak coupling mean field treatment which we will presented in section \ref{sec:MFtheory}. Note that the LLP treatment is quite general. In the main text we present it for the continuum Fr\"ohlich model \eqref{eq:HFroh} and in appendix \ref{sec:LLPlattice} we present a generalization of this method for lattice polarons.

To identify the conserved polaron momentum, a unitary transformation is applied which translates bosons (phonons) by an amount chosen such that the impurity is shifted to the origin of the new frame. Translations of the bosons are generated by the total phonon-momentum operator,
\begin{equation}
\h{\vec{P}}_\ph = \int d^3 \vec{k} ~ \vec{k} \ad_{\vec{k}} \a_{\vec{k}}.
\end{equation}
Next we would like to restrict ourselves to a \emph{single} impurity, which in second-quantized notation means that $\int d^3 \vec{r} ~ \psd(\vec{r}) \ps(\vec{r})=1$ in the relevant subspace. The position operator of the impurity is given by $\vec{R} = \int d^3 \vec{r} ~  \vec{r} ~ \psd(\vec{r}) \ps(\vec{r})$, and the LLP transformation thus reads
\begin{equation}
\h{U}_\text{LLP} = e^{i \h{S}} ,\qquad \h{S} = \vec{R} \cdot \h{\vec{P}}_\ph.
\label{eq:LLPdef}
\end{equation}

To calculate the transformed Fr\"ohlich Hamiltonian, let us first discuss transformations of the phonon operators. Because $\ad_{\vec{k}}$ changes the phonon momentum by $\vec{k}$, and $\vec{R}$ can be treated as a $\mathbb{C}$-number from the point of view of phonons, it follows that
\begin{equation}
\h{U}^\dagger_\text{LLP} \ad_{\vec{k}} \h{U}_\text{LLP} = \ad_{\vec{k}} e^{-i \vec{k} \cdot \vec{R}}.
\end{equation}
For the impurity we also want to calculate the transformation of creation and annihilation operators in momentum space, defined in the usual way as
\begin{equation}
\psd_{\vec{q}} = (2 \pi)^{-3/2} \int d^3 \vec{r} ~ e^{- i \vec{q} \cdot \vec{r}} ~ \psd(\vec{r}).
\end{equation}
Now it is easy to show that the LLP transformation corresponds to a shift in momentum for the impurity,
\begin{equation}
\h{U}^\dagger_\text{LLP} \psd_{\vec{q}}  \h{U}_\text{LLP} = \psd_{\vec{q} + \h{\vec{P}}_\ph}.
\end{equation}

The transformation $\hat{U}_\text{LLP}$ defines the Fr\"ohlich Hamiltonian in the polaron frame,
\begin{multline}
\tilde{\mathcal{H}} := \h{U}^\dagger_\text{LLP} \H \h{U}_\text{LLP} =g_\IB n_0 + \int d^3 {\vec{k}} \left[  \omega_k \ad_{\vec{k}} \a_{\vec{k}} + V_{\vec{k}} \l \ad_{-\vec{k}} + \a_{\vec{k}} \r  \right] +\\
+ \frac{1}{2 M}  \int d^3 \vec{q}  ~ \psd_{\vec{q}} \ps_{\vec{q}} ~ \Bigl( \vec{q} - \int d^3\vec{k} ~\vec{k} \ad_k \a_k \Bigr)^2.
\label{eq:HfrohLLPfull}
\end{multline}
It is easy to see that for a single impurity atom $\vec{q}$ corresponds to the conserved total momentum of the polaron. From now on we will omit the $\psd_{\vec{q}} \ps_{\vec{q}}$ part and write only the phonon part of the Hamiltonian.

Elimination of the impurity degrees of freedom from the problem comes at the cost of introducing a non-linearity for phonon operators. It describes interactions between the phonons, mediated by the mobile impurity. The corresponding coupling constant is given by the (inverse) impurity mass $M^{-1}$. Thus we conclude that, in order to characterize the polaron problem, \emph{two} dimensionless coupling constants $\alpha$ and $m_\B/M$ are important.

\section{Weak coupling mean-field theory}
\label{sec:MFtheory}
Next we present the mean-field (MF) solution of the polaron problem in the weak coupling limit, originally suggested by Lee, Low and Pines (LLP) \cite{Lee1953} as a variational method. In a nutshell, the ground state of the LLP Hamiltonian \eqref{eq:HfrohLLPfull} is approximated by a variational ansatz of coherent phonon states. This state is closely related to the state in Eq.\eqref{eq:gsLangFirsov} obtained for an immobile impurity, but importantly the conservation of total momentum is taken into account. Let us note that, initially, Lee, Low and Pines considered their MF ansatz as an all-coupling approach. However, as will be clarified in chapter \ref{chap:RGapproach}, it does not capture well the strong coupling polaron regime at large couplings $\alpha$. 

First, to understand under which conditions the system is weakly coupled, let us study the Fr\"ohlich Hamiltonian in the polaron frame Eq.\eqref{eq:HfrohLLPfull} more closely. We observe that in two cases \eqref{eq:HfrohLLPfull} is exactly solvable. The first case corresponds to small interaction strength $\alpha \ll 1$. In this case we can approximate $V_{\vec{k}} \approx 0$. Hence $[\H_{\vec{q}}, \ad_{\vec{k}} \a_{\vec{k}}]=0$, and the Hamiltonian is diagonalized in the phonon-number basis (in $\vec{k}$-space). The second case corresponds to the large mass $M \rightarrow \infty$, when the phonon non-linearity can be discarded. Let us estimate under which conditions this is justified. To this end we consider the case $\vec{q}=0$ for simplicity and note that the characteristic phonon momentum is $\xi^{-1}$. We may thus neglect the last term in Eq.\eqref{eq:HfrohLLPfull} provided that $1/ (2 M \xi^{2}) \ll c / \xi$, where $c/\xi = E_\ph$ is the characteristic phonon energy, see Eq.\eqref{eq:defEph}. Thus we arrive at
\begin{equation}
\alpha \ll 1 \qquad \text{or} \qquad M \gg m_\B / \sqrt{2} \qquad  \qquad \text{(weak coupling)}.
\label{eq:defWeakCoupling}
\end{equation}
We note that analysis in Ref. \cite{Grusdt2015DSPP} also showed that perturbation theory
becomes asymptotically exact in the limit of small impurity masses $M \to 0$.

To solve for the ground state of Eq.\eqref{eq:HfrohLLPfull} in the weak coupling regime, we start from the solution for $M \to \infty$. We solved that case exactly by a coherent displacement operator in Sec.\ref{sec:LangFirsov}. In the case of a finite mass $M < \infty$ we can use an analogous wavefunction as a variational MF ansatz,
 \begin{equation}
| \psi_{\rm MF} \rangle = \exp \l \int d^3 \vec{k} ~  \alpha_{\vec{k}}^\MF \ad_{\vec{k}} - \hc \r \ket{0} = \prod_{\vec{k}} \ket{\alpha^\MF_{\vec{k}}},
\label{Psi_MF}
 \end{equation}
 We emphasize that this simple factorizable wavefunction is formulated in the polaron frame, in the lab frame the MF state reads $\ket{\Phi_\MF} = \hat{U}_{\rm LLP} \ket{\psi_\MF} \otimes \psd_{\vec{q}} \ket{0}_\I$.

The coherent amplitudes $\alpha^\MF_{\vec{k}}$ should be chosen such that the variational MF energy functional $ \mathscr{H}_\MF(\vec{q}) = \prod_{\vec{k}} \bra{\alpha^\MF_{\vec{k}}} \H_{\vec{q}} \ket{\alpha^\MF_{\vec{k}}}$ is minimized. This yields  self-consistency equations for the ground state polaron
\begin{equation}
 \alpha^\MF_{\vec{k}} = - \frac{V_k}{\Omega^\MF_{\vec{k}}[\alpha_{\kappa}^\MF]}.
 \label{eq:MFpolaron}
\end{equation}
Here we introduced the renormalized phonon dispersion in the polaron frame, which reads
\begin{equation}
\Omega^\MF_{\vec{k}} = \omega_k + \frac{k^2}{2 M} - \frac{1}{M } \vec{k} \cdot \l \vec{q} - \vec{P}_\ph^\MF [\alpha^\MF_{\vec{k}}] \r,
\label{eq:OmegakDef}
\end{equation}
where $\vec{P}_\ph^\MF$ denotes the MF phonon momentum,
\begin{equation}
\vec{P}_\ph^\MF[\alpha^\MF_{\vec{k}}] = \int d^3\vec{k} ~ \vec{k} |\alpha^\MF_{\vec{k}}|^2.
\label{eq:XiselfCons}
\end{equation}

In the following paragraphs, we will discuss different aspects of the weak coupling MF solution. After simplifying the MF self-consistency equations, we calculate the polaron mass and its energy. From now on, for simplicity, we consider only real-valued scattering amplitudes $V_k\in \mathbb{R}$, which is the case relevant to our model -- see Eq. \eqref{eq:Vktilde} -- and without loss of generality we assume that $\vec{q} = (q,0,0)^T$ is directed along $\vec{e}_x$ with $q > 0$.

\subsection{Self-consistency equation}
\label{subsubsec:MFselfConsEq}
The infinite set of self-consistency equations \eqref{eq:MFpolaron} can be reduced to \emph{a single} equation for $\vec{P}_\ph^\MF  = P_\ph^\MF \vec{e}_x$, which is directed along the $x$-axis by symmetry. Using spherical coordinates one can perform angular integrals in Eq.\eqref{eq:XiselfCons} analytically and we find
\begin{multline}
 P_\ph^\MF = \frac{2 \pi M^2}{(q- P_\ph^\MF)^2} \int_0^{\Lambda_0} dk ~ V_k^2 ~k \bigg[ \frac{2 \l \omega_k + \frac{k^2}{2M} \r \frac{k}{M}\l q- P_\ph^\MF \r}{\l \omega_k + \frac{k^2}{2M} \r^2 - \l \frac{k}{M} (q- P_\ph^\MF)\r^2} + \\ +
  \log \l \frac{\omega_k + \frac{k^2}{2M}-\frac{k}{M} \l q- P_\ph^\MF \r}{\omega_k + \frac{k^2}{2M}+\frac{k}{M} \l q- P_\ph^\MF \r} \r  \bigg] .
 \label{eq:XiselfConsLong}
\end{multline}
This equation can be solved numerically to obtain $ P_\ph^\MF$.

It is natural to expect that the phonon momentum never exceeds the total system momentum, i.e. $P_\ph^\MF \leq q$. To show this rigorously, we note that Eq.\eqref{eq:XiselfConsLong} is of the form $P_\ph^\MF=f(q-P_\ph^\MF)$ where $f$ is some function. Furthermore we read off from Eq.\eqref{eq:XiselfConsLong} that $f$ is anti-symmetric, $f(P_\ph^\MF-q)=-f(q-P_\ph^\MF)$. 
We will now assume that a solution $P_\ph^\MF>q$ exists and show that this leads to a contradiction. To this end let us first note that for vanishing interactions, $P_\ph^\MF=0 < q$ is the solution. Then, assuming that the solution $P_\ph^\MF$ depends continuously on the interaction strength $g_\IB\sim V_k$, we conclude that at some intermediate value of $g_{\IB}$ the solution has to be $P_{\ph 0}^\MF=q$. This leads to a contradiction since $P_{\ph 0}^\MF=f(q-P_{\ph 0}^\MF)=f(0)=-f(0)=0 \neq q$. 

\begin{figure}[b!]
\centering
\epsfig{file=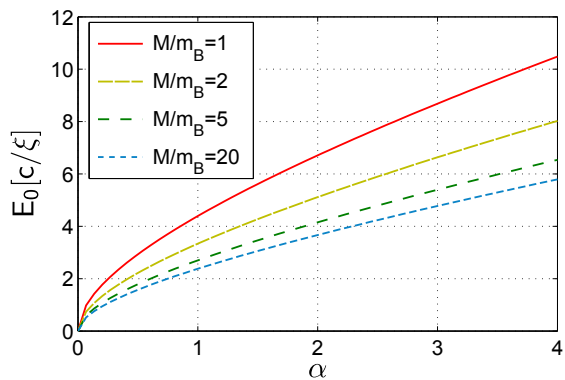, width=0.5\textwidth}
\caption{
The polaron ground state energy $E_0$, calculated from the regularized mean-field theory Eq.\eqref{eq:E0MF}, is shown as a function of the coupling constant $\alpha$. Various impurity-boson mass ratios, $M/m_\text{B}=20$ (bottom curve), $M/m_\text{B}=5$, $M/m_\text{B}=2$ and $M/m_\text{B}=1$ (top curve) are considered. Other parameters are $n_0=1 \times \xi^{-3}$, $q=0$ and we used a sharp momentum cut-off at $\Lambda_0=2 \times 10^{3} / \xi$ where Eq.\eqref{eq:E0MF} is sufficiently converged to its value at $\Lambda_0 = \infty$.}
\label{fig:Ep-aIB}
\end{figure}

\subsection{Polaron energy}
\label{subsubsec:MFpolaronEnergy}
The first physical observable which we easily obtain from the MF solution is the variational polaron energy. At the saddle-point, i.e. for $\alpha^\MF_{\vec{k}}$ from Eq.\eqref{eq:MFpolaron}, the energy functional is given by
\begin{equation}
E_0(q) |_\MF = \frac{q^2}{2 M} - \frac{( P_\ph^\MF )^2}{2M} + g_\IB n_0 - \int^{\Lambda_0} d^3 \vec{k} ~ \frac{V_k^2}{\Omega_{\vec{k}}^\MF}.
\label{eq:MFenergyNonReg}
\end{equation}
Note that we introduced a sharp momentum cut-off at $\Lambda_0$ in the integral on the right hand side. This is necessary because the integral in Eq.\eqref{eq:MFenergyNonReg} is UV-divergent and scales linearly with $\Lambda_0$. We will discuss in detail in chapter \ref{chap:UVregularization} why such UV divergences appear and how they should be regularized. There we derive the regularized result,
\begin{equation}
E_0(q) |_\MF^{\rm reg} = \frac{q^2}{2 M} - \frac{( P_\ph^\MF )^2}{2M} +\frac{2 \pi a_\IB n_0}{m_\rd}  
 - \int^{\Lambda_0} d^3 \vec{k} ~ \frac{V_k^2}{\Omega_{\vec{k}}^\MF}  + \frac{4 a_\IB^2}{m_\rd} \Lambda_0 n_0.
 \label{eq:E0MF}
\end{equation}

In FIG.\ref{fig:Ep-aIB} we show the resulting polaron energies as a function of the interaction strength and using different mass ratios $M/m_\text{B}$ (after regularization of the UV divergence). 

The perturbative result in Eq.\eqref{eq:E0pert} is correctly reproduced by MF polaron theory because $P_\ph^\MF = \mathcal{O}(\alpha)$, see Eq.\eqref{eq:MFenergyNonReg}. Moreover we see that in the spherically symmetric case of vanishing polaron momentum $\vec{q}=0$ we find $P_\ph^\MF=0$ and thus, remarkably, MF theory is \emph{equivalent} to lowest-order perturbation theory in this case \cite{Devreese2013}.

\subsection{Polaron mass}
\label{subsubsec:gsPolaronMass}

In the following we show how the phonon momentum $P_\ph$ can be used to calculate the polaron mass $M_\p$. To this end, we note that the polaron velocity is given by $v_\p=q/M_\p$, where we assumed that in the ground state the full momentum of the system, $q$, is carried by the polaron (instead of being emitted). On the other hand, polaron and impurity position coincide at all times, so $v_{\I}=v_\p$. The momentum $q_\I$ carried by the bare impurity can be calculated from the total momentum $q$ by subtracting the phonon momentum $P_\ph$, i.e. $q_\I=q-P_\ph$. This is a direct consequence of the conservation of total momentum. Furthermore, the average impurity velocity is given by $v_\I = q_\I / M$ and we thus arrive at the MF expression
\begin{equation}
 \frac{M}{M^\MF_\p} = 1-\frac{P_\ph^\MF}{q}.
 \label{eq:M*}
\end{equation}

An alternative way of calculating the polaron velocity is to assume that the polaron forms a wavepacket with average momentum $q$. Thus  $v_\p=\partial_qE_0(q) |_\MF$, and using the self consistency equations \eqref{eq:MFpolaron} and \eqref{eq:XiselfCons} a simple calculation yields
\begin{equation}
 v_\p = \frac{d E_0(q) |_\MF}{dq} = \frac{1}{M} \l q- P_\ph^\MF \r.
 \label{eq:vPMF}
\end{equation}
Assuming a quadratic polaron dispersion $E_0(q) |_\MF = q^2/(2 M_\p)$ we obtain \eqref{eq:M*} again and this result completely agrees with the semiclassical  argument given above. Note however that the semiclassical argument in equation (\ref{eq:M*}) is more general because it does not rely on the particular variational state we select. 

It is instructive to compute the polaron mass perturbatively in $\alpha$. To this end we expand the perturbative ground state energy Eq.\eqref{eq:E0pert} to second order in $q$ (around $q=0$) and obtain
\begin{equation}
 M_\p^{-1} = M^{-1} - \frac{8 \pi}{3} M^{-2}  \int_0^{\Lambda_0} dk~\frac{k^4 V_k^2}{\l \omega_k + \frac{k^2}{2M} \r^3} + \mathcal{O}(\alpha^2).
 \label{eq:pertMp}
\end{equation}
The same expansion can be derived straightforwardly from Eq.\eqref{eq:M*} using the MF expression \eqref{eq:XiselfCons} for the total phonon momentum $P_\ph^\MF$.

\begin{figure}[t]
\centering
\epsfig{file=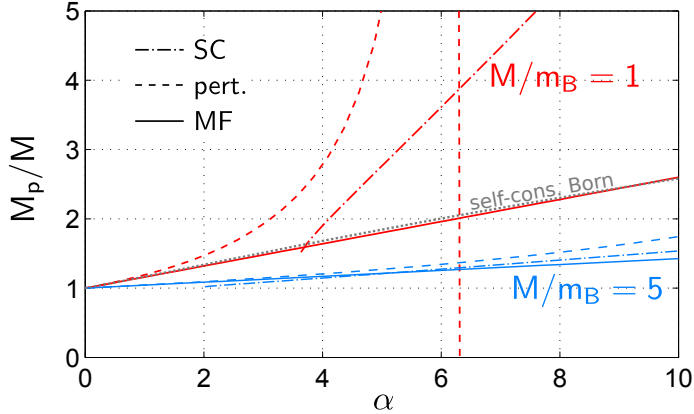, width=0.6\textwidth} $\qquad$
\caption{Polaron mass $M_\text{p}/M$ as a function of the coupling strength $\alpha$ for different mass ratios $M/m_\text{B}=1$ (red) and $M/m_\text{B}=5$ (blue). Other parameters are $q/Mc=0.01$ and $\Lambda_0=2 \times 10^{3} / \xi$. We show MF calculations (solid) and compare them to the perturbative result (dashed) from Eq.\eqref{eq:pertMp} showing an unphysical divergence of the polaron mass. We also show predictions of the  strong coupling approximation (dash-dotted) from Eq.\eqref{eq:MpSC}. The dotted gray line shows the result of the self-consistent Born approximation, see Eq. \eqref{eq:selfConsBornMass}, for $M/m_{\rm B}=1$.}
\label{fig:Mp-LinPlot-aIB}
\end{figure}

In FIG. \ref{fig:Mp-LinPlot-aIB} we show the dependence of the MF polaron mass on the coupling strength $\alpha$. We find a linear dependence 
\begin{equation}
M^\MF_\p - M\propto \alpha.
\end{equation}
When decreasing the bare impurity mass, the ratio $M/M^\MF_\text{p}$ decreases as well, i.e. the smaller the bare impurity mass the larger the renormalization of the polaron mass. This is in accordance with our expectation that the polaron becomes strongly coupled when the mass ratio $m_\text{B}/M$ becomes large. Comparison to the perturbative polaron mass in FIG.\ref{fig:Mp-LinPlot-aIB} shows strong qualitative differences to MF, in a striking contrast to the overall polaron energy. Around some critical coupling strength, in this case $\alpha \approx 6$, the perturbative result shows an unphysical divergence of the polaron mass.

\section{Strong coupling Landau-Pekar approach}
\label{sec:LandauPekar}
Next we review the strong coupling polaron wavefunction, originally introduced by Landau and Pekar \cite{Landau1946,Landau1948} (see also G. R. Allcock in Ref.\cite{Kuper1962}). We will summarize its application to the problem of an impurity in a BEC, which was discussed in Ref. \cite{Casteels2011}. From the very beginning it is worth pointing out that the strong coupling approximation does not capture correctly the UV divergence of the Fr\"ohlich model of BEC polarons. This leads us to conclude that -- in contrast to common assumptions -- the Landau-Pekar wavefunction does \emph{not} always provide an exact treatment of the strong coupling regime (even though it has been demonstrated to be accurate in the case of optical phonons in solid state systems).

The idea of the strong coupling treatment is to assume a factorizable wavefunction for the impurity and the phonons, which is justified for strong interactions. This assumption is natural in the case of a heavy impurity, when phonons can follow the impurity adiabatically. Similar to the Born-Oppenheimer approximation used to describe molecular binding, where electrons instantly adjust to the positions of the atomic nuclei, the phonons can instantly adjust to a change of the position $\hat{\vec{r}}$ of the heavy impurity. We can derive the regime of validity of this approximation by comparing the typical impurity energy $1/2 M \xi^2$ to characteristic phonon energies $c/\xi$. The impurity can be considered as slow when $1/2 M \xi^2 \ll c / \xi$, leading to the condition $M \gg m_\text{B}$.

As the impurity mass is decreased, fluctuations of the impurity position may become important. However, they are suppressed when $V_k$ is large (i.e. when $\alpha$ is large). Therefore the wavefunction can be taken as factorizable when either the impurity is heavy or the coupling constant $\alpha$ is sufficiently large,
\begin{equation}
\alpha \gg 1 \qquad \text{or} \qquad M \gg m_\B \qquad  \qquad \text{(strong coupling)}.
\label{eq:defStrongCoupling}
\end{equation}
Note that this implies that strong and weak coupling regimes overlap in the case of large impurity mass\footnote{Here we use the term \emph{strong coupling} to denote the entire regime of parameters where the polaron wavefunction factorizes. In the solid-state literature on polarons this term is attributed only to a parameter regime where interactions are strong and when the fast impurity follows the slow motion of phonons adiabatically. We thank J.T. Devreese and S. N. Klimin for pointing this out.} $M \gg m_\text{B}$, see FIG.\ref{fig:QualPhaseDiagFROH}.

Mathematically the starting point for the strong coupling treatment is the Fr\"ohlich Hamiltonian Eq.\eqref{eq:HFroh}. The strong coupling wavefunction has a product form,
\begin{equation}
\ket{\Phi_\sc} =   \ket{\Psi_\ph}_a \otimes \underbrace{\int d^3 \vec{r} ~  \psi_\sc(\vec{r})~ \psd(\vec{r}) \ket{0}_\I}_{=: \ket{\psi_\sc}_\I},
\end{equation}
where $\ket{\Psi_\ph}_a$ is the phonon wavefunction and $\ket{\psi_\sc}_\I$ the impurity state. The first step consists of solving the phonon problem for a given, but arbitrary impurity state $\ket{\psi_\sc}_\I$. 

The effective phonon Hamiltonian has a form
\begin{equation}
\H_\text{ph} =  \int d^3 \vec{k} ~  \tilde{V}_{\vec{k}} \l \a_{\vec{k}} + \ad_{-\vec{k}} \r  + \omega_{k} \ad_{\vec{k}} \a_{\vec{k}},
\end{equation}
where the effective interaction is defined by $ \tilde{V}_{\vec{k}}  =V_{\vec{k}} ~_\I \bra{\psi_\sc} e^{i \vec{k}\cdot \vec{r}} \ket{\psi_\sc}_\I$, with $\tilde{V}_{- \vec{k}} = \tilde{V}^*_{\vec{k}}$. This Hamiltonian corresponds to an \emph{im}mobile polaron and can easily be solved by coherent states as described in Sec.\ref{sec:LangFirsov},
\begin{equation}
\ket{\Psi_\ph}_a  = \prod_{\vec{k}} \ket{\beta_{\vec{k}}}, \qquad \qquad \beta_{\vec{k}} = - \frac{\tilde{V}_{\vec{k}}}{\omega_k}.
\end{equation}
The phonon energy is given by $E_\text{ph}[\ket{\psi_\sc}_\I] = - \int d^3 \vec{k} ~ |\tilde{V}_{\vec{k}}|^2 / \omega_k$, and it depends on the particular choice of impurity wavefunction. This corresponds to an effective potential seen by the impurity, to which it can slowly adjust.

To make further progress, a variational ansatz is made for the impurity wavefunction. Motivated by the idea that the effective potential due to the phonon cloud can localize the impurity, one introduces a normalized Gaussian wavefunction to approximate the polaron ground state,
\begin{equation}
\psi_\sc(\vec{r}) = \pi^{-3/4} \lambda^{-3/2} e^{-\frac{r^2}{2 \lambda^2}}.
\label{eq:psiVarSC}
\end{equation}
Here $\lambda$ is a variational parameter, describing the spatial extend of the impurity wavefunction. 

\subsection{Polaron energy}
\label{subsec:SCpolaronEnergy}
The total energy
\begin{equation}
E_0 |_\sc = g_\IB n_0 +  \bra{\psi_\sc} \frac{\nabla^2}{2 M} \ket{\psi_\sc}  - \int d^3 \vec{k} ~ |\tilde{V}_{\vec{k}}|^2 / \omega_k
\end{equation} 
can be expressed in terms of $\lambda$. It was shown in \cite{Casteels2011} that for the BEC polaron
\begin{equation}
E_0(\lambda) |_\sc = g_\IB n_0 + \frac{3}{4 M \lambda^2} - \frac{\sqrt{2}}{\sqrt{\pi}} \alpha \mu \frac{1}{M \lambda \xi} \left[ 1 - \sqrt{\pi} \frac{\lambda}{\xi} e^{\lambda^2 / \xi^2} \l 1 - \text{erf} \l \lambda/\xi \r \r \right],
\label{eq:EscLambda}
\end{equation}
which can be easily  minimized numerically. Here $\rm erf(x)$ denotes the error function and the dimensionless quantity $\mu$ is defined as $\mu = M m_\B / (4 m_\rd^2)$.

A word of caution is in order about this energy and the strong coupling approach in general. The polaronic contribution $E_{\rm p}$ to the energy in Eq.\eqref{eq:EscLambda}, defined as  $E_{\rm p} = E_0 - g_\IB n_0$, is UV convergent, in fact the cut-off $\Lambda_0 = \infty$ was used in the calculations. At first sight this might seem desirable, however as will be shown in chapter \ref{chap:UVregularization} consistency requires the polaronic energy $E_{\rm p}$ to have a diverging energy towards negative infinity. This was indeed what we found in the MF case, see Eq.\eqref{eq:MFenergyNonReg}. 

Thus the resulting variational energy \eqref{eq:EscLambda} of the strong coupling wavefunction is higher than the true ground state energy by a large cut-off dependent amount (infinite when the UV cut-off $\Lambda_0$ is sent to infinity)! The reason is that interactions with high-momentum phonons are exponentially suppressed in the effective scattering amplitude $\tilde{V}_{\vec{k}}$, in contrast to the bare amplitude $V_{\vec{k}}$. Hence the strong coupling energy can not be compared one-by-one to properly regularized polaron energies, and more reliable strong coupling impurity wavefunctions $\psi_\sc(\vec{r})$ should be employed. 

\subsection{Polaron mass}
\label{subsec:SCpolaronMass}
By modifying the variational wavefunction \eqref{eq:psiVarSC} to a wavepacket describing finite polaron momentum, an expression for the polaron energy was derived \cite{Casteels2011}. Using the equations in Ref. \cite{Casteels2011} we arrive at the following expression for the strong coupling polaron mass,
\begin{equation}
M_\p^\sc = M + m_\text{B} \frac{\alpha}{ 3 \sqrt{2 \pi}} \l 1 + \frac{m_\text{B}}{M} \r^2 \left[ -2 \frac{\lambda}{\xi} + \sqrt{\pi} \l 1 + 2 \frac{\lambda^2}{\xi^2} \r e^{\lambda^2 / \xi^2} \l 1 - \text{erf} \l \lambda / \xi \r \r \right].
\label{eq:MpSC}
\end{equation}

A surprising feature of equation \eqref{eq:MpSC} is that the strong coupling polaron mass follows the same power-law $M_\p^\sc - M \propto \alpha$ as MF theory for large $\alpha$ (this can be checked by expanding $\lambda$ in powers of $\alpha^{-1}$). Thus in the strong coupling regime both predictions differ only in the \emph{prefactor} of this power-law. We will show some numerical results for the strong coupling polaron mass in the following chapters, and refer interested readers to Ref.\cite{Casteels2011} for further discussion.

\section{Feynman path integral approach}
\label{sec:Feynman}
So far we were only concerned with limiting cases of either strong or weak coupling. An important question is how these two regimes are connected, and how the polarons look like at intermediate couplings. To address this question, Feynman applied his path-integral formalism and developed a variational all-coupling polaron theory \cite{Feynman1955} which we will now summarize.

Feynman's conceptual starting point is the imaginary-time path-integral for the Fr\"ohlich Hamiltonian \eqref{eq:HFroh}, with a single impurity. For a system at finite temperature $k_B T = 1/ \beta$ the corresponding partition function $\mathcal{Z}(\beta) = e^{- \beta F}$ is related to the free energy $F$. It has the form
\begin{equation}
e^{- \beta F} = \int \mathcal{D} \vec{r}(\tau)  \prod_{\vec{k}} \mathcal{D} \alpha_{\vec{k}}(\tau) e^{ - \mathcal{S}[\vec{r}(\tau),\alpha_{\vec{k}}(\tau)]}.
\end{equation}
Because $\mathcal{S}[\vec{r}(\tau),\alpha_{\vec{k}}(\tau)]$ is quadratic in the phonons, they can be integrated out analytically (see textbooks \cite{Altland2010,Wen2004}). The free energy is then given by $e^{- \beta F} = \int \mathcal{D} \vec{r}(\tau) e^{ - \mathcal{S}[\vec{r}(\tau)]}$, with an effective action $\mathcal{S}[\vec{r}(\tau)]$ including retardation effects due to phonon-phonon interactions. For the Bogoliubov-Fr\"ohlich Hamiltonian \eqref{eq:HFroh} one obtains \cite{Peeters1986,Tempere2009}
\begin{equation}
\mathcal{S}[\vec{r}(\tau)] = \int_0^\beta d\tau ~ M \frac{\dot{\vec{r}}^2(\tau)}{2} - \int d^3 \vec{k} ~  \frac{|V_k|^2}{2} \int_0^\beta d\tau \int_0^\beta d\sigma ~ \mathcal{G}(\omega_k, |\tau - \sigma|) e^{ i \vec{k} \cdot \l \vec{r}(\tau) -  \vec{r}(\sigma) \r },
\label{eq:effPolaronPartition}
\end{equation}
where the Green's function of the Bogoliubov phonons reads
\begin{equation}
\mathcal{G}(\omega, u) = \frac{\cosh \l \omega \l u - \beta/2 \r \r }{\sinh \l \beta \omega / 2 \r}.
\end{equation}

\subsection{Jensen-Feynman variational principle}
\label{subsec:FeynmanVariationalPrinc}
Calculating the full path integral of the remaining action is still a difficult task because of retardation effects. To proceed, Feynman introduced a variational principle. To this end he replaced the true action $\mathcal{S}[\vec{r}(\tau)]$ by a simpler variational model action $\mathcal{S}_0[\vec{r}(\tau)]$. Because of the convexity of the exponential function, it holds $\langle e^f \rangle \geq e^{\langle f \rangle}$ according to Jensen's inequality. Using $ e^{- \mathcal{S}_0[\vec{r}(\tau)]} $ as a positive weight, with 
\begin{equation*}
\langle f[\vec{r}(\tau)] \rangle_0 =  e^{\beta F_0} \int \mathcal{D} \vec{r}(\tau) e^{- \mathcal{S}_0[\vec{r}(\tau)]}  ~ f[\vec{r}(\tau)], \qquad  \quad e^{- \beta F_0} = \int \mathcal{D} \vec{r}(\tau) e^{- \mathcal{S}_0[\vec{r}(\tau)]},
\end{equation*}
this leads to 
\begin{flalign*}
\int \mathcal{D}\vec{r}(\tau) e^{- \mathcal{S}[\vec{r}(\tau)]} &= \int \mathcal{D}\vec{r}(\tau)  e^{- \mathcal{S}_0[\vec{r}(\tau)]} ~ \exp \left[ - \l \mathcal{S}[\vec{r}(\tau)] - \mathcal{S}_0[\vec{r}(\tau)] \r \right] \\
& \geq \int \mathcal{D}\vec{r}(\tau)  e^{- \mathcal{S}_0[\vec{r}(\tau)]} ~  \exp \left[  - \langle \mathcal{S}[\vec{r}(\tau)] - \mathcal{S}_0[\vec{r}(\tau)] \rangle_0 \right].
\end{flalign*}

In terms of the free energy the last result reads,
\begin{equation}
F \leq F_0 + \frac{1}{\beta} \langle \mathcal{S} - \mathcal{S}_0 \rangle_0.
\label{eq:varFreeEnergy}
\end{equation}
In order to find an optimal variational action $\mathcal{S}_0$ the right hand side of this equation has to be minimized.

\subsection{Feynman's trial action}
\label{subsec:FeynmanTrialAction}
Feynman started by applying the simplest trial action, which includes only the kinetic energy of the impurity,
\begin{equation}
\mathcal{S}_0^{(0)} =  \int_0^\beta d\tau ~ M \frac{\dot{\vec{r}}^2(\tau)}{2} .
\end{equation}
He showed \cite{Feynman1955} that $\mathcal{S}_0^{(0)}$ yields a ground state energy $E_0$ ($\beta \to \infty$) which corresponds to perturbation theory in the coupling constant $\alpha$. 

Next, Feynman included a static potential $V(\vec{r})$ in the action, the strength of which had to be determined variationally,
\begin{equation}
\mathcal{S}_0^{\rm (sc)} =  \int_0^\beta d\tau ~ M \frac{\dot{\vec{r}}^2(\tau)}{2} +  \int_0^\beta d\tau ~ V(\vec{r}(\tau)).
\label{eq:trialActionS1}
\end{equation}
He suggested to use either a Coulomb potential, $V_{\rm C}(\vec{r}) = Z / r$, or a harmonic potential, $V_{\rm h}(\vec{r}) = k \vec{r}^2$, with $k$ and $Z$ being variational parameters. Using the harmonic variational potential Feynman reproduced Landau and Pekar's strong coupling result with a Gaussian trial wavefunction $\psi_{\rm sc}(\vec{r})$ for the impurity (cf. Eq.\eqref{eq:psiVarSC}). In the case of the Coulomb potential he found an exponential trial wavefunction $\psi_{\rm sc}(\vec{r}) \simeq e^{- \kappa r}$.

The main shortcoming of the trial action \eqref{eq:trialActionS1} is that the translational symmetry is explicitly broken by the potential. Note that the same is true for Landau and Pekar's strong coupling theory. However, to describe correctly the cross-over from a translationally invariant, weakly interacting impurity to the self-trapped strong coupling polaron, a fully translationally invariant trial action should be chosen. 

Feynman achieved this by considering a mobile impurity which is coupled to a fictitious mobile mass $m$. The idea is that the second mass $m$ models the cloud of phonons which follows the impurity in the polaron state. To obtain an efficient trial action $\mathcal{S}_0$ for the impurity alone, the second mass is then integrated out. This can be done analytically if a harmonic coupling with spring constant $m w^2$ is considered. Then $m$ and $w$ are treated as variational parameters.

Thus, the starting point is the full action
\begin{equation}
\mathcal{S}_0[\vec{r}(\tau), \vec{x}(\tau)] = \int_0^\beta d\tau ~ M \frac{\dot{\vec{r}}^2(\tau)}{2} + \int_0^\beta d\tau ~ m \frac{\dot{\vec{x}}^2(\tau)}{2} +  \int_0^\beta d\tau ~ \frac{1}{2} m w^2 |\vec{r}(\tau) - \vec{x}(\tau)|^2,
\label{eq:FeynmanFullTrialAction}
\end{equation}
where $\vec{x}(\tau)$ describes the trajectory of the second mass $m$. Integrating out the second coordinate $\vec{x}(\tau)$ yields Feynman's famous trial action for the polaron \cite{Feynman1955,Tempere2009},
\begin{equation}
\mathcal{S}_0^{\rm (F)}[\vec{r}(\tau)] =  \int_0^\beta d\tau ~ M \frac{\dot{\vec{r}}^2(\tau)}{2} + \frac{m w^3}{8} \int_0^\beta d\tau \int_0^\beta d\sigma ~ \mathcal{G}(w, |\tau - \sigma|) |\vec{r}(\tau) - \vec{r}(\sigma)|^2.
\label{eq:SFeynmanAllCoupling}
\end{equation}

The evaluation of the variational free energy $F_{\rm var} = F_0+ \frac{1}{\beta} \langle \mathcal{S} - \mathcal{S}_0 \rangle_0$, see Eq.\eqref{eq:varFreeEnergy}, can be performed analytically. In Refs.\cite{Tempere2009,Casteels2013a} it was shown that for the Bogoliubov-Fr\"ohlich polaron one obtains
\begin{multline}
F_{\rm var} = \frac{3}{\beta} \left\{ \log \left[ 2  \sinh \l \frac{\beta \Omega}{2} \r \right] - \log \left[ 2  \sinh \l \frac{\beta w}{2} \r \right]  - \frac{1}{2 } \frac{m}{M + m} \left[ \frac{\beta \Omega}{2} \coth \l \frac{\beta \Omega}{2} \r - 1 \right]  \right\}  - \\
-\frac{3}{2 \beta} \log \left[ \frac{M+m}{M} \right] - \int d^3 \vec{k} ~ |V_k|^2 \int_0^\beta du ~ \l 1- \frac{u}{\beta}\r \mathcal{G}(\omega_k,u) \mathcal{M}_{m,\Omega}(\vec{k},u).
\end{multline}
Here $\Omega = w \sqrt{1 + m / M}$ and the memory function is defined as
\begin{equation}
\mathcal{M}_{m,\Omega}(\vec{k},u) = \exp \left[ - \frac{k^2}{2 (M + m )} \l u - \frac{u^2}{\beta}  + \frac{m}{M}  \frac{\cosh \l \Omega \beta / 2 \r - \cosh \l \Omega \l \beta / 2 - u \r \r }{ \Omega \sinh \l \Omega \beta / 2 \r  }  \r \right].
\end{equation}

In the case of (Einstein) optical phonons, where $\omega_k = \omega$ is independent of $k$, Feynman's all-coupling theory \eqref{eq:SFeynmanAllCoupling} yields very accurate results \cite{Peeters1985Feynman}. However, as we will discuss below in chapter \ref{chap:ResultsForExperiments}, large discrepancies between Feynmann's variational solution and Monte-Carlo results are found for the Bogoliubov-Fr\"ohlich polaron. The main shortcoming of Feynman's ansatz in this case is that it can not describe all phonon modes accurately since it only has two variational parameters. A possible way around this limitation would be to consider a trial quadratic action in the spirit of Eq.\eqref{eq:FeynmanFullTrialAction} but assume auxiliary masses $m_k$ with corresponding $k$-dependent spring constants $w_k$. A dependence of the masses $m_k$ on the Matsubara frequency could also be considered.

\subsection{Polaron mass}
\label{subsec:FeynmanEffMass}
So far we only discussed how the polaron ground state energy (or, for finite temperatures, its free energy $F$) can be calculated using the path-integral formalism. Now we turn to the effective mass of the polaron. One way of calculating the effective polaron mass is to derive the polaron energy $E(\vec{q})$ for a given total system momentum $\vec{q}$. Feynman notes, however, that he could not find a generalization of his variational path integral formalism to this case \cite{Feynman1955}. Indeed we see that after applying the Lee-Low-Pines transformation to obtain the total polaron momentum as a conserved quantity, see Sec.\ref{sec:LLP}, the resulting Hamiltonian is no longer quadratic in the phonon operators. Thus phonons can not be integrated out easily anymore. 

Alternatively Feynman derived the polaron mass $M_{\rm p}$ for small momenta from the asymptotic form of the partition function,
\begin{equation}
 \int \mathcal{D} \vec{r}(\tau)  e^{ - \mathcal{S}[\vec{r}(\tau)]} = \exp \l - E_0 T -  \frac{1}{2} M_{\rm p} \vec{X}^2 / T \r
\end{equation}
for $T \to \infty$ and with the boundary conditions $\vec{r}(0)=0$ and $\vec{r}(T)=\vec{X}$. The so obtained expression for the polaron mass, formulated in $d$ spatial dimensions in Ref.\cite{Casteels2012}, reads
\begin{equation}
M_{\rm p} = M + \frac{1}{d} \int d^d k~ k^2 |V_k|^2 \int_0^\infty du~ e^{-\omega_k u} u^2  \lim_{\beta \to \infty} \mathcal{M}_{m,\Omega}(\vec{k},u)
\label{eq:polaronMassFeynman}
\end{equation}
for zero temperature. 

Feynman notes that the variational mass parameter $m$ in the trial action \eqref{eq:SFeynmanAllCoupling} gives a reasonable approximation for the actual polaron mass, $M_{\rm p} \approx m$. However, in the case of the Bogoliubov-Fr\"ohlich model \eqref{eq:HFroh} this approximation is rather poor even in the weak coupling limit \cite{Grusdt2015RG,Tempere2009,Casteels2012}. It is thus advisable to always use the full expression \eqref{eq:polaronMassFeynman} when calculating the polaron energy.

\section{Monte Carlo approaches}
\label{sec:quantMC}
One of the first Monte Carlo (MC) approaches was developed by Becker et al. in Ref. \cite{Becker1983}. The authors started from Feynman's expression for the polaron partition function \eqref{eq:effPolaronPartition} where phonons have already been integrated out. Then they used a classical MC method to perform the remaining multi-dimensional integrals. Their results for electrons interacting with optical phonons were found to be in excellent agreement with Feynman's variational calculations \cite{Peeters1985Feynman}. As for the Feynman's method, the disadvantage of using the effective action Eq.\eqref{eq:effPolaronPartition} with phonons being integrated out is that the polaron momentum $\vec{q}$ can not be modified. 

This difficulty was overcome by Prokof'ev and Svistunov \cite{Prokofev1998} who developed a diagrammatic quantum Monte Carlo method, see also Ref.\cite{Mishchenko2000}. They were able to calculate the full polaron dispersion relation $E_0(\vec{q})$, which can be used to compute the effective polaron mass. Recently the diagrammatic MC procedure was also applied to the Bogoliubov-Fr\"ohlich polaron model \eqref{eq:HFroh}, where substantial deviations from Feynman's path integral approach were found \cite{Vlietinck2015}. We will discuss these results below in chapter \ref{chap:ResultsForExperiments} and compare them to the RG method, which we introduce in the next chapter.

In the diagrammatic quantum MC method \cite{Prokofev1998,Mishchenko2000} the imaginary-time impurity Green's function is calculated exactly (i.e. with a controlled stochastical error),
\begin{equation}
G(\vec{q},\tau) = \bra{0} e^{\H \tau} \ps_{\vec{q}} e^{- \H \tau} \psd_{\vec{q}} \ket{0}, \qquad \tau \geq 0.
\end{equation}
Here $\vec{q}$ is the polaron momentum, $\H$ is the Fr\"ohlich Hamiltonian \eqref{eq:HFroh} and $\psd_{\vec{q}}$ denotes the Fourier transform of the impurity field operator $\psd(\vec{r})$. From the asymptotic behavior at large imaginary times,
\begin{equation}
G(\vec{q},\tau \to \infty) \longrightarrow Z(\vec{q}) \exp \left[ - \l E_0(\vec{q}) - \mu \r \tau \right],
\end{equation}
the polaron dispersion relation $E_0(\vec{q})$ and the quasiparticle weight $Z(\vec{q})$ can be obtained.

To calculate $G(\vec{q},\tau)$ it is written as an infinite-order perturbation series of Feynman diagrams. They consist of the free impurity propagator $G^{(0)}(\vec{p},\tau_2 - \tau_1) = - \theta(\tau_2 - \tau_1) e^{-\l p^2/2M - \mu \r \l \tau_2 - \tau_1 \r}$ (with $\mu$ a tunable chemical potential and $\theta(\tau)$ the step function), the free phonon propagator $D^{(0)}(\vec{k}, \tau) = - \theta(\tau) e^{- \omega_k \tau}$ and the impurity-phonon scattering amplitude $V_{\vec{k}}$. Thus, formally it may be written as
\begin{equation}
G(\vec{q},\tau) = \sum_{m=0}^\infty \sum_{\substack{\rm diagrams \\ \xi_m}} \int_{\substack{\rm four-\\ \rm momenta}} d^4 p_1 ... d^4 p_m ~ F(\vec{q}; \tau ; \xi_m; p_1 , ... , p_m),
\label{eq:prokofevExpression}
\end{equation}
where $m$ denotes the order of a specific diagram with topology $\xi_m$. The four-momenta $p_j = (\tau_j,\vec{p}_j)$ label the phonon momenta at imaginary time $\tau_j$. 

Prokof'ev and Svistunov \cite{Prokofev1998} developed a MC method which allows to evaluate an expression of the form \eqref{eq:prokofevExpression} stochastically. Eq.\eqref{eq:prokofevExpression} involves integrals with a variable number $m$ of integration variables, which is not included in usual MC methods for the evaluation of multi-dimensional integrals. Therefore they had to develop a Metropolis-like prescription allowing to change between different diagram orders during the MC calculation.

\newpage
\chapter{Renormalization group approach}
\label{chap:RGapproach}

\begin{figure}[b!]
\centering
\epsfig{file=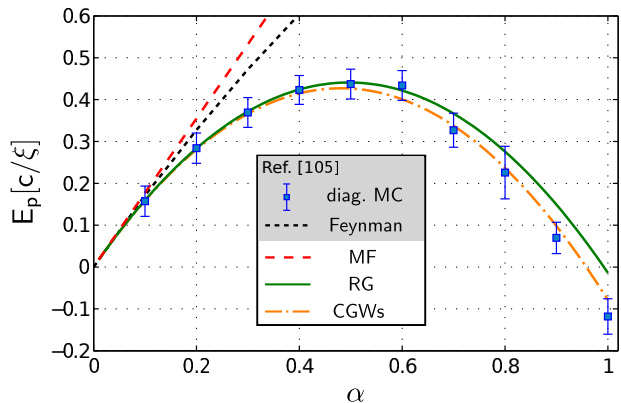, width=0.55\textwidth}
\caption{The polaronic contribution $E_\p = E_0 - 2 \pi a_\IB n_0 m_\text{red}^{-1}$ to the ground state energy of an impurity immersed in a BEC is shown. It corresponds to the ground state energy of the Fr\"ohlich Hamiltonian after the linear power-law divergence predicted by MF theory is regularized. Results from Feynman path-integral calculations and diagrammatic Monte Carlo (MC) calculations -- both taken from Ref. \cite{Vlietinck2015} -- are compared to MF theory, the RG approach and to variational correlated Gaussian wavefunctions (CGWs) \cite{Shchadilova2014}. The results are sensitive to the value of the UV momentum cut-off $\Lambda_0$, here $\Lambda_0=2000 / \xi$. Other parameters are $M/m_\text{B}=0.26316$ and $q=0$. The figure was taken from Ref.\cite{Grusdt2015RG}.}
\label{fig:polaronRGenergies}
\end{figure}

Extending theoretical description of Fr\"ohlich polarons to intermediate coupling strengths requires taking into account correlations between phonons arising from their interaction with the impurity. Such correlations are absent in both the variational mean-field (MF) wavefunction (see Sec.\ref{sec:MFtheory}) and the Landau-Pekar strong coupling wavefunction (see Sec.\ref{sec:LandauPekar}). Effects of quantum fluctuations are at the heart of intermediate coupling polaron physics. Here we present an all-coupling theory for the Fr\"ohlich polaron, based on the renormalization group (RG) approach \cite{Grusdt2015RG}. We apply this method to describe mobile impurity atoms immersed in a BEC using the Bogoliubov-Fr\"ohlich Hamiltonian \eqref{eq:HfrohDiscrete} as a starting point.

When the coupling strength $\alpha \gtrsim 1$ becomes large and the mass ratio $M/m_{\rm B} \lesssim 1/\sqrt{2}$ becomes small, MF theory becomes inaccurate. Recent quantum Monte Carlo (MC) calculations \cite{Vlietinck2015} have shown extremely large deviations of the ground state energy from MF predictions already for small couplings. Comparison of MC results with the mean-field calculations for the Bogoliubov-Fr\"ohlich model is shown in FIG.\ref{fig:polaronRGenergies}. While the MC method \cite{Prokofev1998} is considered to be the most reliable approach for analyzing polaron problems, the analytical insights gained from it are limited. 

The RG approach presented in this chapter accomplishes two main tasks. Firstly it describes the polaron quantitatively all the way from weak to strong coupling. It does so much more efficiently than other analytical approaches, thus providing a numerically convenient method. Secondly, it yields important new analytical insights. In particular in the case of the Bogoliubov-Fr\"ohlich Hamiltonian (in three dimensions) the Monte Carlo analysis exhibited a  logarithmic UV divergence. Analytic insights gained from the RG method allow us to understand the physical origin of this divergence (see Chapter \ref{chap:UVregularization}). As shown in FIG.\ref{fig:polaronRGenergies} the RG approach yields polaron energies which are in excellent agreement with MC results in the weak and intermediate coupling regimes.

The RG method introduced in this chapter builds upon the MF approach to the polaron problem, see Sec.\ref{sec:MFtheory}, by including quantum fluctuations on top of the MF state. In this way correlations between phonons at different momenta are taken into account. The RG formalism relies on a hierarchy of energy scales of phonons with different momenta. This allows to 
implement an iterative construction of the ground state, in which the Hamiltonian is diagonalized with respect to a shell of fast phonons, while modifying the Hamiltonian for the lower energy phonons at the same time. This philosophy is similar to the Born-Oppenheimer approximation used to describe formation of a molecule out of two atoms. Within this method one solves the Schr\"odinger equation for fast electrons assuming that slow ions are stationary. The electron energy obtained from this analysis becomes an effective potential for ions. The simplest mathematical description of an adiabatic decoupling of fast and slow degrees of freedom consists of formulating a product wavefunction which has to be determined self-consistently. However, one can go beyond this simplistic treatment by solving the dynamics of fast degrees of freedom for a given configuration of slow variables. In this way correlations are built into the ground state wavefunction. This is precisely what is done in the RG approach to solving the polaron problem. In this case Schrieffer-Wolff type unitary transformations are used to decouple fast phonons in a high-energy momentum shell from slow phonons at smaller momenta. In practice this is achieved perturbatively in the ratio of fast and slow phonon frequencies. Step by step, this approach gives access to the polaron's ground state wavefunction which  accurately describes the phonon correlations.

In this chapter we apply the RG method to the Bogoliubov-Fr\"ohlich model describing an impurity immersed in a BEC, see Refs.\cite{Grusdt2015RG,Grusdt2015DSPP}. Previously, the intermediate coupling regime of this problem has been studied using Feynman's path integral approach \cite{Tempere2009,Catani2012,Casteels2012} and the numerical MC method \cite{Vlietinck2015}. Beyond the Fr\"ohlich model self-consistent T-matrix calculations \cite{Rath2013}, variational analysis \cite{Li2014,Levinsen2015} and perturbative analysis \cite{Christensen2015} have been performed. Nevertheless, no clear picture of the physics of the Bogoliubov-Fr\"ohlich model at intermediate couplings has been obtained so far. By comparing ground state energies, see FIG.\ref{fig:polaronRGenergies}, substantial deviations of Feynman path-integral results \cite{Tempere2009} from MC predictions were found in \cite{Vlietinck2015}. 

In Ref.\cite{Grusdt2015RG}, relying on the analytical insights from the RG method, the key discrepancy between these predictions was identified: The dependence on the UV cut-off. The RG approach demonstrates analytically that the ground state energy $E_0$ of the Bogoliubov-Fr\"ohlich model diverges logarithmically with the UV cut-off $\Lambda_0$, with the energy going to large negative values. This is true even after the linear power-law divergence of the MF contribution to the polaron energy has been properly regularized. The RG prediction is verified by the variational CGW approach and by diagrammatic MC calculations. On the other hand, the Feynman method predicts a UV convergent ground state energy \cite{Vlietinck2015}.

This chapter is organized as follows. In Sec.\ref{Sec:ModelAndRGcouplings} we introduce the coupling constants which will be renormalized during the shell-by-shell decoupling of phonons. In Sec.\ref{sec:RGformalism} we present the RG formalism and derive the RG flow equations. The ground state energy of the polaron and the logarithmic divergence in the Bogoliubov-Fr\"ohlich model are discussed in Sec.\ref{sec:GSenergyPolaron}. Further equilibrium properties of the polaron ground state are derived from the RG protocol in Sec.\ref{sec:GSpolaronProps}. In Sec.\ref{sec:varApproach} we discuss the variational correlated Gaussian wavefunction (CGW) approach and show how it relates to the RG method.

\section{Fr\"ohlich model and renormalized coupling constants}
\label{Sec:ModelAndRGcouplings}
Our starting point is the Fr\"ohlich Hamiltonian \eqref{eq:HfrohLLPfull} after the Lee-Low-Pines transformation. As discussed in Sec.\ref{sec:MFtheory} the Hamiltonian can be solved approximately using MF theory. We want to build upon the MF solution and include quantum fluctuations on top of it. To this end we apply a unitary transformation
\begin{equation}
\hat{U}_\MF = \exp \l \int^{\Lambda_0} d^3 \vec{k} ~  \alpha_{\vec{k}} \ad_{\vec{k}} -  \hc \r,
\label{eq:UMFfluc}
\end{equation}
which displaces phonon operators by the MF solution,
\begin{equation}
\hat{U}_\MF^\dagger \a_{\vec{k}} \hat{U}_\MF = \a_{\vec{k}} + \alpha_{\vec{k}}.
\end{equation}
As a result, from Eq.\eqref{eq:HfrohLLPfull}, the following polaron Hamiltonian is obtained,
\begin{equation}
\tilde{\cal H}_q = \hat{U}_\MF^\dagger \H_q \hat{U}_\MF =  E_0 |_\MF + \int^{\Lambda_0} d^3 \vec{k} ~ \Omega^\MF_{\vec{k}} \ad_{\vec{k}} \a_{\vec{k}}  +
\int^{\Lambda_0} d^3 \vec{k} d^3 \vec{k}'  ~ \frac{\vec{k} \cdot \vec{k}' }{2 M} :\G_{\vec{k}} \G_{\vec{k}'}:.
\label{H_fluct}
\end{equation}
Here $E_0 |_\MF$ is the MF ground state energy, see Eq.\eqref{eq:MFenergyNonReg}, we defined $\G_{\vec{k}} = \Gd_{\vec{k}}$ as
\begin{equation}
\G_{\vec{k}} := \alpha_{\vec{k}} ( \a_{\vec{k}} + \ad_{\vec{k}} ) + \ad_{\vec{k}} \a_{\vec{k}},
\end{equation}
and $: ... :$ stands for normal-ordering. 

The absence of terms linear in $\a_{\vec{k}}$ in Eq.\eqref{H_fluct} reflects the fact that $\alpha_{\vec{k}}$ correspond to the mean-field (saddle point) solution of the problem. We emphasize that  \eqref{H_fluct} is an exact representation of the original Fr\"ohlich Hamiltonian, where operators $\a_{\vec{k}}$ now describe quantum fluctuations around the MF solution. In the equations above, as well as in the rest of this chapter, $\int^\Lambda d^3 \vec{k}$ stands for a three dimensional integral over a spherical region containing momenta of length $|\vec{k}| < \Lambda$. 

From Eq.\eqref{H_fluct} we notice that the only remaining coupling constant is the (inverse) impurity mass $M^{-1}$. This should be contrasted to the original Hamiltonian \eqref{eq:HfrohLLPfull} before applying the MF shift, where the interaction strength $\alpha$ and the inverse mass $M^{-1}$ define coupling constants. Both are required to determine whether the polaron is in the weak, strong or intermediate coupling regime, see FIG.\ref{fig:QualPhaseDiagFROH}. In the following section \ref{subsec:DimAnRGSec} we will carry out a simple dimensional analysis and show that non-linear terms in \eqref{H_fluct} are marginally irrelevant, allowing an accurate description by the RG (which is perturbative in $M^{-1}$ in every RG step).

To facilitate subsequent RG analysis we generalize the form of the polaron Hamiltonian \eqref{H_fluct} already at this point. To this end we anticipate all required coupling constants generated by the RG flow. The generalized form of the Hamiltonian (\ref{H_fluct}) is
\begin{equation}
 \tilde{\cal H}_q(\Lambda) = E_0 |_\MF + \Delta E + \int^{\Lambda} d^3 \vec{k}~ \l  \Omega_{\vec{k}} \ad_{\vec{k}} \a_{\vec{k}} + W_{\vec{k}} ( \ad_{\vec{k}} + \a_{\vec{k}} ) \r
+\frac{1}{2} \int^{\Lambda} d^3 \vec{k} d^3 \vec{k}'  ~ k_\mu \mathcal{M}_{\mu \nu}^{-1} k_\nu' : \G_{\vec{k}} \G_{\vec{k}'} :.
\label{H_RGuniversal} 
\end{equation}
Note that in equation (\ref{H_RGuniversal}) we introduced $W_{\vec{k}}$ terms that are linear in 
phonon operators $a_k$. Although absent in (\ref{H_fluct}) they will be generated in the course of the RG flow. Ref. \cite{Grusdt2015DSPP} discusses a more sophisticated version of renormalization protocol in which a shift of all $a_k$ operators is introduced in every step of the RG to eiliminate linear terms. In these 

In the expression \eqref{H_RGuniversal} the coupling constant $\mathcal{M}_{\mu \nu}(\Lambda)$ and the RG energy shift $\Delta E(\Lambda)$ depend on the cut-off $\Lambda$ which gets modified during the RG process. Note that interactions are now characterized by a general tensor $\mathcal{M}^{-1}_{\mu\nu}$, where anisotropy originates from the total momentum of the polaron, $\vec{q} = q \vec{e}_x$, breaking rotational symmetry of the system. Indices $\mu=x,y,z$ label cartesian coordinates and summation over repeated indices is implied. Due to the cylindrical symmetry of the problem the mass tensor has the form $\mathcal{M}= \text{diag} ( \mathcal{M}_\parallel , \mathcal{M}_\perp, \mathcal{M}_\perp)$, and we will find different flows for the longitudinal and transverse components. While $\mathcal{M}$ can be interpreted as the (tensor-valued) renormalized mass of the impurity, it should not be confused with the mass of the polaron. For a detailed discussion see Appendix \ref{apdx:renImpMass}. 

The first integral in Eq.\eqref{H_RGuniversal} describes the quadratic part of the renormalized phonon Hamiltonian. It is also renormalized in comparison with the original expression $\Omega_{\vec{k}}^\MF$ in Eq.\eqref{H_fluct},
\begin{equation}
\Omega_{\vec{k}} = \omega_k + \frac{1}{2} k_\mu \mathcal{M}_{\mu \nu}^{-1} k_\nu - \frac{\vec{k}}{M} \cdot \l \vec{q} - \vec{P}_\ph  \r ,
\label{eq:Ok}
\end{equation}
where the momentum carried by the phonon-cloud, $\vec{P}_\ph$, acquires corrections to the MF result $\vec{P}_\ph^\MF$ in the process of the RG flow. In addition there is a term linear in the phonon operators, weighted by
\begin{equation}
W_{\vec{k}} =  \left[  \l \vec{P}_{\ph} - \vec{P}_{\ph}^\MF \r \cdot \frac{\vec{k}}{M} + \frac{k_\mu  k_\nu}{2} \l \mathcal{M}_{\mu \nu}^{-1} - \frac{\delta_{\mu \nu}}{ M}  \r \right] \alpha_{\vec{k}}.
\end{equation}

By comparing Eq.\eqref{H_RGuniversal} to Eq.\eqref{H_fluct} we obtain initial conditions for the RG, starting at the original UV cut-off $\Lambda_0$ where $ \tilde{\cal H}_q(\Lambda_0) = \tilde{\cal H}_q$,
\begin{equation}
\mathcal{M}_{\mu \nu}(\Lambda_0) = \delta_{\mu \nu} M, \quad \vec{P}_\ph(\Lambda_0) = \vec{P}_\ph^\MF, \quad \Delta E (\Lambda_0)=0.
\end{equation}

\section{Renormalization group formalism for the Fr\"ohlich model}
\label{sec:RGformalism}

\begin{figure}[t!]
\centering
\epsfig{file=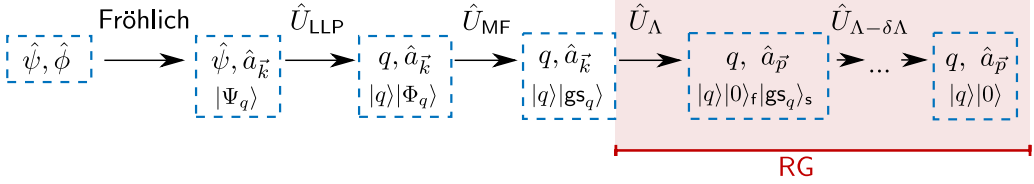, width=0.9\textwidth}
\caption{Sketch of the RG treatment of the Fr\"ohlich polaron problem (in a BEC): After introducing Bogoliubov phonons to describe excitations of the weakly interacting BEC, a Fr\"ohlich Hamiltonian is obtained. It can be further simplified by applying the Lee-Low-Pines transformation, making conservation of total momentum $\vec{q}$ explicit. Next the MF polaron solution is introduced, allowing to transform into a frame describing quantum fluctuations around the MF polaron. To find the ground state including quantum fluctuations, an RG procedure is applied which consists of a series of unitary transformations $\hat{U}_\Lambda$. They lead to a factorization of the ground state in subsequent momentum shells.}
\label{fig:RGsketch}
\end{figure}

Now we proceed to present details of the RG formalism for the Fr\"ohlich polaron. To keep track of all basis transformations we summarize our treatment of the Bose polaron problem in FIG.\ref{fig:RGsketch}. The essence of the RG is to separate a shell of fast phonon modes, with momenta in a thin shell $\Lambda - \delta \Lambda< |\vec{k}| < \Lambda$, and decouple them from the remaining slow phonon modes. This renormalizes the remaining Hamiltonian for slow phonons with momenta $|\vec{p}| < \Lambda- \delta \Lambda$. Such approach is justified by the separation of time-scales associated with slow and fast phonons: In the spirit of the Born-Oppenheimer approximation, slow phonons appear as quasi-static classical variables from the point of view of fast phonons. In practice this means that we may use $1/\Omega_{\vec{k}}$, with $\vec{k}$ a fast-phonon momentum, as a small parameter. This allows us to solve for the ground state of fast phonons, now depending on the slow variables, which introduces entanglement between different phonon modes in the polaron ground state. In practice the RG procedure can be implemented as a consecutive series of unitary transformations $\hat{U}_\Lambda$. We will derive their form explicitly in \ref{sec:GSRGformulation} below.

\subsection{Dimensional analysis}
\label{subsec:DimAnRGSec}

We start our discussion of the RG approach by performing dimensional analysis of different terms in the Hamiltonian. This analysis will not only help us understand the behavior of the RG flow equations obtained in the next subsection, but will also highlight distinct energy regimes, in which phonons contribute differently to polaron properties.

In the following subsection we will perform a momentum-shell RG procedure, i.e. the cut-off $\Lambda$ will be reduced and quantum fluctuations at larger momenta will be decoupled from the problem successively. This gives rise to $\Lambda$-dependent coupling constants, but in this subsection we will ignore such dependence for simplicity. To understand the importance of various terms in Eq. \eqref{H_RGuniversal}, we  assign a scaling dimension $\gamma$ to the fluctuation field, $\a_{\vec{k}} \sim \Lambda^\gamma$, such that contribution of the quadratic term 
$\int^\Lambda \Omega_k a_k^\dagger a_k$ has dimension one at all energy scales. 
\begin{flalign}
\int^\Lambda d^d \vec{k} ~ \ad_{\vec{k}} \a_{\vec{k}} \Omega_{\vec{k}}  \stackrel{!} {\sim} \Lambda^0 = 1 \\
[\Omega_{|k|=\Lambda}] \cdot \Lambda^d \cdot \Lambda^{2\gamma} \sim 1
\label{eq:defEngineering}
\end{flalign}
In this step our analysis is presented for the general case of $d$ spatial dimensions.

A unique property of the Bogoliubov-Fr\"ohlich Hamiltonian is the crossover of the phonon dispersion from quadratic behavior ($\Omega_{\vec{k}} \sim k^2$) for large momenta to linear behavior ($\Omega_{\vec{k}} \sim k$) for small momenta. The crossover takes place around $k_0 \approx \min \l \xi^{-1}, \mathcal{M}_\parallel c \r$. For heavy impurities $k_0\approx1/\xi$ and the crossover is due to the Bogoliubov dispersion, whereas for light impurities $k_0 \approx M c$ and the crossover is caused by impurity fluctuations (we approximated $\mathcal{M}_\parallel \approx M$). To understand the physics at mid-energies, i.e. before crossover to the linear dispersion regime takes place, we need to perform dimensional analysis based on the \emph{quadratic} dispersion. For the actual ground state properties on the other hand, we will examine the regime of \emph{linear} dispersion.

\begin{table}[t!]
\centering
\begin{tabular}{  p{4.2cm} | p{2cm} | p{2.8cm} | p{2.7cm} }
 operator &  $\Lambda \gtrsim 1/\xi$ & $ 1/\xi \gtrsim \Lambda \gtrsim M c$  & $\Lambda \lesssim Mc, 1/\xi$ \\
 \hline
 $\a_{\vec{k}}$ & $\Lambda^{-d/2-1}$ & $\Lambda^{-d/2-1}$  & $\Lambda^{-(d+1)/2}$ \\
 \hline
 $ \int d^d k ~ d^d k' ~k k' \alpha_k \alpha_{k'} ~ a_k a_{k'} $ & $\Lambda^{d-4}$ & $\Lambda^{d-3}$ & $\Lambda^d$ \\
  $ \int d^d k ~ d^d k' ~k k' \alpha_k ~ a_k a_{k'}^2  $ & $\Lambda^{d/2-3}$ & $\Lambda^{d/2-5/2}$ & $\Lambda^{d/2}$ \\
  $ \int d^d k ~ d^d k' ~k k' a_{k'}^2  a_{k}^2$ & $\Lambda^{-2}$ & $\Lambda^{-2}$ & $\Lambda^0$
\end{tabular}
\caption{Dimensional analysis of the Hamiltonian \eqref{H_RGuniversal} in different energy regimes. In the three columns to the right the scaling of various operators with momentum cut-off $\Lambda$ is shown for the three distinct cases described in the main text. The first line shows the engineering dimension of the fluctuations, determined by making the bare phonon dispersion scale as $\int d^d k ~ \Omega_k a_k^2 \sim \Lambda^0$, see Eq.\eqref{eq:defEngineering}. }
\label{tab:dimAn}
\end{table}

In the following we distinguish three different regimes. In the linear dispersion low-energy regime we have $k \lesssim M c, \xi^{-1}$ such that
\begin{equation}
\Omega_{\vec{k}} \sim k, \qquad V_k \sim k^{1/2}, \qquad \alpha_{\vec{k}} \sim k^{-1/2}.
\end{equation}
In the quadratic light-impurity regime it holds $Mc \lesssim k \lesssim \xi^{-1}$ and we obtain
\begin{equation}
\Omega_{\vec{k}} \sim k^2, \qquad V_k \sim k^{1/2}, \qquad \alpha_{\vec{k}} \sim k^{-3/2}.
\label{eq:defLightImp}
\end{equation}
Finally in the quadratic heavy-impurity regime where $\xi^{-1} \lesssim k$ (no matter whether $k \gtrless M c$) it holds
\begin{equation}
\Omega_{\vec{k}} \sim k^2, \qquad V_k \sim 1, \qquad \alpha_{\vec{k}} \sim k^{-2}.
\end{equation}

In Table \ref{tab:dimAn} we summarize the resulting scaling dimensions of different terms of the Hamiltonian (\ref{H_fluct}). We observe a markedly different behavior in the regimes of linear and quadratic phonon dispersion. For intermediate energies (i.e. the low-energy sector of the quadratic dispersion regime) we find that all quantum fluctuations are relevant in three spatial dimensions, $d=3$, with the exception of the two-phonon term in the light-impurity case \eqref{eq:defLightImp}. Even for $d > 6$ the quartic phonon term is always relevant. For low energies in contrast, in all spatial dimensions quantum fluctuations are mostly irrelevant, only the quartic term is marginal. Its pre-factor is given by the inverse mass $M^{-1}$ which we expect to become small due to dressing with high-energy phonons, making the term marginally irrelevant (this will be shown explicitly later). Therefore we expect that the linear regime $\Lambda \ll 1/\xi$ is generically well described by the MF theory on a qualitative level. On a quantitative level we also expect that corrections can be captured accurately by the RG protocol introduced below, which is perturbative in $M^{-1}$ at every step.

\subsection{Formulation of the RG}
\label{sec:GSRGformulation}
Now we turn to the derivation of the RG flow equations for the coupling constants $\mathcal{M}_{\mu\nu}^{-1}(\Lambda)$ and $\vec{P}_\ph(\Lambda)$. After introducing the basic idea of our scheme we will carry out the technical part of the calculations. Then we summarize the resulting RG flow equations. In this section we also derive the RG flow of the polaron ground state energy $\Delta E(\Lambda)$ in the process of decoupling phonons step by step. We discuss the resulting polaron energy renormalization in the next section \ref{sec:GSenergyPolaron}. Effects of renormalization on other important observables (e.g. the polaron mass) will be discussed later in Section \ref{sec:GSpolaronProps}.

\subsubsection{RG step -- motivation}
\label{subsubsec:RGstepMoti}
Generally the idea of the RG procedure is to make use of the separation of time scales, which usually translates into different corresponding length scales. In our case, momentum space provides us with a natural order of energy scales through the phonon dispersion $\Omega_{\vec{k}}$, see Eq.\eqref{eq:Ok}. The latter is mostly dominated by the bare Bogoliubov dispersion $\omega_k$, which ultimately allows us to perform the RG. 

To make use of separation of time scales, we may formally split the Hamiltonian into the slow (labeled \s) and the fast parts (labeled $\f$), as well as the coupling term between the two (labeled $\sf$) \cite{Altland2010},
\begin{equation}
\H = \H_\f + \H_\sf + \H_\s.
\end{equation}
In our case, $\H_\f$ contains only phonons with momenta $\vec{k}$ from a high-energy shell $\Lambda - \delta \Lambda \leq k \leq \Lambda$, where $\Lambda$ is the sharp momentum cut-off and $\delta \Lambda \rightarrow 0$ is the momentum-shell width. $\H_\s$ on the other hand contains the remaining phonons with momenta $\vec{p}$, where $p < \Lambda - \delta \Lambda$. 

In order to integrate out fast degrees of freedom, we decouple the latter from the slow Hamiltonian. In practice this is achieved by applying a unitary transformation $\hat{U}_\Lambda$, which should be chosen such that the resulting Hamiltonian is diagonal in the fast phonon number operators,
\begin{equation}
[\Ud_\Lambda \H \U_\Lambda , \ad_{\vec{k}} \a_{\vec{k}} ] =0.
\label{eq:ULambdaEq}
\end{equation}
While solving this equation exactly for $\U_\Lambda$ is a hopeless task in general, we may at least do so perturbatively. Here we make use of the separation of time scales. As explained above, $\Omega_{\vec{p}} / \Omega_{\vec{k}}  \ll 1$ is automatically the case for most of the slow phonons. The coupling $|| \H_\sf ||$, however, has to be sufficiently weak for our method to work. For this reason, our RG protocol should be understood as a perturbative RG in the coupling defined by the impurity mass $M^{-1}$. 

We will present the perturbative solution of Eq.\eqref{eq:ULambdaEq} for $\U_\Lambda$ below. In the end, when we evaluate the decoupled Hamiltonian in the fast-phonon ground state $\ket{0}_\f$, we obtain 
\begin{equation}
~_\f \bra{0} \Ud_\Lambda \H \U_\Lambda \ket{0}_\f = \H_\s + \delta \H_\s + \mathcal{O}(\Omega_{\vec{k}})^{-2}.
\end{equation}
This is the essence of the RG-step: (i) We find the transformation that diagonalizes the fast phonons and take the fast phonons eigenstate. This effectively reduces the cut-off for the remaining slow-phonon Hamiltonian, $\Lambda \rightarrow \Lambda - \delta \Lambda$. (ii) We calculate the effect of fast phonons in their ground state on the new slow degrees of freedom, which results in the RG flow
\begin{equation}
\H_\s \rightarrow \H_\s + \delta \H_\s = \H_\s'.
\label{eq:formalRenormalization}
\end{equation}
We can now start with the renormalized Hamiltonian $\H_\s'$ and integrate out its highest energy shell in the next RG step. Following this procedure, provided $\H_\s'$ has the same algebraic form as the original Hamiltonian $\H$, we obtain differential RG flow equations for the coupling constants.

\subsubsection{RG step -- formal calculation}
\label{subsubsec:RGstep}
Now we turn to the actual calculation, the individual steps of which were discussed in the last paragraph. In the course of doing this analysis we will prove that (to the considered order in $(\Omega_{\vec{k}})^{-1}$) no more than the two coupling constants $\mathcal{M}_{\mu \nu}^{-1}$ and $\vec{P}_\ph$ introduced in Eq.\eqref{H_RGuniversal} are required. Readers who are not interested in the technical details can skip this section and proceed directly to the RG flow equations presented in the following section.

To begin with, we bring the Hamiltonian \eqref{H_RGuniversal} into a more transparent form by evaluating the normal-ordering operator $: ... :$ in the last term. To this end we make use of the identity
\begin{equation}
:\G_{\vec{k}} \G_{\vec{k}'}: = \G_{\vec{k}} \G_{\vec{k}'} - \delta \l \vec{k} - \vec{k}' \r \left[  \G_{\vec{k}} + |\alpha_{\vec{k}} |^2 \right]
\end{equation}
which we insert in the initial Hamiltonian at $\Lambda=\Lambda_0$. This yields the universal Hamiltonian
\begin{multline}
\tilde{\mathcal{H}}_q(\Lambda) = E_0 |_\MF + \Delta E - \int^{\Lambda_0} d^3 \vec{k} ~ \frac{k^2}{2 M} |\alpha_{\vec{k}}|^2 
+ \frac{1}{2} \int^\Lambda d^3 \vec{k} ~ d^3 \vec{k}' ~ k_\mu \mathcal{M}_{\mu \nu}^{-1} k_\nu'  ~ \G_{\vec{k}} \G_{\vec{k}'} \\
+ \int^\Lambda d^3 \vec{k} \left[ \ad_{\vec{k}} \a_{\vec{k}} \Omega^\MF_{\vec{k}}  + \l  (P_\ph - P_\ph^\MF)_\mu \frac{k_\mu}{M}  - \frac{k^2}{2 M} \r \G_{\vec{k}} \right].
\label{eq:HquantFluc2}
\end{multline}
In the following we will perform the RG procedure described in the last paragraph on this Hamiltonian \eqref{eq:HquantFluc2}.

We proceed by extracting the fast-phonon Hamiltonian from the general expression \eqref{eq:HquantFluc2},
\begin{multline}
\H_\f = \int_\f d^3 \vec{k} \left[ \ad_{\vec{k}} \a_{\vec{k}} \Omega^\MF_{\vec{k}}  + \l  (P_\ph - P_\ph^\MF)_\mu \frac{k_\mu}{M}  + \frac{1}{2} k_\mu W_{\mu \nu} k_\nu \r\G_{\vec{k}} \right] \\
+ \int_\f d^3 \vec{k} ~ \frac{k_\mu  \mathcal{M}_{\mu \nu}^{-1} k_\nu }{2} |\alpha_{\vec{k}}|^2 +  \int_\f d^3 \vec{k} ~ d^3 \vec{k}' ~ \frac{k_\mu \mathcal{M}_{\mu \nu}^{-1} k_\nu'}{2}   : \G_{\vec{k}} \G_{\vec{k}'} : ,
\label{eq:HfUniversalFormRGstep}
\end{multline}
Here, and throughout this section, we use the definition
\begin{equation}
W_{\mu \nu} := \mathcal{M}_{\mu \nu}^{-1} - \delta_{\mu \nu} M^{-1}.
\end{equation}
to make our expressions more handy. Note that in Eq.\eqref{eq:HfUniversalFormRGstep}, unlike in Eq.\eqref{eq:HquantFluc2}, we wrote the phonon-phonon interactions in a normal-ordered form again. As a consequence we observe an energy shift (first term in the second line of \eqref{eq:HfUniversalFormRGstep}), which seems to reverse the effect of the corresponding term in Eq.\eqref{eq:HquantFluc2} where we started from. Importantly, however, here the \emph{renormalized} mass $\mathcal{M}_{\mu \nu}$ appears instead of the bare mass $M$, thus yielding a non-vanishing overall energy renormalization. Below we will furthermore show that the normal-ordered double-integral in the second line of Eq.\eqref{eq:HfUniversalFormRGstep} yields only corrections of order $\mathcal{O}(\delta \Lambda^2)$,  and may thus be neglected in the limit $\delta \Lambda \rightarrow 0$ considered in the RG.

The slow phonon Hamiltonian $\H_\s$ is simply given by Eq.\eqref{eq:HquantFluc2} after replacing $\Lambda \rightarrow \Lambda - \delta \Lambda$ in the integrals. Finally, for the coupling terms we find
\begin{equation}
\H_\sf = \int_\f d^3 \vec{k} \int_\s d^3 \vec{p} ~ k_\mu \mathcal{M}_{\mu \nu}^{-1} p_\nu \G_{\vec{k}} \G_{\vec{p}},
\end{equation}
where use was made of the symmetry $\mathcal{M}_{\mu \nu}^{-1} = \mathcal{M}_{\nu \mu}^{-1}$. Next we will decouple fast from slow phonons. To this end we make an ansatz for the unitary transformation $\U_\Lambda$ as a displacement operator for fast phonons,
\begin{equation}
\U_\Lambda =  \exp \l \int_\f d^3 \vec{k} ~ \left[ \F_{\vec{k}}^\dagger \a_{\vec{k}} - \F_{\vec{k}} \ad_{\vec{k}} \right] \r.
\label{eq:defU}
\end{equation}
Importantly, we assume the shift $\F_{\vec{k}}$ to depend solely on slow phonon operators, i.e.
\begin{equation}
[\F_{\vec{k}} , \a_{\vec{k}}] = [\F_{\vec{k}} , \a^\dagger_{\vec{k}}] = 0.
\end{equation}
The effect of $\hat{U}_\Lambda$ on fast phonons is simply
\begin{equation}
\Ud_\Lambda  \a_{\vec{k}} \U_\Lambda = \a_{\vec{k}} - \F_{\vec{k}}.
\label{eq:UactionRG}
\end{equation}
As a first consequence, normal-ordering of fast phonon operators is unmodified such that 
\begin{equation}
~_\f \bra{0} \int_\f d^3 \vec{k} d^3\vec{k}'~ \frac{k_\mu \mathcal{M}_{\mu \nu}^{-1} k_\nu'}{2}   : \G_{\vec{k}} \G_{\vec{k}'} : \ket{0}_\f = \mathcal{O}(\delta \Lambda^2),
\label{eq:0Uf0}
\end{equation}
where we used that $\F_{\vec{k}}$ is a smooth function of $\vec{k}$. Since the decoupling unitary $\hat{U}_\Lambda$ is chosen such that $\ket{0}_\f$ (with $\a_{\vec{k}} \ket{0}_\f = 0$) is the fast phonon ground state, we may neglect terms from Eq.\eqref{eq:0Uf0}. This is the reason why fast phonon-phonon interaction terms in Eq.\eqref{eq:HfUniversalFormRGstep} can be discarded.

The operator $\F_{\vec{k}}$ can be determined from the condition in Eq.\eqref{eq:ULambdaEq}, i.e. we demand that terms linear in fast phonon operators $\a_{\vec{k}}$ vanish. To this end, let us perform a series expansion in the fast phonon frequency $\Omega_{\vec{k}}$ and note that $\F_{\vec{k}} = \mathcal{O}(\Omega_{\vec{k}})^{-1}$ (for $\Omega_{\vec{k}} = \infty$, fast and slow phonons are decoupled already before applying $\hat{U}_\Lambda$). This allows us to make use of the following identity, valid for any slow-phonon operator $\hat{O}_\s$ (with $\hat{O}_\s = \mathcal{O}(\Omega_{\vec{k}}^0 = 1)$), 
\begin{equation}
\hat{U}_\Lambda^\dagger \hat{O}_\s \hat{U}_\Lambda = \hat{O}_\s + \int_\f d^3 \vec{k}  ~ \left\{  \a_{\vec{k}} [\hat{O}_\s , \F^\dagger_{\vec{k}}]  - \ad_{\vec{k}} [\hat{O}_\s , \F_{\vec{k}}] \right\}  + \mathcal{O}(\Omega_{\vec{k}})^{-2}.
\label{eq:ULambdaOnOs}
\end{equation}
In this way we derive the following equation for $\F^\dagger_{\vec{k}}$, which is a sufficient condition for $\hat{U}_\Lambda^\dagger \H \hat{U}_\Lambda$ to decouple into fast and slow phonons, 
\begin{equation}
 \Omega_{\vec{k}} \F_{\vec{k}}^\dagger = W_{\vec{k}} + \l \alpha_{\vec{k}} - \F_{\vec{k}}^\dagger \r k_\mu \mathcal{M}_{\mu \nu}^{-1} \int_\s d^3 \vec{p} ~  p_\nu \G_{\vec{p}}  + [\H_\s , \F_{\vec{k}}^\dagger] + \mathcal{O}(\Omega_{\vec{k}})^{-2}.
 \label{eq:selfConsEqF}
\end{equation}
As expected, to zeroth order the solution $\F_{\vec{k}}$ of Eq.\eqref{eq:selfConsEqF} vanishes, $\F_{\vec{k}} = \mathcal{O}(\Omega_{\vec{k}})^{-1}$. Higher orders can easily be solved iteratively and we obtain
\begin{multline}
\F_{\vec{k}} =  \frac{1}{\Omega_{\vec{k}}} \left[ W_{\vec{k}} + \alpha_{\vec{k}} k_\mu \mathcal{M}_{\mu \nu}^{-1} 
\int_\s d^3 \vec{p} ~  p_\nu  \G_{\vec{p}}  \right] - 
\frac{1}{ \Omega_{\vec{k}}^2} \left[  \alpha_{\vec{k}} k_\mu \mathcal{M}_{\mu \nu}^{-1}  \int_\s d^3 \vec{p} ~ \Omega^\MF_{\vec{p}} p_\nu \alpha_{\vec{p}} \l \ad_{\vec{p}} - \a_{\vec{p}} \r + \right. \\ \left. +
\l W_{\vec{k}} +
\alpha_{\vec{k}} k_\mu \mathcal{M}_{\mu \nu}^{-1}  \int_\s d^3 \vec{p} ~  p_\nu  \G_{\vec{p}}  \r k_\sigma \mathcal{M}_{\sigma \lambda}^{-1}   \int_\s d^3 \vec{p} ~  p_\lambda  \G_{\vec{p}} \right] + 
\mathcal{O}(\Omega_{\vec{k}})^{-3}.
\label{eq:Fresult}
\end{multline}

Using Eq.\eqref{eq:Fresult} we can calculate the transformed Hamiltonian. It can be written in a compact form
\begin{equation}
\Ud_\Lambda \tilde{\mathcal{H}}_q \U_\Lambda  = \int_\f d^3 \vec{k} ~ \ad_{\vec{k}} \a_{\vec{k}} \l \Omega_{\vec{k}} + \hat{\Omega}_\s(\vec{k}) \r + \delta \H_\s + \H_\s  + \mathcal{O}(\Omega_{\vec{k}})^{-2}.
\label{eq:renHam}
\end{equation}
Here the fast-phonon frequency has been modified due to its coupling to slow degrees of freedom, by the amount
\begin{equation}
\hat{\Omega}_\s(\vec{k}) = k_\mu \mathcal{M}_{\mu \nu}^{-1}  \int_\s d^3 \vec{p} ~ p_\nu \G_{\vec{p}},
\label{eq:Omsk}
\end{equation}
while renormalization of the slow-phonon Hamiltonian reads
\begin{equation}
\delta \H_\s =  \int_\f d^3 \vec{k} ~ \frac{k_\mu  \mathcal{M}_{\mu \nu}^{-1} k_\nu }{2} |\alpha_{\vec{k}}|^2  
- \int_\f d^3 \vec{k} ~ \frac{1}{\Omega_{\vec{k}}} \left[ W_{\vec{k}} + \alpha_{\vec{k}}   k_\mu \mathcal{M}_{\mu \nu}^{-1} \int_s d^3 \vec{p} ~ p_\nu \G_{\vec{p}} \right]^2 + \mathcal{O}(\Omega_{\vec{k}})^{-2}.
\label{eq:renHs}
\end{equation}
This expression includes both, terms depending on the slow phonon operators leading to renormalized coupling constants $\mathcal{M}_{\mu \nu}^{-1}$ and $\vec{P}_\ph$, as well as real numbers describing renormalization of the polaron ground state energy. Finally, let us also mention that in order to calculate Eq. \eqref{eq:renHs}, only the first order terms $\sim \l \Omega_{\vec{k}} \r^{-1}$ are required in $\F_{\vec{k}}$.

\subsection{RG flow equations}
\label{subsubsec:RGflowEQ}
Now we are in a position to derive the RG flow equations for the ground state. To this end we compare the Hamiltonian $\H_\s'$ obtained in the RG step Eq.\eqref{eq:formalRenormalization} to the original one in Eq.\eqref{eq:HquantFluc2}. Using Eq.\eqref{eq:renHs} we find that in the new Hamiltonian the mass term is renormalized, $\mathcal{M}_{\mu \nu}^{-1} \rightarrow \mathcal{M}_{\mu \nu}^{'-1} $ with
\begin{equation}
\mathcal{M}_{\mu \nu}^{'-1} = \mathcal{M}_{\mu \nu}^{-1} - 2 \mathcal{M}_{\mu \lambda}^{-1}  \int_\f d^3 \vec{k} ~ \frac{|\alpha_{\vec{k}}|^2 }{\Omega_{\vec{k}}} k_\lambda k_\sigma  ~\mathcal{M}_{\sigma \nu}^{-1}.
\end{equation}
Rewriting formally $\mathcal{M}_{\mu \nu}^{'-1} = \mathcal{M}_{\mu \nu}^{-1} + \delta \Lambda  \l \partial_\Lambda \mathcal{M}_{\mu \nu}^{-1}(\Lambda) \r$ and employing $d^3 \vec{k} = - d^2 \vec{k} \delta \Lambda$ \footnote{Note the minus sign in $d^3 \vec{k} = - d^2 \vec{k} \delta \Lambda$, accounting for the fact that the RG flows from large to small $\Lambda$.}, we obtain the RG flow equation
\begin{equation}
\frac{\partial \mathcal{M}_{\mu \nu}^{-1} }{\partial \Lambda} = 2 \mathcal{M}_{\mu \lambda}^{-1}  \int_\f d^2 \vec{k} ~ \frac{|\alpha_{\vec{k}}|^2 }{\Omega_{\vec{k}}} k_\lambda k_\sigma  ~\mathcal{M}_{\sigma \nu}^{-1}.
\label{eq:gsFlowM} 
\end{equation}
Analogously the RG flow of the $\mu$-th component of the phonon momentum is obtained,
\begin{equation}
\frac{\partial P_{\ph}^\mu }{\partial \Lambda} = - 2 \mathcal{M}_{\mu \nu}^{-1} \int_\f d^2 \vec{k} ~  \Big[ \l \vec{P}_\ph^\MF - \vec{P}_{\ph} \r \cdot  \vec{k}  + \frac{1}{2} k_\sigma \l \delta_{\sigma \lambda} - M \mathcal{M}_{\sigma \lambda}^{-1} \r k_\lambda \Big] \frac{| \alpha_{\vec{k}} |^2 }{\Omega_{\vec{k}}} k _\nu. 
\label{eq:gsFlowQ}
\end{equation}

\begin{figure}[b!]
\centering
\epsfig{file=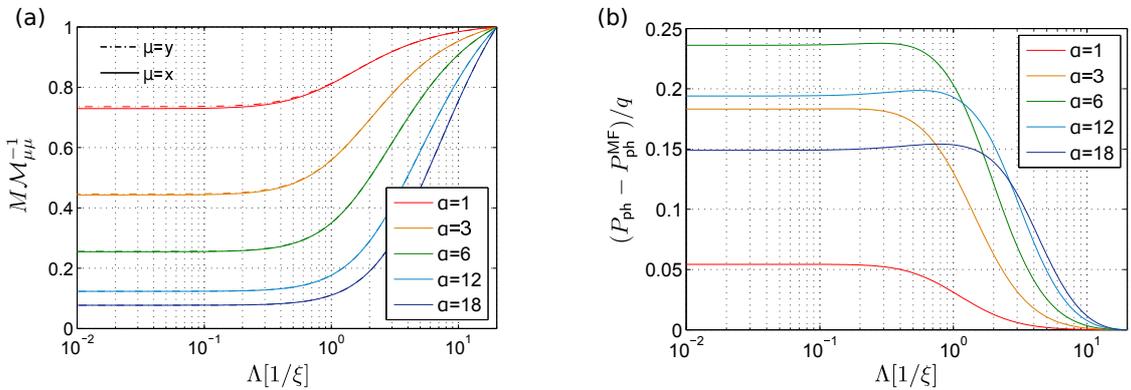, width=\textwidth}
\caption{Typical RG flows of the (inverse) renormalized impurity mass $\mathcal{M}^{-1}$ (a) and the excess phonon momentum $P_\ph-P_\ph^\MF$ along the direction of the system momentum $q$ (b). Results are shown for different coupling strengths $\alpha$ and we used parameters $M/m_\text{B}=0.3$, $q/Mc=0.5$ and $\Lambda_0=20 / \xi$ in $d=3$ dimensions. This figure was taken from Ref.\cite{Grusdt2015RG}.}
\label{fig:RGflowEq}
\end{figure}

\subsection{Solutions of RG flow equations}
\label{sec:GSRGsols}
Now we will discuss solutions of the RG flow equations \eqref{eq:gsFlowM}, \eqref{eq:gsFlowQ}. We find that both the inverse mass $\mathcal{M}_{\mu \nu}^{-1}$ and momentum $\vec{P}_\ph$ are determined mostly by phonons from the intermediate energy region $k \gtrsim 1/\xi$. For smaller momenta the RG flow of the coupling constants effectively stops, in accordance with our expectation based on dimensional analysis in Sec.\ref{subsec:DimAnRGSec}. 

In FIG.\ref{fig:RGflowEq} (a) and (b) a typical RG flow of $\mathcal{M}_{\mu \mu}^{-1} $ and $\vec{P}_\ph (\Lambda)$ is calculated numerically for different values of the coupling constant $\alpha$. In both cases we observe that the coupling constants flow substantially only in the intermediate regime where $\Lambda \approx 1/\xi$. For smaller momenta $\Lambda < 1 /\xi$, as we discussed in \ref{subsec:DimAnRGSec}, all terms in the fluctuation Hamiltonian become irrelevant (or marginal) which manifests itself in the well-converged couplings as $\Lambda \rightarrow 0$. By comparing different $\alpha$, as expected, we observe that corrections to the renormalized impurity mass $\mathcal{M}$ become larger for increasing $\alpha$. Interestingly we observe a non-monotonic behavior for the phonon momentum, which takes a maximum value between $\alpha=6$ and $\alpha=12$ in this particular case. 

\begin{figure}[t!]
\centering
\epsfig{file=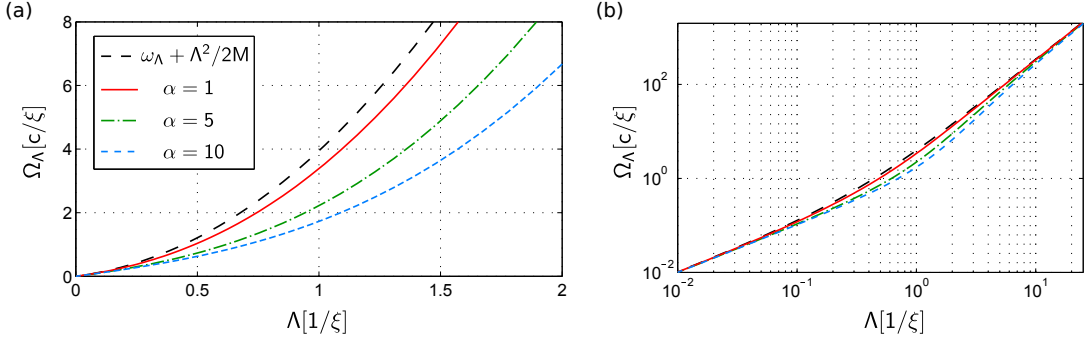, width=\textwidth}
\caption{The renormalized phonon dispersion relation $\Omega_\Lambda$ is compared to the bare dispersion $\omega_\Lambda + \Lambda^2 / 2 M$ for different coupling strengths $\alpha$ in linear (a) and double-logarithmic scale (b). Because the total momentum $q=0$ vanishes there is no direction-dependence of the dispersion. Parameters were $M/m_\text{B}=0.26$ and $\Lambda_0=200 / \xi$.}
\label{fig:RGomegakRenorm}
\end{figure}

In FIG.\ref{fig:RGomegakRenorm} the renormalized phonon dispersion relation $\Omega_k$ is shown as a function of the RG cut-off $k = \Lambda$. Around $\Lambda \approx 1/\xi$ we observe large deviations from the bare dispersion $\omega_k + k^2 / 2 M$. We find that the regime of linear dispersion is extended for large couplings $\alpha$ as compared to the non-interacting case. The slope of the linear regime is given by the speed of sound $c$ and does not change. 

\section{Polaron ground state energy in the renormalization group approach}
\label{sec:GSenergyPolaron}
The first property of the polaron ground state that we will discuss is its energy. In the course of formulating the RG protocol in the last section we have already derived an explicit expression. It consists of the MF term plus corrections from the RG, $E_0^\RG(\Lambda) = E_0 |_\MF + \Delta E(\Lambda)$, which are given by
\begin{equation}
\Delta E(\Lambda) = - \int_\Lambda^{\Lambda_0} d^3 \vec{k} \left\{  \frac{ |\alpha_{\vec{k}}|^2}{2 M} k_\mu  \left[ \delta_{\mu \nu} - M \mathcal{M}_{\mu \nu}^{-1}(k)  \right] k_\nu + \frac{|W_{\vec{k}}|^2}{\Omega_{\vec{k}}}  \right\} + \mathcal{O}(\Omega_{\vec{k}}^{-2}).
\label{eq:DeltaERG}
\end{equation}
To obtain this equation we combined energy shifts from every RG step, see Eq.\eqref{eq:renHs}, with the constant term arising from normal-ordering of the original Hamiltonian, see Eq.\eqref{eq:HquantFluc2}. Note that the fully converged energy is obtained from Eq.\eqref{eq:DeltaERG} by sending $\Lambda \to 0$. 

The resulting polaron energy is plotted in FIG.\ref{fig:polaronRGenergies}. There we found excellent agreement with recent MC calculations \cite{Vlietinck2015}, confirming the validity of the RG approach. However both the RG and MC predict large deviations from MF theory, already for values of the coupling constant $\alpha \approx 1$. The magnitude (of the order of $20 \%$) of the deviations at such relatively small coupling strength is rather surprising. 

\begin{figure}[t!]
\centering
\epsfig{file=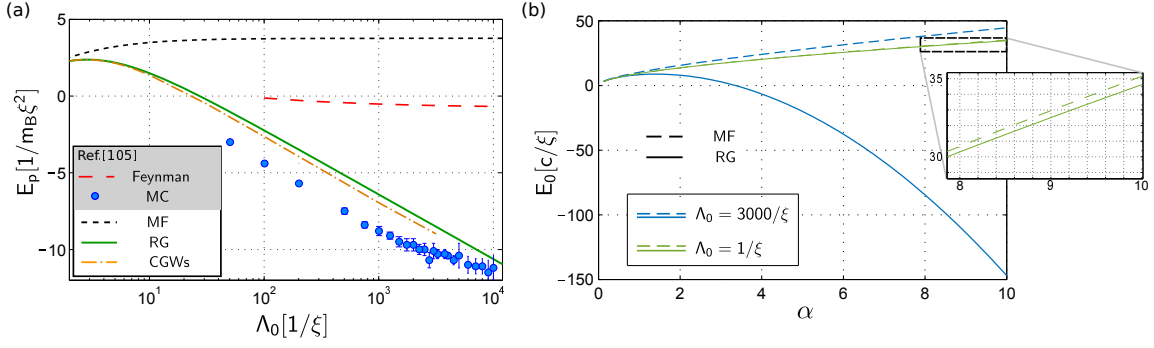, width=\textwidth}
\caption{(a) The polaronic contribution to the ground state energy $E_\p = E_0 - 2 \pi a_\IB n_0 m_\text{red}^{-1}$ in $d=3$ dimensions is shown as a function of the UV momentum cut-off $\Lambda_0$. Note the logarithmic scale. Results from RG, MF theory and correlated Gaussian wavefunctions (CGWs) \cite{Shchadilova2014} are compared to predictions in Ref. \cite{Vlietinck2015} (diagrammatic MC - bullets, Feynman - dashed). The data shows a logarithmic UV divergence of the polaron energy. Parameters are $M/m_\text{B}=0.263158$, $q=0$ and $\alpha=3$. (b) We show the polaron energy $E_0$ as a function of the coupling constant $\alpha$ for two different UV cut-offs $\Lambda_0$. We conclude that the polaron ground state energy $E_0$ depends sensitively on the value of $\Lambda_0$. Parameters are $M/m_\text{B}=0.26$ and $n_0=1\times \xi^{-3}$.}
\label{fig:RGenergyCutOffDependence}
\end{figure}

In FIG.\ref{fig:RGenergyCutOffDependence} (b) we plot the polaron energy $E_0$ again, but for larger values of the coupling constant $\alpha$. For a large value of the UV cut-off $\Lambda_0 = 3000/\xi$ we make a surprising observation that the polaron energy becomes negative, $E_0 < 0$. This is unphysical for the microscopic Hamiltonian \eqref{eq:Hmicro} with repulsive interactions $g_\IB > 0$. This suggests that the approximate Fr\"ohlich model for these values of the coupling constant is not sufficient.

In the following subsection we will analytically derive a log-divergence of the polaron energy within the RG formalism and argue that it is related to zero-point fluctuations of the impurity. In the following chapter we provide a physical explanation of this log-divergence. 

\subsection{Logarithmic UV divergence of the polaron energy}
\label{subsec:LogDiv}

In this section we show that Fr\"ohlich model of BEC polaron has a logarithmic UV divergency of the polaron energy \emph{in addition} to the power law divergence that we pointed out in section \ref{subsubsec:MFpolaronEnergy}. 

Recent work \cite{Vlietinck2015} considered Feynman's variational method and showed that after the power law divergence is regularized, the remaining part of the polaron energy is UV convergent. We summarize these results in FIG.\ref{fig:RGenergyCutOffDependence} (a). Numerical MC results, on the other hand, are less clear. Authors of Ref. \cite{Vlietinck2015} suggested that after regularizing the power law divergence the polaron energy converges around a cut-off $\Lambda_0 \gtrsim 3000/\xi$. When plotted on a logarithmic scale, however, the data shows no clear convergence, see FIG.\ref{fig:RGenergyCutOffDependence} (a). Instead the data suggests 
a \emph{logarithmic UV divergence}, within the errorbars of the MC calculations. In FIG.\ref{fig:RGenergyCutOffDependence} (b) we also compare MC data to the RG prediction Eq.\eqref{eq:DeltaERG}, which is clearly logarithmically divergent. While the overall scale of the RG energy is somewhat different from the numerical MC results, the slopes of the curves, $\partial_{\Lambda_0} E_0(\Lambda_0)$, are in excellent agreement with each other. Only for large $\Lambda_0 \gtrsim 3000/\xi$ deviations are observed, which are of the order of the MC errorbars however.

In the next subsection we will demonstrate logarithmic divergence by an explicit calculation. Our analysis is based on finding the asymptotic form of momentum dependence of $\mathcal{M}(k)$ from RG analysis and putting it into equation (\ref{eq:DeltaERG}). A logarithmic UV divergence for BEC polarons has also been identified using third order perturbation theory \cite{Christensen2015}.

\subsubsection{Asymptotic solution of impurity mass renormalization}
First we consider the spherically symmetric case when $q=0$, where the RG flow equation \eqref{eq:gsFlowM} for the mass is exactly solvable. In this case because of the symmetry $\mathcal{M}_{\mu \nu} = \delta_{\mu \nu} \mathcal{M}$ and the flow equation reads
\begin{equation}
\frac{\partial \mathcal{M}^{-1} }{\partial \Lambda} = \frac{2}{3} \mathcal{M}^{-2}  \int_\f d^2 \vec{k} ~ \frac{|\alpha_{\vec{k}}|^2 }{\Omega_{\vec{k}}} k^2
=  \frac{8 \pi}{3} \mathcal{M}^{-2} \frac{|\alpha_\Lambda|^2}{\Omega_\Lambda} \Lambda^4.
\end{equation}
It is a separable differential equation with the solution
\begin{equation}
\mathcal{M}(\Lambda) = M + \frac{8 \pi}{3} \int_\Lambda^{\Lambda_0} d k ~ \frac{|\alpha_k|^2}{\Omega_k} k^4.
\label{eq:Mspherical}
\end{equation}

From Eq.\eqref{eq:Mspherical} we obtain the asymptotic behavior for $\Lambda, \Lambda_0 \gg \xi^{-1}$.
Using $\alpha_k = -V_k/\Omega^\MF_k$ and the asymptotic expressions for $V_k$ and $\Omega^\MF_k$,
\begin{equation}
\Omega^\MF_k = \frac{k^2}{2 m_\rd}  \l 1 + \mathcal{O}(k^{-1}) \r, 
\qquad V_k^2=n_0 \frac{a_\IB^2}{2 \pi}  
m_{\text{red}}^{-2} + \mathcal{O}(k^{-2}),
\label{eq:asymptoticsOmegaV}
\end{equation}
we arrive at
\begin{equation}
 \mathcal{M}(\Lambda) = M + \frac{32}{3} n_0 a_{\rm IB}^2 m_{\rd} \l \frac{1}{\Lambda} - \frac{1}{\Lambda_0} \r.
 \label{eq:Masympt0}
\end{equation}
Since we assume that this is a perturbative correction to the mass (in the limit $\Lambda,\Lambda_0 \to \infty$), we can rewrite the last expression as
\begin{eqnarray}
1 - \frac{M}{\mathcal{M}(\Lambda)}  = \frac{32}{3} n_0 a_{\rm IB}^2 \frac{m_{\rd}}{M}\, 
\l \,\frac{1}{\Lambda} - \frac{1}{\Lambda_0} \, \r.
\label{eq:Masympt}
\end{eqnarray}
More generally, a perturbative expansion of $\mathcal{M}_{\mu \nu}^{-1}$ in $\Lambda^{-1}$ in the full RG flow equation \eqref{eq:gsFlowM} shows that the last Eq. \eqref{eq:Masympt} is correct even for non-vanishing polaron momentum $q \neq 0$.

\subsubsection{Derivation of the log-divergence from the RG}

The ground state energy in Eq.\eqref{eq:DeltaERG} has two contributions. The first, which has $|W_{\vec{k}}|^2$ in it, is UV convergent because $| W_{\vec{k}} | \propto 1/k$ at large $k$. By using the asymptotic expression for the renormalized impurity mass Eq.\eqref{eq:Masympt}, valid in the UV limit, we find that the second part
\begin{equation}
\Delta E_{\I 0}= - 4 \pi \int_{0}^{\Lambda_0} dk ~ k^2 \frac{|\alpha_k|^2}{2M} k^2 \l 1- \frac{M}{\mathcal{M}(k)} \r
 = - \frac{4 n_0 a_{\IB}^2 }{M } \int_0^{\Lambda_0} dk \l 1- \frac{M}{\mathcal{M}(k)} \r
\label{DeltaE0divergent}
\end{equation}
becomes logarithmically UV divergent 
\begin{eqnarray}
\Delta E_{\text{UV}} = - \frac{128}{3} \frac{m_\rd }{M^2} n_0^2 a_\IB^4  \int_{\xi^{-1}}^{\Lambda_0} dk \,
\l \,\frac{1}{k} - \frac{1}{\Lambda_0} \,\r
\nonumber\\
 \approx - \frac{128}{3} \frac{m_\rd }{M^2} n_0^2 a_\IB^4 \log \l \Lambda_0 \xi \r.
\label{eq:EUVRGmain}
\end{eqnarray}
We find that the prefactor in front of the log predicted by this curve is in excellent agreement with the MC data shown in FIG.\ref{fig:RGenergyCutOffDependence} (a). We also point out the negative sign of the UV divergence, i.e. $\Delta E_{\text{UV}} \to - \infty$ as $\Lambda_0 \to \infty$. This is in contradiction to the fact that the microscopic Hamiltonian \eqref{eq:Hmicro} is positive definite for $g_\IB > 0$, indicating that additional terms besides the approximate Fr\"ohlich Hamiltonian have to be taken into account in the intermediate and strong coupling regimes. We note that similar log-divergences are known to appear e.g. in the Casimir effect in quantum electrodynamics, or in relativistic polaron models \cite{Volovik2014PRD}. 

Before concluding this section we point out that the discussion of UV divergences of the polaron energy is special to the BEC polarons with the $\omega_k$ and $V_k$ given by equations \eqref{eq:BogoliubovDispersion} and \eqref{eq:Vktilde} respectively. In the case of Einstein (optical) phonons with $\omega_k = {\rm const}$ and $V_k \sim 1/k$ the UV convergence of the polaron energy was proven rigorously by Lieb and Yamazaki \cite{Lieb1958}.

\section{Ground state polaron properties from RG}
\label{sec:GSpolaronProps}
In this section we show how properties of the polaron ground state, that are different from the ground state energy, can be calculated from the RG protocol. In particular we discuss the effective polaron mass $M_\p$ (in \ref{subsec:MpCalcRG}), the phonon number $N_\ph$ in the polaron cloud (in \ref{subsec:NphCalcRG}) and the quasiparticle weight $Z$ (in \ref{subsec:ZCalcRG}). In this section the RG-flow equations for these observables are derived. Results will be discussed in chapter \ref{chap:ResultsForExperiments}.

To calculate polaron properties we find it convenient to introduce the following notations for the ground state in different bases used in this text. Our notations are summarized in FIG.\ref{fig:RGsketch}. When $\ket{\Psi_q}$ denotes the ground state of the Fr\"ohlich Hamiltonian Eq.\eqref{eq:HFroh} with total momentum $q$, the corresponding ground state of the Hamiltonian \eqref{eq:HfrohLLPfull} in the polaron frame reads $\ket{q} \otimes \ket{\Phi_q} = \hat{U}^\dagger_\text{LLP} \ket{\Psi_q}$. Analogously, the ground state of the Hamiltonian \eqref{H_fluct} reads $\ket{q} \otimes \ket{\gs_q}=\ket{q} \otimes \hat{U}^\dagger_\MF \ket{\Phi_q}$. To keep our notation simple, we always assume a fixed value of $q$ and introduce the short-hand notation $\ket{\gs} \equiv \ket{\gs_q}$. In the course of the RG, this ground state factorizes in different momentum shells in every single RG step. After the application of the RG unitary transformation $\hat{U}_\Lambda$, the ground state in the new frame reads $\ket{\gs'} := \hat{U}_\Lambda^\dagger \ket{\gs} = \ket{0}_\f \otimes \ket{\gs}_\s$ and factorizes.

\subsection{Polaron Mass}
\label{subsec:MpCalcRG}
First we turn our attention to the polaron mass $M_\p$. As we pointed out in section \ref{subsubsec:gsPolaronMass} it can be determined from the total phonon momentum $q_\ph$, see Eq.\eqref{eq:M*},
\begin{equation}
\frac{M}{M_\p} = 1 - \frac{q_\ph}{q}.
\label{eq:MpUsingq}
\end{equation}
In MF approximation we used $q_\ph = P_\ph^\MF $ defined by Eq.\eqref{eq:XiselfCons}. In the following we will include the effect of quantum fluctuations to derive corrections to the polaron mass. To this end we will calculate corrections to the phonon momentum first, which is defined as
\begin{equation}
q_\ph = \int^{\Lambda_0} d^3\vec{k} ~ k_x \bra{\gs} \U^\dagger_\MF \ad_{\vec{k}} \a_{\vec{k}} \U_\MF \ket{\gs} 
= P_\ph^\MF + \int^{\Lambda_0} d^3\vec{k}~ k_x \bra{\gs} \G_{\vec{k}}  \ket{\gs}.
\label{eq:qPhRG}
\end{equation}
Here $\ket{\gs}$ denotes the ground state in the polaron frame and after introducing quantum fluctuations around MF polaron, i.e. after application of $\hat{U}_\MF$ Eq.\eqref{eq:UMFfluc}.

As shown in the Appendix \ref{apdx:MPfromRGqph}, the phonon momentum $q_\text{ph}$ including corrections from the RG, reads
\begin{equation}
q_\ph =P_\ph(\Lambda \to 0),
\label{eq:qPh}
\end{equation}
where $P_\ph(\Lambda \to 0) = \lim_{\Lambda \rightarrow 0} P_\ph(\Lambda)$ denotes the fully converged RG coupling constant. The last equation justifies the interpretation of $P_\ph$ as the phonon momentum in the polaron cloud.

\subsection{Phonon Number}
\label{subsec:NphCalcRG}
Next we discuss the phonon number $N_\ph$ in the polaron cloud. In the basis of Bogoliubov phonons, before applying the MF shift \eqref{eq:UMFfluc}, the phonon number operator reads $\hat{N}_\ph = \int d^3 \vec{k} ~ \ad_{\vec{k}} \a_{\vec{k}}$. After application of the MF shift Eq.\eqref{eq:UMFfluc} we obtain
\begin{equation}
N_\ph = N_\ph^\MF + \int^{\Lambda_0} d^3 \vec{k} ~ \bra{\gs} \G_{\vec{k}} \ket{\gs},
\label{eq:RGflowNphStart}
\end{equation}
where the MF result reads
\begin{equation}
N_\ph^\MF = \int^{\Lambda_0} d^3 \vec{k} ~ |\alpha_{\vec{k}}|^2.
\end{equation}

The second term on the right hand side of Eq.\eqref{eq:RGflowNphStart} can be evaluated by applying an RG rotation $\U_\Lambda$, and the calculation in Appendix \ref{apdx:MPfromRGqph} leads to the following RG flow equation for the phonon number,
\begin{equation}
\frac{\partial N_\ph}{\partial \Lambda} = 2 \int_\f d^2 \vec{k} ~ \frac{\alpha_{\vec{k}}}{\Omega_{\vec{k}}} \left[ W_{\vec{k}} + \alpha_{\vec{k}} \frac{\vec{k}}{M} \cdot \l \vec{P}_\ph(0) - \vec{P}_\ph \r  \right] + \mathcal{O}(\Omega_{\vec{k}}^{-2}).
\label{eq:RGflowNph}
\end{equation}
It should be supplemented with the initial condition $N_\ph(\Lambda_0) = N_\ph^\MF$. Note that in Eq.\eqref{eq:RGflowNph} the fully converged coupling constant $\vec{P}_\ph(0) = \vec{P}_\ph(\Lambda=0)$ at cut-off $\Lambda=0$ appears and $\vec{P}_\ph \equiv \vec{P}_\ph(\Lambda)$ should be evaluated at the current RG cut-off $\Lambda$.

\subsection{Quasiparticle weight}
\label{subsec:ZCalcRG}
The last observable we discuss here is the polaron quasiparticle weight $Z$, which is a key property characterizing the polaron's spectral function $I(\omega,q)$, see e.g. Refs. \cite{Schirotzek2009,Shashi2014RF}. It is defined by the overlap of the polaron to the bare impurity, $Z=|\bra{\Phi_q} 0 \rangle |^2$ where $\ket{\Phi_q}$ is the phonon ground state in the polaron frame (see FIG.\ref{fig:RGsketch}) and $\ket{0}$ denotes the phonon vacuum in this frame. 

After applying also the MF shift, Eq.\eqref{eq:UMFfluc}, the quasiparticle weight reads 
\begin{equation}
Z = |\bra{\gs}  \prod_{\vec{k}} \ket{- \alpha_{\vec{k}} }|^2,
\label{eq:ZdefRG}
\end{equation}
where we used that 
\begin{equation}
\hat{U}^\dagger_\MF \ket{0} = \prod_{\vec{k}} \ket{- \alpha_{\vec{k}}}.
\label{eq:UMFvac}
\end{equation}
Moreover, $\prod_{\vec{k}}$ includes all momenta $0 < |\vec{k}| < \Lambda_0$ in these expressions.

A characteristic feature of MF theory is that the polaron quasiparticle weight $Z^\MF$ is directly related to its phonon number, 
\begin{equation}
Z^\MF=e^{-N_\ph^\MF}.
\label{eq:ZMF}
\end{equation}
This is a direct consequence of the Poissonian phonon statistics assumed in the MF wavefunction. Indeed we find that it is no longer true for the ground state determined by the RG. 

By introducing unities of the form $\hat{1} = \hat{U}_\Lambda \hat{U}_\Lambda^\dagger$ into Eq.\eqref{eq:ZdefRG} for subsequent momentum shells $\Lambda> \Lambda' =  \Lambda - \delta \Lambda > ...$ we can formulate an RG for the quasiparticle weight. Our calculation is presented in Appendix \ref{apdx:MPfromRGqph} and it leads to the following RG flow equation,
\begin{equation}
\frac{\partial \log Z}{\partial \Lambda} = \int_\f d^2 \vec{k} ~ \left| \alpha_{\vec{k}} - \frac{1}{\Omega_{\vec{k}}} \left[ W_{\vec{k}} - \alpha_{\vec{k}}  k_\mu \mathcal{M}_{\mu \nu}^{-1} \int_\s d^3 \vec{p} ~ p_\nu |\alpha_{\vec{p}}|^2 \right] \right|^2 + \mathcal{O}(\Omega_{\vec{k}}^{-2}).
\label{eq:RGflowZp}
\end{equation}

Comparison of this expression with the RG flow of the phonon momentum Eq.\eqref{eq:RGflowNph} yields
\begin{equation}
\frac{\partial }{\partial \Lambda} \l \log Z + N_\ph \r = 2 \int_\f d^2 \vec{k} ~ \frac{|\alpha_{\vec{k}}|^2}{\Omega_{\vec{k}}} k_x \mathcal{M}_{\parallel}^{-1} 
 \l \int_\s d^3 \vec{p} ~ p_x  |\alpha_{\vec{p}}|^2 + \l P_\ph(0) - P_\ph \r \frac{\mathcal{M}_{\parallel}}{M} \r + \mathcal{O}(\Omega_{\vec{k}}^{-2}),
\label{eq:logZvsNph}
\end{equation}
where we assumed $\vec{q} = q \vec{e}_x$ points along $x$. Thus for $q \neq 0$ we find $Z < e^{-N_\ph}$, i.e. the phonon correlations taken into account by the RG lead to a further reduction of the quasiparticle weight, even beyond an increase of the phonon number.

\section{Gaussian variational approach}
\label{sec:varApproach}
We close this chapter by discussing an alternative approach \cite{Shchadilova2014,Barentzen1975,Nagy1990,Altanhan1993} to describing phonon correlations in the polaron cloud which is closely related to the RG method. This approach is variational and relies on the refinement of the mean-field wavefunction formulated in the polaron frame (i.e. after applying the Lee-Low-Pines transformation, see Sec.\ref{sec:LLP}).

The starting point is the LLP Hamiltonian $\H_{\vec{q}}$, see Eq. \eqref{eq:HfrohLLPfull}. Instead of only displacing all phonon modes coherently, as in the MF case, multimode squeezing between different phonons is included in the variational wavefunction. This is achieved using a so-called correlated Gaussian wavefunction (CGW),
\begin{equation}
\ket{\rm CGW[\beta,Q] } = \hat{\mathcal{D}}[\beta]  \hat{\mathcal{S}}[Q] \ket{0},
\label{eq:defCGW}
\end{equation}
where the multi-mode squeezing is described by
\begin{equation}
\hat{\mathcal{S}}[Q] =  \exp \l \frac{1}{2} \int^{\Lambda_0} d^3 \vec{k} d^3 \vec{k}' ~  Q_{\vec{k},\vec{k}'} \ad_{\vec{k}} \ad_{\vec{k}'} -  \hc \r,
\end{equation}
and a MF-type coherent displacement 
\begin{equation}
\hat{\mathcal{D}}[\beta] = \exp \l \int^{\Lambda_0} d^3 \vec{k} ~  \beta_{\vec{k}} \ad_{\vec{k}} -  \hc \r
\label{eq:UMF_CGW}
\end{equation}
is added. Note that in every step of the RG an infinitesimal amount of similar multi-mode squeezing is generated.

To use \eqref{eq:defCGW} as a variational wavefunction, the energy functional $E_{0}[\beta,Q] = \bra{{\rm CGW}} \H_{\vec{q}} \ket{\rm CGW}$ should be minimized by finding optimal values for all $\beta_{\vec{k}}$ and $Q_{\vec{k},\vec{k}'}$. Recently, a fully self-consistent treatment of the resulting variational problem was suggested \cite{Shchadilova2014}. It was found that in general $\beta_{\vec{k}} \neq \alpha_{\vec{k}}$ differs from the MF coherent amplitude $\alpha_{\vec{k}}$. Because the wavefunction is Gaussian, the Wick-theorem applies and all phonon correlators can be calculated. In particular, 
\begin{equation}
 \hat{\mathcal{S}}^\dagger[Q] \hat{\mathcal{D}}^\dagger[\beta]  \a_{\vec{k}} \hat{\mathcal{D}}[\beta]  \hat{\mathcal{S}}[Q] = \beta_{\vec{k}} + \int d^3 \vec{k}'~ \left[ \cosh Q \right]_{\vec{k},\vec{k}'} \a_{\vec{k}'} + \int d^3 \vec{k}'~ \left[ \sinh Q \right]_{\vec{k},\vec{k}'} \ad_{\vec{k}'},
\end{equation}
see Ref.\cite{Shchadilova2014}. Here $[\cosh Q]_{\vec{k},\vec{k}'}$ denotes the matrix element $\bra{\vec{k}} \cosh Q \ket{\vec{k}'}$ of the matrix function $\cosh Q$, where the matrix $Q$ is defined by its elements $\bra{\vec{k}} Q \ket{\vec{k}'} = Q_{\vec{k},\vec{k}'}$. Dealing with these complicated matrix functions is what makes the variational CGW problem challenging. However a perturbative treatment of the correlations encoded in $Q$ makes the problem manageable \cite{Shchadilova2014}. 

CGWs yield accurate results for the polaron energy, which are very similar to those predicted by the RG. This is shown in FIG. \ref{fig:polaronRGenergies} (cf. Ref.\cite{Shchadilova2014}) for the case of the Bogoliubov-Fr\"ohlich polaron Hamiltonian \eqref{eq:HFroh}. Moreover, starting from the Lee-Low-Pines transformation, they allow to calculate the full polaron dispersion relation $E_0(\vec{q})$ which gives access also to the polaron mass. Some results for ultracold quantum gases will be presented in the following chapter.

Sometimes states of the type \eqref{eq:defCGW} are also referred to as generalized coherent-squeezed states \cite{Lo1993a}. A lot is known about them from quantum optics, see e.g. Refs.\cite{Scully1997,Loudon2000,Meystre2007}. In that case, however, typically only a few modes are considered and matrix functions like $\cosh Q$ can easily be handled. An interesting direction will be to explore whether standard theoretical approaches, used routinely in quantum optics to solve few-mode problems, can be generalized to obtain solutions of the polaron problem efficiently.

\newpage
\chapter{UV Regularization and log-divergence}
\label{chap:UVregularization}

In this chapter we explain how different UV divergences of the polaron energy $E_0$ can be regularized. For the MF polaron we found a power-law UV divergence, see Sec.\ref{subsubsec:MFpolaronEnergy}. From the RG we derived a logarithmic UV divergence which is confirmed by numerical MC calculations, see Sec.\ref{subsec:LogDiv}. Both are specific to the polaron problem in a BEC, thus this chapter can be skipped by readers interested in more generic polaron models. 

The key for regularizing the ground state energy is to note that it has three contributions,
\begin{equation}
E_0 = g_{\rm IB} n_0 + \langle \H_{\rm FROH} \rangle + \langle \H_{\rm 2ph} \rangle,
\end{equation}
see Eq.\eqref{eq:HFroh}. The first term corresponds to the BEC mean-field energy shift. The second term corresponds to the polaronic contribution, given by the Hamiltonian $\H_{\rm FROH}$, and describes impurity-phonon scattering. The third term corresponds to two-phonon scattering events $ \H_{\rm 2ph} = \mathcal{O}(\a^2_{\vec{k}})$, which have been neglected when deriving the Bogoliubov-Fr\"ohlich model, see Sec.\ref{Sec:FroehlichDerivationFull}.

To calculate meaningful polaron energies $E_0$, all results should be expressed in terms of the scattering length $a_{\rm IB}$. It is related to the effective interaction strength $g_{\rm IB}$ through the Lippmann-Schwinger equation (LSE), see Sec.\ref{Sec:LippmannSchwinger}, which solves the two-particle scattering problem. Consistency requires that, to a given order in $a_{\rm IB}$, the LSE for $g_{\rm IB}$ yields UV divergences of the BEC mean-field energy $g_{\rm IB} n_0$ which cancel the UV divergences of $\langle \H_{\rm FROH} \rangle$ and possibly of $\langle \H_{\rm 2ph} \rangle$, see Ref.\cite{Tempere2009}.

\section{Regularization of the power-law divergence}
\label{sec:PowerLawDiv}
In the zeroth order in the interaction $g_{\rm IB}$ we do not include the depletion of the condensate by the impurity. This gives us only the BEC mean-field energy shift,
\begin{eqnarray}
E_0^{(0)} = g_{\rm IB} n_0.
\end{eqnarray}
At this point we realize that it is unphysical to use the bare interaction $g_{\rm IB}$ as the interaction strength. We should solve the LSE to relate the strength of the local pseudopotential, $g_{\rm IB}$, to the scattering length $a_{\rm IB}$. This corresponds to taking the low energy limit of the $T$-matrix, as described in Sec.\ref{Sec:LippmannSchwinger}. Taking the leading order in $a_{\rm IB}$ gives us the regularized energy
\begin{eqnarray}
E_{0}^{(0 {\rm reg})} =  \frac{2 \pi a_{\rm IB}}{m_\rd} n_0.
\end{eqnarray}

Now we add the mean-field part, see Eq.\eqref{eq:MFenergyNonReg},
\begin{equation}
E_\MF = \frac{q^2}{2 M} - \frac{(P_\ph^\MF )^2}{2M} + g_\IB n_0 - \int^{\Lambda_0} d^3 \vec{k} ~ \frac{V_k^2}{\Omega_{\vec{k}}^\MF},
\label{eq:MFenergyNonReg2}
\end{equation}
which gives the ground state energy
\begin{equation}
E_{0}^{(1)} =  E_{0}^{(0 {\rm reg})} + E_\MF.
\end{equation}
The right-most term in Eq.\eqref{eq:MFenergyNonReg2} has the asymptotic behavior
\begin{equation}
- \int^{\Lambda_0} d^3 \vec{k} ~ \frac{V_k^2}{\Omega_{\vec{k}}^\MF} \simeq 
- \Lambda_0 m_{\rm red} n_0 g_{\rm IB}^2 \pi^{-2} =: E_{\rm MF}^{\rm UV},
\label{eq:EmfPowerLawUVdiv}
\end{equation}
which has a power-law UV divergence. We realize that we should not include simultaneous renormalization of $g_{\rm IB} \rightarrow \frac{2\pi a_{\rm IB}}{m_\rd}$ and $E_{\MF}$. This amounts to double counting. Processes that correspond to renormalization of $g_{\rm IB}$ to $\frac{2\pi a_{\rm IB}}{m_\rd}$ (virtual scattering of atoms to high momentum states) are the same ones that are included in $E_{\MF}$. They correspond to condensate particles scattering from $k=0$ to high momenta. 

To remove this double counting we need to subtract
\begin{eqnarray}
E_{0}^{(1 {\rm reg})} = E_{0}^{(0 {\rm reg})}  + E_{\MF} - n_0 V G V.
\label{eq:subtractVGV}
\end{eqnarray}
Using the result (see Sec.\ref{Sec:LippmannSchwinger})
\begin{eqnarray}
VGV  = g_\IB^2 \int \frac{d^3k}{(2\pi)^3} \frac{1}{-\frac{k^2}{2m_\rd}},
\end{eqnarray}
we have after setting $g_{\rm IB} \rightarrow \frac{2\pi a_{\rm IB}}{m_\rd}$ (since we are only working in this order of $a_{\rm IB}$)
\begin{eqnarray}
E_{0}^{(1 {\rm reg})} =  \frac{2 \pi a_{\rm IB}}{m_\rd} n_0 + E_\MF + n_0 \l \frac{2\pi a_{\rm IB}}{m_\rd} \r^2 \int \frac{d^3k}{(2\pi)^3} \frac{1}{\frac{k^2}{2m_{red}}}.
\label{eq:E0MFfullReg}
\end{eqnarray}
Comparison with Eq.\eqref{eq:EmfPowerLawUVdiv} shows that the UV divergence cancels in this expression. From the last equation the regularized MF result in Eq.\eqref{eq:E0MF} is easily obtained.

\section{Explanation of the logarithmic divergence}
\label{sec:LogDiv}

The origin of the logarithmic UV divergence of the ground state energy of the Fr\"ohlich Hamiltonian can be understood 
from the following physical argument. Let us go back to to the lowest order value of the impurity BEC interaction
$E_0^{(0)} = g_{\rm IB} n_0$. As we discussed in the previous section  this expression should be modified so that
instead of the pseudopotential $g_{\rm IB} $ we take the low energy limit of the T-matrix. In the second order in the interaction between impurity and BEC atoms we have 
\begin{eqnarray}
T= g_{\IB} + g_{\IB}^2 \, \int_{|k|<\Lambda_0} \frac{d^3k}{(2\pi)^3} G^0(k) 
\label{Tmx0}
\end{eqnarray}
where $G^0(k)$ is the Green's function of the pair of particles: impurity atom and one host atom (in the center of mass frame)
$
G^0(k) = - {2 m_\rd}/{k^2}
$.
Taking $g_{\IB} = \frac{2\pi a_{\IB}}{m_\rd}$ we find that the second term in (\ref{Tmx0}) has the same power law divergence
as the mean-field energy computed earlier. In the previous section we discussed the reason behind it. T-matrix integrates out virtual scattering of atoms to high energy states in order to obtain scattering amplitude at low energies. And mean-field solution of the polaron problem does essentially the same but in a different language: it includes higher momentum states of Bogoliubov phonons, which at high momenta coincide with the original host atoms.

The RG solution of the Fr\"ohlich model goes beyond the mean-field approach and includes
momentum dependent renormalization of the impurity mass $m_{\text{red}}^*(k)$. Thus we can expect that the RG solution
gives correction to the polaron energy of the type 
\begin{eqnarray}
\Delta E^{\rm Q Fluct}_{0} =  - n_0 g_{\IB}^2 \int \frac{d^3k}{(2\pi)^3} \frac{2 (m^*_\rd(k) -m_\rd)}{k^2}
\end{eqnarray}
In the UV regime the mass renormalization is asymptotically small. Thus, even though integrals involving this small correction may be large, we can write
\begin{eqnarray}
\frac{m^*_\rd(k) -m_\rd}{m_\rd^2} = \frac{\mathcal{M}(k) - M}{M^2} \approx \frac{1}{M} \l 1 - \frac{M}{\mathcal{M}(k)} \r,
\end{eqnarray}
from which we obtain
\begin{eqnarray}
\Delta E^{\rm Q Fluct}_{0} =  - n_0 \left( \frac{ 2 \pi a_{\IB}}{m_\rd} \right)^2 \frac{m_{\rd}^2}{M}
\int \frac{d^3k}{(2\pi)^3} \frac{1}{k^2}  \l 1 - \frac{M}{\mathcal{M}(k)} \r
\end{eqnarray}
This has the same structure of the divergent part as expression (\ref{eq:EUVRGmain}). 

We emphasize that this divergence is more subtle than the power law divergence discussed earlier. The latter could be dealt with by combining the mean-field energy with the 
"background part" of the polaron energy, $g_{\IB} n_0$, and arguing that the divergent parts of the two cancel each other. Proper regularization of the log divergence on the other hand requires proper treatment of two-phonon terms $\langle \H_{\rm 2ph} \rangle$ is required, but a detailed discussion of their effects on the Fr\"ohlich polaron is still lacking. Such terms have been shown to lead to rich physics beyond the Fr\"ohlich model however \cite{Ardila2015}, including the formation of molecules \cite{Rath2013,Li2014}, Efimov trimers \cite{Levinsen2015} and bubble polarons \cite{blinova2013single}.

\newpage
\chapter{Results for experimentally relevant parameters}
\label{chap:ResultsForExperiments}

In this chapter we present numerical results from different theoretical models for experimentally relevant parameters of the Bogoliubov-Fr\"ohlich model. To this end we first discuss the range of validity of this model in Sec.\ref{sec:ExpParameters}. In Sec.\ref{sec:RFspectroscopy} we summarize how the polaron mass $M_{\rm p}$, its energy $E_0$ and the quasiparticle weight $Z$ can be measured experimentally using RF spectroscopy. Then, in Sec.\ref{sec:PolPropsResults}, we present numerical data, except for the ground state energy which we discussed extensively in Chaps. \ref{chap:UVregularization} and \ref{chap:RGapproach}.

\section{Experimental considerations}
\label{sec:ExpParameters}

The microscopic model in Eq.\eqref{eq:Hmicro} applies to a generic mixture of cold atoms. Such systems can be realized experimentally either by mixing atoms of different species \cite{Hadzibabic2002,Roati2002} (including the possibility of choosing different isotopes \cite{truscott2001observation,Schreck2001}), or by using one species but with different meta-stable (e.g. hyperfine) ground states \cite{Weitenberg2011,Fukuhara2013}. The model \eqref{eq:Hmicro} can be used as well to describe a charged impurity immersed in a BEC, as realized e.g. with $~^+ \text{Ba}$ and $~^+ \text{Rb}$ ions \cite{schmid2010dynamics}. Experiments cited above provide only a few examples, and in fact a whole zoo of Bose-Bose \cite{Egorov2013,Spethmann2012,Pilch2009,Lercher2011,mccarron2011dual,Catani2008,Hohmann2015} and Bose-Fermi \cite{Hadzibabic2002,Roati2002,truscott2001observation,Schreck2001,Wu2012a,Park2012,Shin2008,Bartenstein2005,Ferlaino2006,Ferlaino2006err,Inouye2004,Scelle2013,Stan2004,Schuster2012,Ferrier-Barbut2014} mixtures have been realized. 

Now we will discuss whether these experiments can be described by the Fr\"ohlich polaron model \eqref{eq:HFroh}, when interpreting atoms from the minority species as impurities. To this end we derive conditions on the experimental parameters under which the Fr\"ohlich Hamiltonian \eqref{eq:HFroh} can be obtained from the microscopic model \eqref{eq:Hmicro}. In addition we present typical experimental parameters for such a case.

To give readers some familiarity with typical length and energy scales in current experiments with ultracold atoms, we present realistic numbers for the specific case of a Rb-Cs mixture \cite{Spethmann2012,Pilch2009,Lercher2011,mccarron2011dual,Hohmann2015}. For a typical $~^{87}\text{Rb}$ BEC with a density of $n_0 \approx 10^{13} \text{cm}^{-3}$ and $a_\text{BB} \approx 100 a_0$ \cite{Egorov2013,Chin2010} (with $a_0$ the Bohr radius) we obtain values for the healing length $\xi \approx 0.9 \mu \text{m}$ and for the speed of sound $c\approx 0.6 \text{mm}/\text{s}$. The characteristic time scale associated with phonons is thus $\xi / c = 1.5 \text{ms}$. For a $~^{133} \text{Cs}$ impurity the inter-species scattering length is $a_{\text{RbCs}} = 650 a_0$ \cite{mccarron2011dual}, leading to a dimensionless coupling constant $\alpha=0.25$. The mass-ratio of $M/m_\text{B} \approx 1.5$ is of the order of unity.

\subsection{Conditions for the Fr\"ohlich model}
\label{subsec:condFrohModel}
Now we turn to the discussion of condition \eqref{eq:condBogoFroh}, $ |g_\IB| \ll 4 c \xi^2$, which requires sufficiently \emph{weak} impurity-boson interactions $g_\IB$ for the Fr\"ohlich Hamiltonian to be valid. On the other hand, to reach the interesting intermediate coupling regime of the Fr\"ohlich model, coupling constants $\alpha$ larger than one $\alpha \gtrsim 1$ -- i.e. large interactions $g_\IB$ -- are required\footnote{We assume mass ratios $M/m_\text{B} \simeq 1$ of the order of one throughout this text. The case of very light impurities is discussed using perturbation theory in $M$ by Ref.\cite{Grusdt2015DSPP}.}. We will now discuss under which conditions both $\alpha \gtrsim 1$ and Eq. \eqref{eq:condBogoFroh} can simultaneously be fulfilled. 

To this end we express both equations in terms of the experimentally relevant parameters $a_\text{BB}$, $m_\text{B}$ and $M$ which are assumed to be fixed, and we treat the BEC density $n_0$ as well as the impurity-boson scattering length $a_\IB$ as variable parameters. Using the first-order Born approximation result $g_\IB = 2 \pi a_\IB / m_\text{red}$, see Eq.\eqref{eq:aIBgIBaBBgBB}, condition \eqref{eq:condBogoFroh} reads
\begin{equation}
\epsilon := 2 \pi^{3/2} \l 1 + \frac{m_\text{B}}{M} \r  a_\IB \sqrt{a_\text{BB} n_0} \stackrel{!}{\ll} 1.
\label{eq:epsExp}
\end{equation}
Similarly the polaronic coupling constant can be expressed as
\begin{equation}
\alpha = 2 \sqrt{2 \pi} \frac{a_\IB^2 \sqrt{n_0}}{\sqrt{a_\text{BB}}}.
\label{eq:alphaExp}
\end{equation}

Both $\alpha$ and $\epsilon$ are proportional to the BEC density $n_0$, but while $\alpha$ scales with $a_\IB^2$, $\epsilon$ is only proportional to $a_\IB$. Thus to approach the strong coupling regime $a_\IB$ has to be chosen sufficiently large, while the BEC density has to be small enough in order to satisfy Eq.\eqref{eq:epsExp}. 

When setting $\epsilon = \epsilon_\text{max} \ll 1$ and assuming a fixed impurity-boson scattering length $a_\IB$, we find an upper bound for the BEC density,
\begin{equation}
n_0 \leq n_0^\text{max} = \l 1 + m_\text{B}/M \r^{-2} \l \frac{a_\IB / a_0}{100} \r^{-2} \l \frac{a_\text{BB} / a_0}{100} \r^{-1} \epsilon_\text{max}^2 \times 5.45 \times 10^{16} \text{cm}^{-3},
\label{eq:rhoMax}
\end{equation}
where $a_0$ denotes the Bohr radius. For the same fixed value of $a_\IB$ the coupling constant $\alpha$ takes a maximal value
\begin{equation}
\alpha^{\text{max}} = \epsilon_\text{max} \frac{\sqrt{2}}{\pi}  \l 1 + m_\text{B}/M \r^{-1} \frac{a_\IB}{a_\text{BB}},
\end{equation}
which is compatible with condition \eqref{eq:condBogoFroh}.

\begin{table*}[b!]
 \renewcommand{\arraystretch}{1.4}
 \centering
\begin{tabular}{c||c|c|c|c|c|c}
\hline
$a_\text{Rb-K}/a_0$ & 284. & 994. & 1704. & 2414. & 3124. & 3834. \\
$\alpha^\text{max}_\text{Rb-K}$ & 0.26 & 0.91 & 1.6 & 2.2 & 2.9 & 3.5 \\
$n_0^{\text{max}} [10^{14} \text{cm}^{-3}]$ &  2.8 & 0.23 & 0.078 & 0.039 & 0.023 & 0.015 \\
\hline 
$a_\text{Rb-Cs}/a_0$ &  ~650. & 1950. & 3250. & 4550. & 5850. & 7150. \\
$\alpha^\text{max}_\text{Rb-Cs}$ &  0.35 & 1.0 & 1.7 & 2.4 & 3.1 & 3.8 \\
$n_0^{\text{max}} [10^{14}  \text{cm}^{-3}]$ & 0.18 & 0.02 & 0.0073 & 0.0037 & 0.0022 & 0.0015 \\
\hline
\end{tabular}
\caption{Experimentally the impurity-boson scattering length $a_\IB$ can be tuned by more than one order of magnitude using a Feshbach-resonance. We consider two mixtures ($~^{87}\text{Rb} - ~^{41} \text{K}$, top and $~^{87}\text{Rb} - ~^{133} \text{Cs}$, bottom) and show the maximally allowed BEC density $n_0^\text{max}$ along with the largest achievable coupling constant $\alpha^\text{max}$ compatible with the Fr\"ohlich model, using different values of $a_\IB$ and choosing $\epsilon_\text{max}=0.3 \ll1$. This table was taken from Ref.\cite{Grusdt2015RG}.}
\label{tab:alphaMaxNmax}
\end{table*}

\subsection{Experimentally achievable coupling strengths}
\label{subsec:achievableCouplings}
Before discussing how Feshbach resonances allow to reach the intermediate coupling regime, we estimate values for $\alpha^{\text{max}}$ and $n_0^\text{max}$ for typical background scattering lengths $a_\IB$. Despite the fact that these $a_\IB$ are still rather small, we find that keeping track of condition \eqref{eq:epsExp} is important. 

To this end we consider two experimentally relevant mixtures, (i) $~^{87}\text{Rb}$ (majority) -$~^{41}\text{K}$ \cite{Catani2012,Catani2008} and (ii) $~^{87}\text{Rb}$ (majority) -$~^{133}\text{Cs}$ \cite{mccarron2011dual,Spethmann2012}. For both cases the boson-boson scattering length is $a_\text{BB}=100 a_0$ \cite{Egorov2013,Chin2010} and typical BEC peak densities realized experimentally are $n_0=1.4 \times 10^{14} \text{cm}^{-3}$ \cite{Catani2008}. In the first case (i) the background impurity-boson scattering length is $a_\text{Rb-K}=284 a_0$ \cite{Chin2010}, yielding $\alpha_\text{Rb-K} = 0.18$ and $\epsilon=0.21<1$. By setting $\epsilon_\text{max}=0.3$ for the same $a_{\text{Rb-K}}$, Eq.\eqref{eq:rhoMax} yields an upper bound for the BEC density $n_0^\text{max}=2.8 \times 10^{14} \text{cm}^{-3}$ below the value of $n_0$, and a maximum coupling constant $\alpha^{\text{max}}_{\text{Rb-K}} = 0.26$. For the second mixture (ii) the background impurity-boson scattering length $a_\text{Rb-Cs}=650 a_0$ \cite{mccarron2011dual} leads to $\alpha_{\text{Rb-Cs}} = 0.96$ and $\epsilon = 0.83 < 1$. Setting $\epsilon_\text{max}=0.3$ for the same value of $a_{\text{Rb-Cs}}$ yields $n_0^\text{max} = 0.18 \times 10^{14} \text{cm}^{-3}$ and $\alpha_{\text{Rb-Cs}}^{\text{max}}=0.35$. We thus note that already for small values of $\alpha \lesssim 1$, Eq.\eqref{eq:epsExp} is often \emph{not} fulfilled and has to be kept in mind.

The impurity-boson interactions, i.e. $a_\IB$, can be tuned by the use of an inter-species Feshbach resonance \cite{Chin2010}, available in a number of experimentally relevant mixtures \cite{Pilch2009,Park2012,Ferlaino2006,Ferlaino2006err,Inouye2004,Stan2004,Schuster2012}. In this way, an increase of the impurity-boson scattering length by more than one order of magnitude is realistic. 

In Table \ref{tab:alphaMaxNmax} we show the maximally achievable coupling constants $\alpha^\text{max}$ for various impurity-boson scattering lengths. We consider the two mixtures from above ($~^{87}\text{Rb} - ~^{41} \text{K}$ and $~^{87}\text{Rb} - ~^{133} \text{Cs}$), where broad Feshbach resonances are available \cite{Catani2012,Pilch2009,Ferlaino2006,Ferlaino2006err}. We find that coupling constants $\alpha \approx 1$ in the intermediate coupling regime can be realized, which are compatible with the Fr\"ohlich model and respect condition \eqref{eq:condBogoFroh}. The required BEC densities are of the order $n_0 \sim 10^{13} \text{cm}^{-3}$, which should be achievable with current technology. Note that if Eq.\eqref{eq:condBogoFroh} would not be taken into account, couplings as large as $\alpha \sim 100$ would be possible, but then $\epsilon \sim 8 \gg 1$ indicates the importance of the phonon-phonon scatterings shown in FIG.\ref{fig:IBvortices} (c,d). 

In summary we conclude that a faithful realization of the intermediate coupling ($\alpha \approx 1$) Fr\"ohlich polaron in ultracold quantum gases should be possible with current technology.

\section{RF spectroscopy}
\label{sec:RFspectroscopy}

Radio-frequency (RF) spectroscopy \cite{Gupta2003,Stewart2008} provides one of the most powerful tools for the investigation of polarons. Essentially it consists of a measurement of the full spectral function $I(\omega,q)$ of the impurity, when an RF pulse is used to change the hyperfine state
of impurity atoms from a state that does not interact with the BEC to the interacting one. 
For other theoretical discussions of this problem see Refs. \cite{Rath2013,Shashi2014RF,Kain2014,Grusdt2015dRG}. The data presented below was obtained from time-dependent mean-field calculations \cite{Shashi2014RF}, see also Chap.\ref{chap:polaronBO}, but the RG method can be used as well to calculate polaron spectra \cite{Grusdt2015dRG}. In the case of Fermi-polarons, RF spectra have been successfully measured \cite{Schirotzek2009}. This experimental technique can be directly carried over to Bose polarons, and RF spectra in Bose-Bose mixtures have indeed been measured \cite{Wild2012} already.

\begin{figure}[t!]
\centering
\epsfig{file=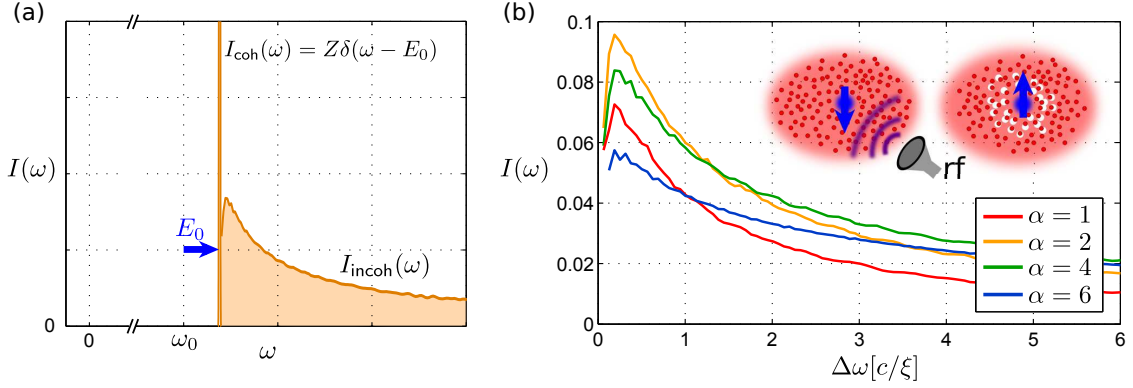, width=0.99\textwidth}
\caption{In this section we discuss RF spectra of an impurity immersed in a BEC, as shown in the inset of (b). The interacting impurity is described by the Fr\"ohlich Hamiltonian. (a) The characteristic properties of an impurity RF spectrum, taken for a given impurity momentum $q$, allow to extract equilibrium properties of the polaron. (b) Inverse RF spectra calculated from MF theory \cite{Shashi2014RF} are shown for different coupling strengths $\alpha$ at $\vec{q}=0$. Only the incoherent part $I_\text{incoh}(\Delta \omega)$ is plotted as a function of the frequency off-set $\Delta \omega$ from the coherent polaron peak. Other parameters are $M/m_\text{B}=0.26$ and $\Lambda_0 \approx 200/\xi$ (a soft UV cut-off was used to avoid unphysical high-frequency oscillations resulting from a sharp cut-off).}
\label{fig:polaronRFintro}
\end{figure}

Momentum dependent RF spectra $I(\omega,q)$ allow to extract a number of equilibrium properties of the polaron, as summarized in FIG.\ref{fig:polaronRFintro} (a). The full spectrum $I(\omega,q)$ moreover contains information about excitations of the polaron. For example, a power-law tail has been identified at high energies \cite{Wild2012,Shashi2014RF} and can be related to universal two-body physics \cite{Tan2008,Tan2008a}. 

\subsection{Basic theory of RF spectroscopy}
\label{sec:RFspectra}
In the following we discuss the RF spectrum of a Fr\"ohlich polaron in a BEC. To this end we consider an atomic mixture with a minority ($=$ impurities) and a majority species ($=$ BEC). We assume that the impurities are sufficiently dilute, allowing to treat them independently of each other. This may safely be assumed when their average distance $n_\I^{-1/3}$ (where $n_\I$ denotes their average density) is well below the typical polaron radius, i.e. for $n_\I^{-1/3} \ll \xi$. 

We assume that impurity atoms have two spin states $\sigma=\uparrow, \downarrow$ which interact with the bosons with different interaction strengths $g_\IB^{\sigma}$. In practice they correspond to different hyperfine states of the atoms. In the absence of the BEC the two spin states are off-set by an energy difference $\omega_0$. The Hamiltonian for this system reads
\begin{equation}
\H_\text{RF} = \frac{\omega_0}{2}  \sigma_z + \ket{\uparrow} \bra{\uparrow} \otimes \H(g_\IB^\uparrow) + \ket{\downarrow} \bra{\downarrow} \otimes \H(g_\IB^\downarrow) + \biggl( \Omega  e^{i \omega t} \ket{\downarrow}  \bra{\uparrow}+ \hc \biggr) , 
\label{eq:HamiltonianRF}
\end{equation}
where $\H(g_\IB^\sigma)$ denotes the Fr\"ohlich Hamiltonian \eqref{eq:HFroh} for the interaction strength $g_\IB^\sigma$. For simplicity we consider the case when $g_\IB^{\downarrow}=0$ in the discussion below. The last term in Eq.\eqref{eq:HamiltonianRF} describes the RF-field induced transition, which is driven with strength $\Omega$ at frequency $\omega$. It conserves the total momentum $q$ because the momentum transfer of the RF beams $k_{\rm RF} \ll 1/\xi$ can be safely neglected. 

The RF spectrum can be obtained from the absorption cross-section $P_{\rm abs}(\omega)$ of the RF beam in the atomic cloud. Consider the case when initially all impurity atoms are in the state $\ket{\downarrow}$ at momentum $q$. The absorption cross-section of an RF beam is then related to the transition rate into the second spin-state $\ket{\uparrow}$,
\begin{equation}
P_{\rm abs}(\omega) \propto \Gamma_{\uparrow, \downarrow} = 2 \pi |\Omega|^2 I(\omega,q).
\end{equation}
According to Fermi's golden rule the spectral function $I(\omega,q)$ is given by
\begin{equation}
I(\omega,q) = \sum_n |\bra{\psi_\uparrow^n} \uparrow \rangle \bra{\downarrow} \psi_\downarrow^0 \rangle |^2 \delta \l \omega - ( E_\uparrow^n - E_\downarrow^0 ) \r.
\label{eq:polaronSpectrum}
\end{equation}
Here $\ket{\psi^{0}_\uparrow}$ is the initial non-interacting state with energy $E_\downarrow^0$ and $\ket{\psi^{n}_\uparrow}$ label all $n=0,1,...$ interacting eigenstates with energies $E_\uparrow^n$.

The RF-spectroscopy described above, where a non-interacting state is flipped into an interacting one, is referred to as \emph{inverse} RF. In the \emph{direct} RF protocol, in contrast, the interacting ground state $\ket{\psi_\uparrow^0}$ is coupled to non-interacting states $\ket{\psi_\downarrow^n}$. In this case, Eq.\eqref{eq:polaronSpectrum} still holds after exchanging spin-labels in the expression ($\uparrow $ to $\downarrow$ and $\downarrow$ to $\uparrow$). 

In the rest of this section, for concreteness, we will only be concerned with the inverse RF spectrum. Its qualitative properties are the same as those of the direct RF protocol. In particular, both spectra allow to measure characteristic polaron properties. We note that, from an experimental point of view, the inverse RF protocol has the conceptual advantage of being less sensitive to finite polaron life-time: Before applying the spin-flip, the non-interacting (or weakly interacting) impurity is stable. Strongly interacting impurities close to a Feshbach-resonance, on the other hand, can have a finite lifetime due to energetically lower molecular impurity-boson bound states, see Refs.\cite{Rath2013,Levinsen2015}.

\subsection{Basic properties of RF spectra}
\label{sec:RFspectraProps}
A typical RF spectrum is shown in FIG.\ref{fig:polaronRFintro}. The Bogoliubov-Fr\"ohlich polaron is a stable quasiparticle with an infinite lifetime, which corresponds to a coherent delta-function peak (amplitude $Z$) in the spectrum. The finite intrinsic line width of the RF transition, as well as other experimental limitations \cite{Schirotzek2009}, are discarded here. 

The coherent peak is located at the polaron ground state energy, $\omega = \omega_0 + E_0(q)$. Moreover, because the impurity is coupled to phonons by the impurity-boson interaction, we obtain a broad spectrum of phonon excitations at larger frequencies $\omega - \omega_0 > E_0(q)$. These features can be identified in FIG. \ref{fig:polaronRFintro} (a). Thus the spectral function takes the form
\begin{equation}
I(\omega,q) = Z \delta \l \omega - (\omega_0 + E_0) \r + I_{\text{incoh}}(\omega,q).
\label{eq:IwqDef}
\end{equation}
This can be shown from Eq.\eqref{eq:polaronSpectrum} using a Lehmann expansion, see \cite{Shashi2014RF,Abrikosov1965}. 

Thus a measurement of the position of the polaron peak is sufficient to obtain the polaron energy $E_0(q)$. By comparing RF spectra at different polaron momenta $q$ the dispersion relation $E_0(q)$ can be measured, which gives access also to the polaron mass $M_\text{p}$. 

The spectral weight $Z$ of the coherent part $I_{\text{coh}}(\omega,q)=Z \delta \l \omega-(\omega_0 + E_0(q)) \r$ is given by the quasiparticle weight $Z$. The quasiparticle weight describes the amount of free-impurity character of the polaron,
\begin{equation}
Z=|\bra{\rm free ~particle} \rm polaron \rangle|^2.
\end{equation}
The incoherent part $I_{\text{incoh}}(\omega,q)$ is non-vanishing only for $\omega - \omega_0 > E_0(q)$, and it fulfills the following sum-rule (which follows from Eqs.\eqref{eq:polaronSpectrum}, \eqref{eq:IwqDef})
\begin{equation}
\int d \omega ~ I_{\text{incoh}}(\omega,q) = 1 - Z.
\label{eq:SumRuleIwq}
\end{equation}
Using this sum-rule the quasiparticle weight $Z$ can be determined from the absorption spectrum $P_{\rm abs}(\omega)$, even when the non-universal prefactor relating $P_{\rm abs}(\omega)$ and $I(\omega)$ is unknown.

The RF spectrum of an impurity can serve as a fingerprint of polaron formation, and it was used to proof polaron formation of impurities in a Fermi sea \cite{Schirotzek2009}. Let us summarize the required signatures to claim that a polaron has formed. First of all the presence of a coherent (delta)-peak in the RF spectrum proofs the existence of a long-lived quasiparticle in the system. Secondly, to distinguish a non-interacting impurity from a polaron -- i.e. a dressed impurity -- the spectral weight $Z$ of the coherent peak has to be measured. Only when $Z < 1$ the quasiparticle has a phonon cloud characteristic for the polaron. Alternatively, because of the sum-rule Eq.\eqref{eq:SumRuleIwq}, it is sufficient to show the existence of an incoherent tail in the RF spectrum to conclude that $Z < 1$.

\section{Properties of polarons}
\label{sec:PolPropsResults}

As shown in the last section, many polaron properties can be measured experimentally using e.g. RF spectroscopy. Now we present numerical results and compare different theoretical models. We start by discussing the polaron mass, proceed with the phonon number, and close with the quasiparticle weight. To compare different theoretical approaches, sometimes we use coupling strengths $\alpha \gg 1$ beyond what is experimentally achievable.

\subsection{Polaronic Mass}
\label{subsec:MpResRG}

\begin{figure}[b!]
\centering
\epsfig{file=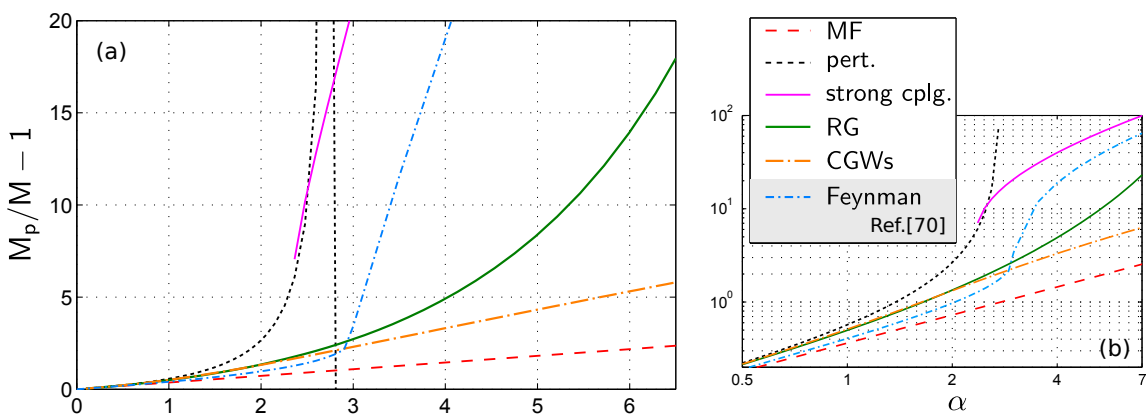, width=\textwidth}
\caption{The polaron mass $M_\p$ (in units of $M$) is shown as a function of the coupling strength $\alpha$. We compare the RG method to MF, strong coupling theory \cite{Casteels2011}, correlated Gaussian wavefunctions (CGWs) \cite{Shchadilova2014} and Feynman's variational path-integral approach. We are grateful to Wim Casteels for providing his results of Feynman path-integral calculations \cite{Casteels2012}. We used parameters $M/m_\text{B}=0.26$, $\Lambda_0=200 / \xi$ and set $q/Mc=0.01$. In (b) the same data is shown as in (a) but in a double-logarithmic scale. The figure was taken from Ref.\cite{Grusdt2015RG}.}
\label{fig:MpCFtempere}
\end{figure}

FIG.\ref{fig:MpCFtempere} shows the polaron mass, calculated using several different approaches. In the weak coupling limit $\alpha \to 0$ the polaron mass can be calculated perturbatively in $\alpha$, and the lowest-order result is shown in the figure. Around $\alpha \approx 3$ the perturbative result diverges and perturbation theory is no longer valid. We observe that in the limit $\alpha \to 0$ all approaches follow the same line which asymptotically approaches the perturbative result. The only exception is the strong coupling Landau-Pekar approach, which only yields a self-trapped polaron solution above a critical value of $\alpha$, see Sec.\ref{subsec:SCpolaronMass}.

For larger values of $\alpha$, MF theory sets a lower bound for the polaron mass. Naively this would be expected, because MF theory does not account for quantum fluctuations due to couplings between phonons of different momenta. These fluctuations require additional correlations to be present in beyond-MF wavefunctions and should lead to an increased polaron mass. Indeed, for intermediate couplings $\alpha \gtrsim 1$ the RG approach predicts a polaron mass $M_\p^\RG > M_\p^\MF$ which is considerably different from the MF result. 

In FIG.\ref{fig:MpCFtempere} we present another interesting aspect of our analysis, related to the nature of the cross-over \cite{Gerlach1988,Gerlach1991} from weak to strong coupling polaron regimes. While Feynman's variational approach predicts a rather sharp transition, the RG results show no sign of any discontinuity. Instead they suggest a smooth cross-over from one into the other regime, as expected on general grounds \cite{Gerlach1988,Gerlach1991}. It is possible that the sharp crossover obtained using Feynman's variational approach is an artifact of the limited number of parameters used in the variational action.

\begin{figure}[b!]
\centering
\epsfig{file=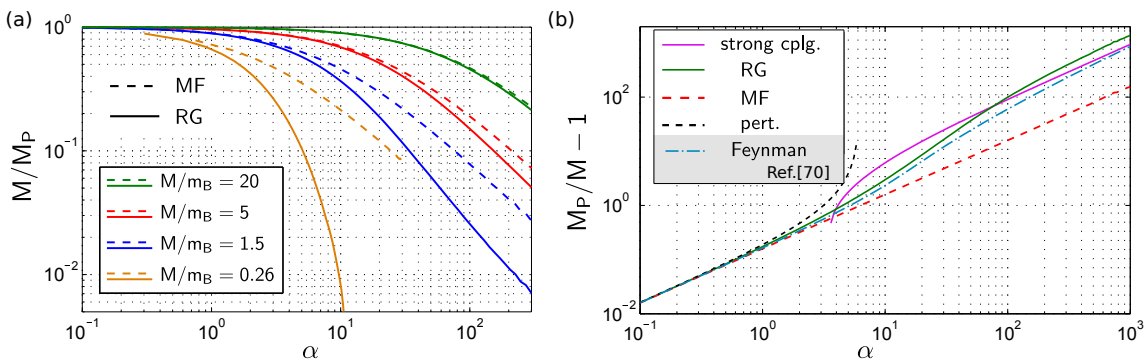, width=\textwidth}
\caption{(a) The inverse polaron mass $M/M_\p$ is shown as a function of the coupling strength $\alpha$, for various mass ratios $M/m_\text{B}$. We compare MF (dashed) to RG (solid) results. The parameters are $\Lambda_0=2000 / \xi$ and we set $q/Mc=0.01$ in the calculations. (b) The polaron mass $M_\p/M-1$ is shown as a function of the coupling strength for an impurity of mass $M=m_\text{B}$ equal to the boson mass. We compare MF, perturbation and strong coupling theories to the RG as well as to Feynman path-integral results by Wim Casteels \cite{Casteels2012}. We used parameters $\Lambda_0=200 / \xi$ and $q/Mc=0.01$. The figure was taken from Ref.\cite{Grusdt2015RG}.}
\label{fig:MpSC}
\end{figure}

In FIG.\ref{fig:MpCFtempere} the polaron mass is calculated in the strongly coupled regime for rather large $\alpha$ while the mass ratio $M/m_\text{B}=0.26$ is very small. It is also instructive to see how the system approaches the integrable limit $M \to \infty$ when the problem becomes exactly solvable, see Sec.\ref{sec:MFtheory}. FIG.\ref{fig:MpSC} (a) shows the (inverse) polaron mass as a function of $\alpha$ for different mass ratios $M/m_\text{B}$. For $M \gg m_\text{B}$, as expected, the corrections from the RG are negligible and MF theory is accurate. 

When the mass ratio $M/m_\text{B}$ approaches unity, we observe deviations from the MF behavior for couplings above a critical value of $\alpha$ which depends on the mass ratio. Remarkably, for very large values of $\alpha$ the mass predicted by the RG follows the same power-law as the MF solution, albeit with a different prefactor. This can be seen more clearly in FIG.\ref{fig:MpSC} (b), where the case $M/m_\text{B}=1$ is presented. This behavior can be explained from strong coupling theory. As shown in \cite{Casteels2011} the polaron mass in this regime is predicted to be proportional to $\alpha$, as is the case for the MF solution. However prefactors entering the weak coupling MF and the strong coupling masses are different.

To make this more precise, we compare the MF, RG, strong coupling and Feynman polaron masses for $M/m_\text{B}=1$ in FIG.\ref{fig:MpSC} (b). We observe that the RG smoothly interpolates between the strong coupling and the weak coupling MF regime. While the MF solution is asymptotically recovered for small $\alpha \to 0$ (by construction), this is not strictly true on the strong coupling side. Nevertheless, the observed value of the RG polaron mass in FIG.\ref{fig:MpSC} (b) at large $\alpha$ is closer to the strong coupling result than to the MF theory. 

\begin{figure}[t!]
\centering
\epsfig{file=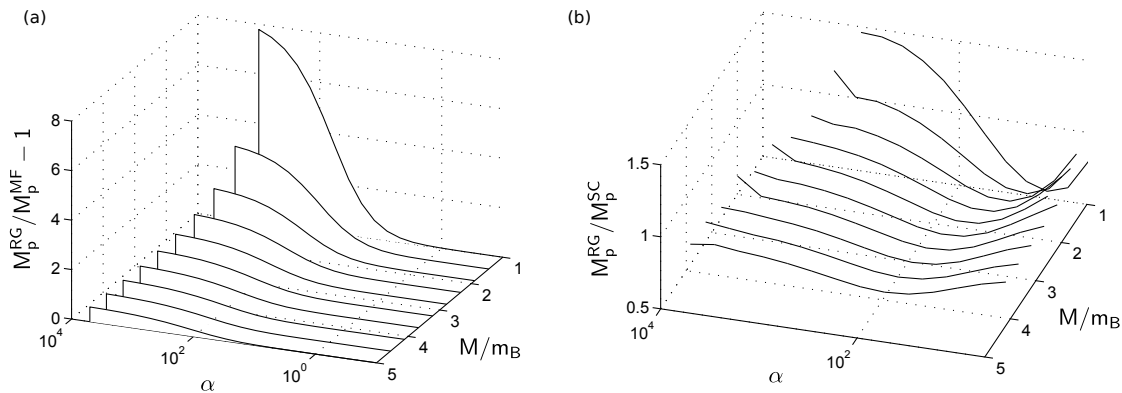, width=\textwidth}
\caption{Transition from weak to strong coupling regime: (a) The ratio of the RG effective polaron mass $M_\p^{\RG}$ to the weak coupling MF prediction $M_\p^\MF$ is shown as a function of the mass ratio $M/m_\text{B}$ and the coupling strength $\alpha$. (b) The ratio of the RG effective polaron mass $M_\p^{\RG}$ to the strong coupling Landau-Pekar prediction $M_\p^\text{SC}$ \cite{Casteels2011} is shown as a function of the mass ratio $M/m_\text{B}$ and the coupling strength $\alpha$. We used parameters $\Lambda_0=200 / \xi$ and $q/Mc=0.01$. Note that the strong coupling solution exists only for sufficiently large values of $\alpha$.}
\label{fig:MpSCtoMFtransition}
\end{figure}

In FIG.\ref{fig:MpSCtoMFtransition} we investigate the relation between RG and weak and strong coupling results more closely. As expected, the MF result is accurate for large mass ratios or small $\alpha$, and large deviations are observed otherwise, see FIG.\ref{fig:MpSCtoMFtransition} (a). The strong coupling result, on the other hand, is not as well reproduced by the RG for large $\alpha$, but deviations are much smaller than for the MF theory in this regime, see \ref{fig:MpSCtoMFtransition} (b). 

Now we return to the discussion of the polaron mass for systems with a small mass ratio $M/m_\text{B}<1$. In this case FIG.\ref{fig:MpSC} (a) suggests that there exists a large regime of intermediate coupling, where neither strong coupling approximation nor MF can describe the qualitative behavior of the polaron mass. This is demonstrated in FIG.\ref{fig:MpCFtempere}, where the RG predicts values for the polaron mass midway between MF and strong coupling, for a wide range of coupling strengths. We find that in this intermediate regime, to a good approximation, the polaron mass increases exponentially with $\alpha$, over more than a decade. In this intermediate coupling regime, the impurity is constantly scattered between phonons, leading to strong correlations between them. Here it acts as an exchange-particle mediating interactions between phonons. These processes change the behavior of the polaron completely, until the impurity mass becomes so strongly modified by phonons that a MF-like behavior of the renormalized impurity is restored in the strong coupling regime.

We conclude that measurements of the polaron mass rather than the binding energy should be a good way to discriminate between different theories describing the Fr\"ohlich polaron at intermediate couplings. Quantum fluctuations manifest themselves in a large increase of the effective mass of polarons, in strong contrast to the predictions of the MF approach based on the wavefunction with uncorrelated phonons. Experimentally both the quantitative value of the polaron mass, as well as its qualitative dependence on the coupling strength can provide tests of the RG theory. The mass of the Fermi polaron has successfully been measured using collective oscillations of an atomic cloud \cite{Nascimbene2009}, and similar experiments should be possible with Bose polarons in the near future.

\subsection{Phonon Number}
\label{subsec:NphResRG}

\begin{figure}[t!]
\centering
\epsfig{file=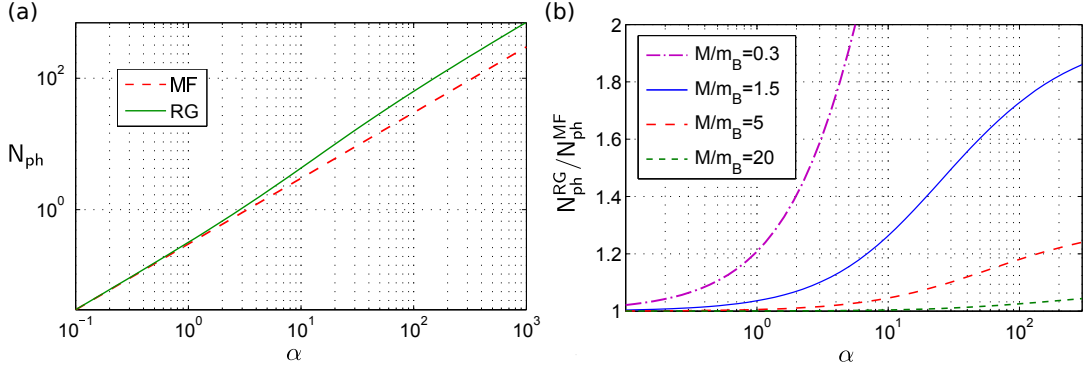, width=\textwidth}
\caption{(a) The phonon number is plotted as a function of the coupling $\alpha$ on a double-logarithmic scale, using RG and MF theory. Parameters were $M/m_\text{B}=1$,  $q=0.01 Mc$ and $\Lambda_0=2000/\xi$. (b) The ratio between RG phonon number $N_\ph^\RG$ in the polaron cloud and the MF result $N_\ph^\MF$ is shown as a function of the coupling constant $\alpha$ for various mass ratios. Parameters are $q=0.01 Mc$ and $\Lambda_0=2000/\xi$.}
\label{fig:NphRGres}
\end{figure}

In FIG.\ref{fig:NphRGres} (a) we plot the phonon number in the polaron cloud for one specific example. We observe that for $\alpha \lesssim 1$ RG and MF are in good agreement with each other. For couplings $\alpha > 1$ the RG predicts more phonons than MF theory, as expected from the presence of quantum fluctuations leading to additional dressing. The qualitative behavior of the phonon number, however, does not change for larger couplings. For very large $\alpha$ we find the same power-law as predicted by MF theory, but with a different prefactor. This is another indicator of the smooth transition from weak to strong coupling regime.

To make this smooth cross-over even more apparent, we calculated the ratio of the RG phonon number to the MF prediction in FIG.\ref{fig:NphRGres} (b). The ratio starts to grow around $\alpha = 1$ and eventually it saturates at very larger values of $\alpha \approx 10^2$. The ratio of RG phonon number to MF theory at the largest couplings $\alpha$ increases with decreasing mass ratio. In the integrable limit $M \to \infty$ no deviations can be observed at all.

\subsection{Quasiparticle weight}
\label{subsec:ZResRG}

\begin{figure}[b!]
\centering
\epsfig{file=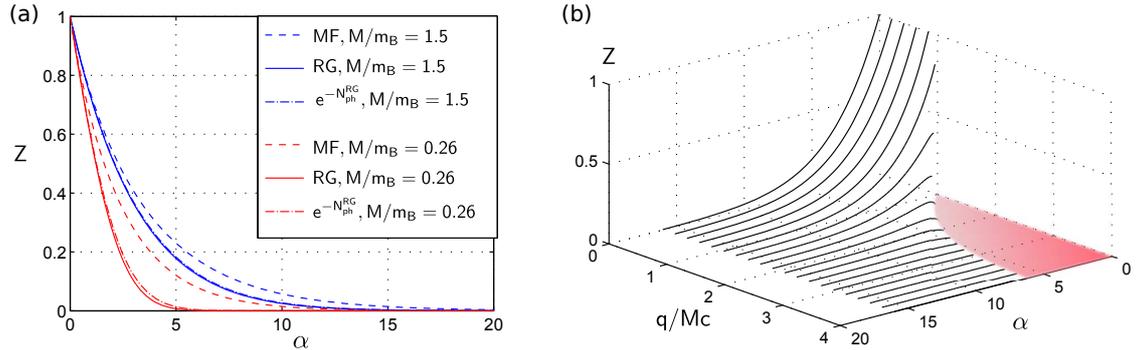, width=\textwidth}
\caption{(a) The RG quasiparticle weight $Z$ is compared to MF predictions and to the MF-type expression $\exp \l - N_\ph^\RG \r$ for different mass ratios. Parameters are $q=0.75 Mc$ and $\Lambda_0=2000/\xi$. (b) The quasiparticle weight $Z$ of the polaron peak (calculated from RG) is shown as a function of the polaron momentum $q$ and the coupling strength $\alpha$. For $q \geq M c$ and sufficiently small $\alpha$ the polaron becomes supersonic and the quasiparticle weight $Z=0$ vanishes (red shaded area). Parameters are $M/m_\text{B}=1.53$ and $\Lambda_0=200/\xi$.}
\label{fig:ZRGres}
\end{figure}

In FIG.\ref{fig:ZRGres} (a) we show how the quasiparticle weight $Z$ depends on the coupling strength $\alpha$ for two different mass ratios. By plotting the data from FIG.\ref{fig:ZRGres} (a) on a logarithmic scale, we found that while MF yields an exponential decay of $Z$, the RG predicts faster than exponential decay. In FIG.\ref{fig:ZRGres} (a) we also compare $Z$ to the MF-type expression $e^{- N_\ph}$. Note however that we calculate $N_\ph$ using the RG in this case. For the smaller mass ratio $M/m_\text{B}=0.26$ we observe slight deviations of $Z$ from this expression, indicating that the RG polaron includes correlations between phonons going beyond the Poissonian statistics of the MF state. 

In FIG.\ref{fig:ZRGres} (b) it is also shown how the quasiparticle weight depends on the polaron momentum $q$. For $q<Mc$ the polaron is subsonic for all couplings $\alpha$ and the quasiparticle weight $Z$ decays as a function of $\alpha$. For $q \geq Mc$ and for sufficiently small couplings $\alpha$ we find a supersonic polaron where $Z=0$. For $\alpha=0$ the quasiparticle weight $Z$ jumps discontinuously from $Z=1$ at $q < Mc $ to $Z=0$ for $q > Mc$. FIG.\ref{fig:ZRGres} (b) indicates that while $Z(q)$ decreases on the subsonic side when approaching the supersonic polaron, it still jumps discontinuously at the critical polaron momentum $q_c$. For large $\alpha$, however, the polaron quasiparticle weight $Z$ is exponentially suppressed and it is hard to distinguish a smooth cross-over from a sharp transition. For large $q > M c$ we find that the function $Z(\alpha)$ takes a maximum value at finite $\alpha > 0$. It is suppressed at smaller couplings due to the proximity to the supersonic polaron, and at larger couplings due to the additional interactions.

\newpage
\chapter{Example of a Dynamical Problem: Bloch oscillations of Bose polarons}
\label{chap:polaronBO}
One of the most exciting new frontiers in the area of polaron problems is dynamical out of equilibrium phenomena. Recently dynamical problems involving mobile impurities coupled to a bath of bosons have become a topic of major interest, both theoretically and experimentally. For example, in cuprate compounds an onset of superconductivity well above the critical temperature $T_c$ was observed when the system was driven out-of-equilibrium by applying teraherz radiation that resonantly excited phonon modes \cite{Fausti2011}. While out of equilibrium solid state systems remain rather confusing due to many possible phenomena happening at the same time, ultracold gases provide an alternative, conceptually simpler setting with an extra advantage of fully tunable model parameters. For example, by preparing impurity atoms inside a superfluid the dynamics of polaron formation has been studied in Refs. \cite{Fukuhara2013,Catani2012}. In another recent experiment the relaxation dynamics of impurity atoms in an optical lattice has been investigated, with decoherence due to the interaction of impurity atoms with the surrounding phonons of a BEC \cite{Scelle2013}.

Describing full non-equilibrium dynamics of polarons is theoretically a challenging task. In one-dimensional systems the time-evolving block decimation (TEBD) method can be used, which has lead to a study of impurity transport in strongly interacting quantum gases \cite{Bruderer2010,Johnson2011}. In another approach impurity dynamics in higher-dimensional systems were calculated using a Gutzwiller ansatz \cite{Wernsdorfer2010}. In the rest of this chapter we review a powerful time-dependent variational approach \cite{Shashi2014RF}, which builds upon the Lee-Low-Pines mean-field theory presented in Sec.\ref{sec:MFtheory}. We illustrate the approach by applying it to describe Bloch oscillations of lattice polarons \cite{Grusdt2014BO}, which allows to derive analytical results for transport properties (see Sec.\ref{sec:PolaronBO}). We also note that one can go beyond a mean-field treatment and include dynamics of quantum fluctuations, by generalizing the RG method of Chap.\ref{chap:RGapproach}  to non-equilibrium problems \cite{Grusdt2015dRG}. Multiband effects on lattice polarons were also recently discussed in the context of ultracold atoms \cite{Yin2015}.

\section{Time-dependent mean-field approach}
\label{sec:tDependentMF}
Now we introduce the time-dependent variational MF approach \cite{Shashi2014RF} for the Bogoliubov-Fr\"ohlich Hamiltonian \eqref{eq:HFroh}. Our starting point is the Hamiltonian $\H_{\vec{q}}$ in the polaron frame, see Eq.\eqref{eq:HfrohLLPfull}, i.e. after applying the LLP transformation (see Sec.\ref{sec:LLP}). To describe the phonon dynamics in this frame, we consider a time-dependent variational wavefunction of the from
\begin{equation}
\ket{\psi_{\vec{q}}(t)} = e^{- i \chi_{\vec{q}}(t) } \exp \l \int d^3 \vec{k} ~  \alpha_{\vec{k}}(t) \ad_{\vec{k}} - \hc \r \ket{0} = e^{- i \chi_{\vec{q}}(t) } \prod_{\vec{k}} \ket{\alpha_{\vec{k}}(t)}.
\label{eq:defTdepMF}
\end{equation}
This ansatz is almost identical to the MF wavefunction in Eq.\eqref{Psi_MF}, except that the coherent amplitudes $\alpha_{\vec{k}}(t)$ are time-dependent and we included a time-dependent phase factor $\chi_{\vec{q}}(t)$. 

From the variational wavefunction \eqref{eq:defTdepMF} observables of interest can be easily calculated. For example the phonon number $N_{\rm ph}(t)$ is given by
\begin{equation}
N_\ph(t) = \int d^3 \vec{k} ~ |\alpha_{\vec{k}}(t)|^2.
\end{equation}
Also the RF spectra presented in Sec.\ref{sec:RFspectra} can be obtained from the time-dependent wavefunction \eqref{eq:defTdepMF}, when the initial condition $\ket{\psi_{\vec{q}}(0)} = \ket{0}$ is employed. As shown in Ref.\cite{Shashi2014RF} they are given by the Fourier-transform of the time-dependent overlap $\bra{0} \psi_{\vec{q}}(t) \rangle$,
\begin{flalign}
I(\omega, \vec{q}) &= \text{Re} \frac{1}{\pi} \int_0^\infty dt ~ e^{i \omega t} \bra{0} \psi_{\vec{q}}(t) \rangle \\
&=  \text{Re} \frac{1}{\pi} \int_0^\infty dt ~ e^{i \omega t}  \exp \left[ i t \frac{q^2}{2M} - i \chi_{\vec{q}}(t) -\frac{1}{2} N_\ph^\MF(t) \right].
\end{flalign}

\subsection{Equations of motion -- Dirac's time-dependent variational principle}
\label{subsec:EOMdirac}
To derive equations of motion for the coherent amplitudes $\alpha_{\vec{k}}(t)$ and the phase $\chi_{\vec{q}}(t)$ in Eq.\eqref{eq:defTdepMF}, we employ Dirac's time-dependent variational principle, see e.g. Ref. \cite{Jackiw1980}. It states that, given a (possibly time-dependent) Hamiltonian $\H(t)$, the dynamics of a quantum state $\ket{\psi(t)}$ can be obtained from the variational principle
\begin{equation}
\delta \int dt ~ \mathcal{L} =0, \qquad  \quad \mathcal{L} = \bra{\psi(t)} i \partial_t - \H(t) \ket{\psi(t)}.
\label{eq:DiracVarPrinciple}
\end{equation}
Here $\mathcal{L}$ denotes a Lagrangian action.

When using a general variational ansatz $\ket{\psi(t)} = \ket{\psi[x_j(t)]}$ defined by a set of some time-dependent variational parameters $x_j(t)$, we obtain their dynamics from the Euler-Lagrange equations of the classical Lagrangian $\mathcal{L}[x_j,\dot{x}_j,t]$. We note that there is a global phase degree of freedom: when $\ket{\psi(t)}$ is a solution of \eqref{eq:DiracVarPrinciple}, then so is $e^{-i \chi(t)} \ket{\psi(t)}$ because the Lagrangian changes as $\mathcal{L} \rightarrow \mathcal{L} + \partial_t \chi(t)$. To determine the dynamics of $\chi_q(t)$ in Eq.\eqref{eq:defTdepMF} we note that for the exact solution $\ket{\psi_{\text{ex}}(t')}$ of the Schr\"odinger equation we should have
\begin{equation}
\int_0^t dt' ~  \mathcal{L}(t')=0,
\label{eq:EOMphase_SM}
\end{equation}
for all times $t$, i.e. $\mathcal{L}=0$. This equation can then be used to determine the dynamics of the overall phase for variational states.

The equations of motion for $\alpha_{\vec{k}}(t)$ can now be derived from the Lagrangian
\begin{equation}
\mathcal{L}[\alpha_{\vec{k}},\alpha_{\vec{k}}^*,\dot{\alpha}_{\vec{k}},\dot{\alpha}_{\vec{k}}^*,t] = \partial_t \chi_q - \mathscr{H}[\alpha_{\vec{k}},\alpha_{\vec{k}}^*]  - \frac{i}{2} \int d^3 \vec{k} \l \dot{\alpha}_{\vec{k}}^*\alpha_{\vec{k}} - \dot{\alpha}_{\vec{k}} \alpha_{\vec{k}}^* \r,
\label{eq:LtMFapdx}
\end{equation}
where we used the following identity valid for coherent states $\ket{\alpha}$,
\begin{equation}
\bra{\alpha} \partial_t \ket{\alpha} = \frac{1}{2} \l \dot{\alpha} \alpha^* - \dot{\alpha}^* \alpha \r.
\end{equation}
The variational energy functional $\mathscr{H}$ in Eq.\eqref{eq:LtMFapdx} evaluates to
\begin{equation}
 \mathscr{H}[\alpha_{\vec{k}},\alpha_{\vec{k}}^*] =  \frac{q^2}{2M} - \frac{( \vec{P}_\ph[\alpha_{\vec{k}}] )^2}{2M} + g_\IB n_0 
+ \int d^3 \vec{k} ~ \left[  |\alpha_{\vec{k}}|^2 \Omega_{\vec{k}}[\alpha_{\vec{k}}]  + V_k \l \alpha_{\vec{k}} + \alpha_{\vec{k}}^* \r \right].
\label{eq:HMFapdx}
\end{equation}
Here the renormalized phonon dispersion $\Omega_{\vec{k}}[\alpha_{\vec{k}}]$ and the phonon momentum $\vec{P}_\ph[\alpha_{\vec{k}}]$ depend explicitly on the amplitudes $\alpha_{\vec{k}}$. They are given by the MF expressions,
\begin{flalign}
\Omega^\MF_{\vec{k}} &= \omega_k + \frac{k^2}{2 M} - \frac{1}{M } \vec{k} \cdot \l \vec{q} - \vec{P}_\ph [\alpha_{\vec{k}}] \r,\\
\vec{P}_\ph[\alpha_{\vec{k}}] &= \int d^3\vec{k} ~ \vec{k} |\alpha_{\vec{k}}|^2,
\label{eq:MFeomDefs}
\end{flalign}
see Eqs.\eqref{eq:OmegakDef} and \eqref{eq:XiselfCons}. 

Using the Euler-Lagrange equations associated with the Lagrangian \eqref{eq:LtMFapdx} the following equations of motion follow:
\begin{equation}
i \partial_t \alpha_{\vec{k}}(t) = \Omega_{\vec{k}}[\alpha_{\vec{\kappa}}(t)] \alpha_{\vec{k}}(t) + V_k.
\label{eq:EOMalphakApdx}
\end{equation}
From the condition that $\mathcal{L}=0$ we obtain
\begin{equation}
\partial_t \chi_q(t) = \frac{q^2}{2 M} - \frac{ ( \vec{P}_\ph[\alpha_{\vec{\kappa}}(t)] )^2}{2 M}+ g_\IB n_0 + \text{Re} \int d^3 \vec{k} ~ V_{\vec{k}} \alpha_{\vec{k}}(t).
\label{eq:EOMchiqApdx}
\end{equation}

\section{Bloch oscillations of polarons in lattices}
\label{sec:PolaronBO}
In this chapter we show how the time-dependent MF approach can be used to describe polaron dynamics in a lattice \cite{Grusdt2014BO}. In particular we investigate an impurity hopping between neighboring lattice sites while interacting with the surrounding phonons. This situation can be realized experimentally with electrons interacting with phonons in a crystal, or using an ultracold impurity atom confined to a deep optical lattice and immersed in a BEC \cite{Fukuhara2013,Scelle2013}. The numerical results presented below correspond to the cold atoms set-up, although the theoretical analysis is generic. For a review of lattice polarons in solid state systems, see Refs.\cite{Holstein1959,Mott1987,Mott1990,Ranniger2006}.

\subsection{Model}
\label{subsec:LatticePolaronModel}
Our starting point is a polaron Hamiltonian after the application of a lattice-version of the Lee-Low-Pines transformation, for details see Appendix \ref{sec:LLPlattice}.
\begin{equation}
\H_{q}(t) = \int d^3 \vec{k} \Bigl[ \omega_k \ad_{\vec{k}} \a_{\vec{k}} + \overline{V}_k \l \ad_{\vec{k}} + \a_{\vec{k}} \r \Bigr]  - 2 J \cos \l a (q - F t) - a \int d^3 \vec{k} ~k_x \ad_{\vec{k}} \a_{\vec{k}} \r.
\label{eq:HqLattice}
\end{equation}
Here we consider an impurity in a one-dimensional lattice, interacting with a three-dimensional bath of phonons. We restrict our analysis to the tight-binding nearest neighbor approximation that gives rise to the dispersion relation $-2 J \cos (q a)$ for a free impurity. Here $J$ is the hopping strength and $a$ the lattice constant. The scattering amplitude $\overline{V}_k$ has a natural UV cut-off at the inverse size of the Wannier function. In this formalism we also included the effect of a force $F$ acting on the impurity, which leads to a constant change of the impurity quasimomentum $q(t)$,
\begin{equation}
q(t) = q - F t.
\end{equation}
For a detailed derivation of the Hamiltonian \eqref{eq:HqLattice} in the case of ultracold quantum gases we refer the interested reader to Ref.\cite{Grusdt2014BO}. Let us emphasize that Eq.\eqref{eq:HqLattice}, as well as the following analysis, can easily be generalized to describe e.g. phonons which are restricted to a lattice Brillouin zone or an impurity in a higher-dimensional lattice.

In the following we consider the dynamics of an impurity which is initially non-interacting. Then, at time $t=0$, we switch on its interactions with the surrounding phonons, which are initially assumed to be in the zero-temperature vacuum state $\ket{0}$. To describe dynamics of the impurity within the polaron frame, we can decompose the initial wavefunction in its Fourier components,
\begin{equation}
\ket{\psi(0)}= \sum_{q \in \BZ} f_q \ket{q} \otimes \ket{0}.
\end{equation}
Here $\ket{q}$ denotes the impurity eigenstate with lattice quasimomentum $q$ and $\ket{0}$ is the phonon vacuum. The Fourier amplitudes $f_q$ are determined by the initial impurity wavefunction $\psi_j^{\rm in}$,
\begin{equation}
 f_q = \frac{1}{\sqrt{L/a}} \sum_j e^{i q a j} \psi_j^\text{in},
 \label{eq:fqDefFT}
\end{equation}
where $L$ is the length of the lattice and $j=...,-1,0,1,...$ denote lattice sites.

Most importantly, because of the discrete translational invariance of the problem, the Fourier components $f_q$ are conserved during the time-evolution. After some time $t$ the wavefunction (in the polaron frame) reads
\begin{equation}
\ket{\psi(t)}= \sum_{q \in \BZ} f_q \ket{q} \otimes  \underbrace{\mathcal{T} e^{- i \int_0^t d\tau~  \H_q(\tau)} \ket{0}}_{= \ket{\psi_q(t)}}.
\label{eq:PsitLabFrame}
\end{equation}
Therefore the dynamics of the phonon cloud $\ket{\psi_q(t)}$ at a given quasimomentum $q$ of the polaron is determined by the Hamiltonian $\H_q(t)$ in the polaron frame. In the following section we derive the solution $\ket{\psi_q(t)}$ approximately using the time-dependent MF approach presented in the last section.

Once we have found the solution for $\ket{\psi_a(t)}$ we can use Eq.\eqref{eq:PsitLabFrame} to derive all observables of interest. To derive transport properties of the impurity in the presence of phonons, we need the impurity trajectory. It can be obtained from the impurity density distribution $n_j(t)$, which is determined by
\begin{equation}
n_j (t)=  \frac{a}{L} \sum_{q_2,q_1 \in \BZ} e^{i a (q_2 - q_1) j} A_{q_2,q_1}(t) f^*_{q_2} f_{q_1}.
\label{eq:njViaAq1q2}
\end{equation}
Here we defined the time-dependent overlap 
\begin{equation}
A_{q_2,q_1}(t)=\bra{\Psi_{q_2}(t)} \Psi_{q_1}(t) \rangle,
\label{eq:defNonEqGreen}
\end{equation}
which is thus a quantity of key interest.

\subsection{Time-dependent mean-field description}
\label{subsec:tDepMFlatticePolaron}
As in the case of the Bogoliubov-Fr\"ohlich polaron in the continuum, see Sec.\ref{sec:tDependentMF}, a time-dependent MF description of the lattice polaron dynamics can be constructed. To this end we make the same ansatz as in Eq.\eqref{eq:defTdepMF}
\begin{equation}
\ket{\psi_{q}(t)} = e^{- i \chi_{q}(t) } \prod_{\vec{k}} \ket{\alpha_{\vec{k}}(t)}.
\label{eq:defTdepMFlattice}
\end{equation}
Equations of motion are obtained as described in Sec.\ref{subsec:EOMdirac},
\begin{flalign}
 i \partial_t \alpha_{\vec{k}}(t) &=\overline{\Omega}_{\vec{k}}[\alpha_{\vec{\kappa}}(t)] ~  \alpha_{\vec{k}}(t) + \overline{V}_k,  \label{eq:DMF_EOM} \\
 \partial_t \chi_q &= \frac{i}{2} \int d^3\vec{k} \l \dot{\alpha}_{\vec{k}}^*\alpha_{\vec{k}} - \dot{\alpha}_{\vec{k}} \alpha_{\vec{k}}^* \r+ \mathscr{H}_{q-F t}[\alpha_{\vec{\kappa}}(t)].
\label{eq:DMF_EOM_phases}
\end{flalign}

In the lattice case, from Eq.\eqref{eq:HqLattice} we obtain the renormalized phonon dispersion
\begin{equation}
\overline{\Omega}_{\vec{k}}[\alpha_{\vec{\kappa}}]  = \omega_k + 2 J e^{- C[\alpha_{\vec{\kappa}}]}  \Bigl[ \cos \l a (q-Ft)  -S[\alpha_{\vec{\kappa}}] \r  - \cos \l a (q-Ft)- a k_x -S[\alpha_{\vec{\kappa}}] \r \Bigr],
\label{eq:effPhononDispersion}
\end{equation}
where we defined
\begin{equation}
 C[\alpha_{\vec{\kappa}}] = \int d^3 \vec{k} |\alpha_{\vec{k}}|^2 (1- \cos (a k_x)), \qquad S[\alpha_{\vec{\kappa}}] = \int d^3 \vec{k} |\alpha_{\vec{k}}|^2 \sin (a k_x). 
 \label{eq:SCqdef}
\end{equation}
Moreover, the MF energy functional reads
\begin{equation}
\mathscr{H}_{q- F t}[\alpha_{\vec{\kappa}}(t)]= - 2 J e^{-C[\alpha_{\vec{\kappa}}]} \cos \l a (q-F t) -S[\alpha_{\vec{\kappa}}] \r +\int d^3 \vec{k} ~ \left[ \omega_k |\alpha_{\vec{k}}|^2 + V_k \l \alpha_{\vec{k}} + \alpha_{\vec{k}}^* \r \right].
 \label{eq:HMfunctionalMainText}
\end{equation}


\subsection{Adiabatic approximation and polaron dynamics}
\label{subsec:adiabApprox}
To gain analytical insight into the time-dependent MF solution, we make use of the adiabatic approximation. The starting point is a ground state polaron in a lattice, which we can describe using Lee-Low-Pines MF theory. By minimizing the variational energy in Eq.\eqref{eq:HMfunctionalMainText} for a given quasimomentum $q$ and setting $F=0$ we obtain the MF coherent amplitudes,
\begin{equation}
\alpha_{\vec{k}}^\MF(q) = -  \overline{V}_k / \overline{\Omega}_{\vec{k}}[\alpha_{\vec{\kappa}}^\MF(q)].
\label{eq:MFselfCons}
\end{equation}
In practice this infinite set of self-consistency equations can be reduced to only two equations for $S^\MF(q) = S[\alpha_{\vec{k}}^\MF(q)]$ and $C^\MF(q) = C[\alpha_{\vec{k}}^\MF(q)]$. Properties of this MF solution were discussed in detail in Ref.\cite{Grusdt2014BO}.

The idea of the adiabatic approximation is to assume that the phonon cloud $\alpha_{\vec{k}}(t)$ adiabatically follows its ground state $\alpha_{\vec{k}}^\MF(q - F t)$, when $q$ changes as a consequence of the non-vanishing force $F$. That is, we make the ansatz
\begin{equation}
\ket{\Psi_q(t)} \approx e^{- i \chi_q(t)} \prod_{\vec{k}} \ket{\alpha_{\vec{k}}^\MF(q-Ft)}.
\end{equation}

As a simple application of the adiabatic approximation, consider a stationary initial Gaussian impurity wavepacket which is centered around $X_0$ at time $t=0$. Then the adiabatic approximation can be used to show that the dynamics of the wavepacket is given by
\begin{equation}
X(t) = X_0 + \left[ E_0^\MF(F t) - E_0^\MF(0) \right] / F.
\label{eq:XtAd}
\end{equation}
Here $E_0^\MF(q) = \mathscr{H}_{q}[\alpha^\MF_{\vec{\kappa}}]$ is the polaron ground state energy. Therefore the impurity trajectory resembles the polaron dispersion relation in the adiabatic limit. The corresponding impurity trajectory (or polaron trajectory) is periodic in time, with the frequency given by $\omega_{\rm B} = a F$. Thus we conclude that the polaron undergoes coherent Bloch oscillations, and the Bloch oscillation frequency coincides with the result for a non-interacting impurity.

More importantly, we can describe corrections to the adiabatic approximation using a simple bilinear Hamiltonian. To this end we apply a time-dependent unitary basis transformation,
\begin{equation}
\hat{U}(q-F t) = \prod_{\vec{k}} \exp \l \alpha_{\vec{k}}^\MF(q - F t) \ad_{\vec{k}} - \l \alpha_{\vec{k}}^\MF (q- F t)\r^* \a_{\vec{k}} \r.
\label{eq:UMFflucLatticet}
\end{equation}
In the so-obtained frame, $\a_{\vec{k}}$'s describe quantum fluctuations around the time-dependent MF solution. Their dynamics is goverened by the effective time-dependent Hamiltonian
\begin{equation}
\tilde{\mathcal{H}}(t) = \int d^3 \vec{k} ~ \overline{\Omega}_{\vec{k}}(q- F t) \ad_{\vec{k}} \a_{\vec{k}} + i F \int d^3 \vec{k} ~ \l \partial_q \alpha^\MF_{\vec{k}} (q - F t ) \r \left[  \ad_{\vec{k}} -   \a_{\vec{k}}  \right] + \mathcal{O}(J^* \a^2) .
\label{eq:HflucMF}
\end{equation}
Here we introduced $J^*(q - F t ):=J \exp \left[ - C^\MF(q - F t ) \right]$ and we neglected terms of order $\mathcal{O}(J^* \a^2)$. 

The non-adiabatic corrections $\propto F$ on the right hand side of this equation determine the emission rate of phonons due to periodic oscillations in the system at the frequency $\omega_{\rm B}$. This leads to an incoherent drift current of the impurity along the applied force, which adds to the coherent Bloch oscillations predicted in Eq.\eqref{eq:XtAd}. The corresponding drift velocity $v_{\rm d}$ of the impurity can be calculated from the phonon emission rate $\gamma_\ph$, by relating the emitted power $P_\gamma$ to the energy gain per time $F v_{\rm d}$ when the impurity slides down the lattice. Assuming that only phonons which are resonant with the Bloch oscillation frequency $\omega_{\rm B}$ are emitted, $P_\gamma = \omega_{\rm B} \gamma_\ph$, we obtain
\begin{equation}
v_{\rm d} = a \gamma_\ph.
\end{equation}

The phonon emission rate can be calculated from Eq.\eqref{eq:HflucMF} using Fermi's golden rule to treat the time-dependent non-adiabatic corrections. As a result we obtain an analytic expression for the drift velocity,
\begin{equation}
v_\text{d}(F) = S_{d-2} 8 \pi  \frac{J_0^{*2}}{aF^2}  \frac{ k^{d-1} \overline{V}_k^2}{(\partial_k \omega_k)} \l 1 - \text{sinc} (a k)\r + \mathcal{O}(J_0^*)^3.
\label{eq:gmaPh}
\end{equation}
This result assumed a slightly generalized model where phonons in $d$ spatial dimensions are considered (instead of $d=3$ as e.g. in Eq.\eqref{eq:HflucMF} above). In Eq.\eqref{eq:gmaPh} the value of $k$ has to be determined from the condition that $\omega_k = \omega_\text{B}$. Moreover, $J_0^* := \lim_{J \rightarrow 0} J^*(q)$ is the renormalized polaron hopping in the heavy impurity limit (which is independent of $q$), and $S_n=(n+1) \pi^{(n+1)/2} / \Gamma(n/2+3/2)$ denotes the surface area of an $n$-dimensional unit sphere. $\text{sinc}(x)$ is a shorthand notation for the function $\sin(x) / x$. For a detailed derivation of Eq.\eqref{eq:gmaPh} we refer the reader to Ref.\cite{Grusdt2014BO}.

In the following we will compare the analytical results presented in this section to a full numerical treatment of the time-dependent MF equations \eqref{eq:DMF_EOM}, \eqref{eq:DMF_EOM_phases}.

\begin{figure}[b!]
\centering
\epsfig{file=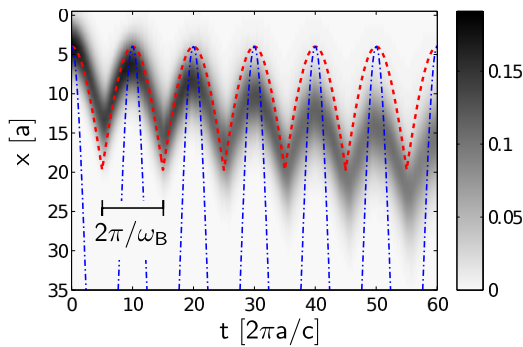, width=0.45\textwidth}
\caption{Impurity density $n_j(t)$ (gray scale) with $j a = x$ for a heavily dressed impurity, taken from Ref.\cite{Grusdt2014BO}. The polaron dynamics, starting from phonon vacuum, is compared to the result from the adiabatic approximation (red, dashed) as well as the trajectory of a non-interacting impurity wavepacket (blue, dashed-dotted).} 
\label{fig:polaronTrajectory}
\end{figure}

\subsection{Polaron transport properties}
\label{subsec:polaronTransportResults}
\begin{figure}[b!]
\centering
\epsfig{file=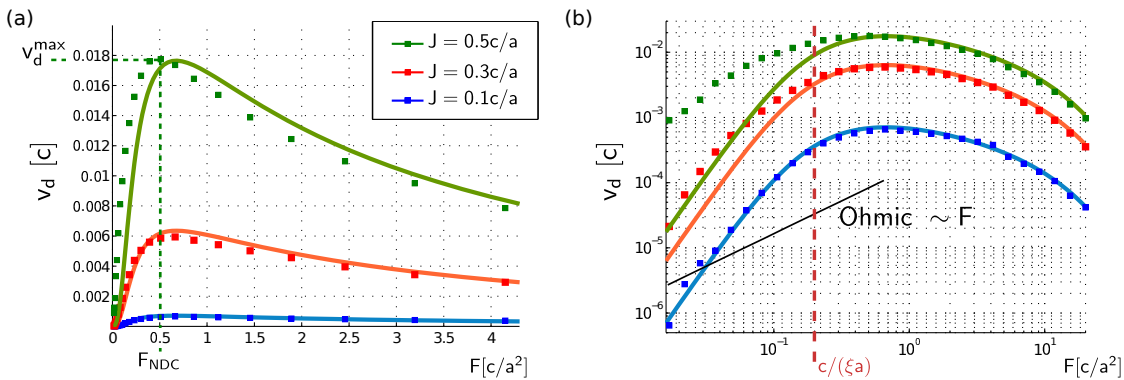, width=0.99\textwidth}
\caption{The dependence of the polaron drift velocity $v_\text{d}$ on the driving force $F$ is shown. The data points were obtained from impurity trajectories calculated by solving for the time-dependent mean-field wavefunction of phonons in a three-dimensional BEC. A fixed interaction strength is used and various hoppings are considered (top: $J=0.5c/a$, middle: $J=0.3c/a$, bottom: $J=0.1 c/a$). For each $J$ also the result from the analytical model Eq.\eqref{eq:gmaPh} of polaron transport (solid lines) is shown, which is free of any fitting parameters. In (b) the same data is shown as in (a) (legend from (a) applies), but on a double-logarithmic scale. In the lower left corner an Ohmic power-law dependence $v_{\rm} \sim F$ is indicated (thin solid line). The figure is taken from Ref.\cite{Grusdt2014BO} where further explanations can be found.}
\label{fig:vd_F}
\end{figure}

To obtain polaron transport properties, the time-dependent MF equations \eqref{eq:DMF_EOM}, \eqref{eq:DMF_EOM_phases} can be solved numerically. In FIG.\ref{fig:polaronTrajectory} we show an example for a trajectory of a strongly interacting impurity. We compare it to the result of the adiabatic approximation and to the trajectory of a non-interacting impurity. First we notice a strong suppression of the oscillation amplitude compared to the free impurity. This effect is captured by the adiabatic approximation and may be explained from the strong renormalization of the polaron mass. In addition, we observe a drift of the polaron along the lattice. This corresponds to an incoherent current, induced by the force $F$, which is absent for a non-interacting impurity. In FIG.\ref{fig:vd_F} the corresponding current-force relation $v_{\rm d}(F)$ obtained from the time-dependent MF theory is shown, where $v_{\rm d}$ denotes the drift velocity. In the following we will review the theoretical progress in understanding polaron transport properties, and put our time-dependent MF results in FIG.\ref{fig:vd_F} into context with other works.

Historically, the study of the interplay between coherent Bloch oscillations of a single particle and inelastic scattering (e.g. on thermal phonons) was pioneered in the solid state physics context by Esaki and Tsu \cite{Esaki1970}, who derived a phenomenological relation between the driving force $F$ and the net (incoherent) current $v_\text{d}$. They used the relaxation-time approximation which assumes that randomly distributed  scattering events bring the system back into its ground state instantly. Between the scattering events the evolution of the system is assumed to be fully coherent. The resulting Esaki-Tsu relation reads
\begin{equation}
v_\text{d}(F) = 2 J a \frac{\omega_\text{B} \tau}{1 + \l \omega_\text{B} \tau \r^2}, \qquad \omega_{\rm B} = a F,
\label{eq:EsakiTsu}
\end{equation}
where $\tau$ denotes the relaxation time $\tau$ (i.e. the average duration between two consecutive scattering events). In particular, it predicts a generic Ohmic regime for weak driving, i.e. $v_\text{d} \sim F$.

The precision of ultra-cold atom experiments allowed a detailed verification of the Esaki-Tsu relation in thermal gases \cite{Ott2004}. In such a setting, with coupling to an incoherent thermal bath, theoretical analysis  \cite{Ponomarev2006,Kolovsky2008} showed that the relation \eqref{eq:EsakiTsu} is valid. However, it is not clear a priori whether the result by Esaki and Tsu applies also to the polaron problem inside a BEC where interactions should be treated on a fully coherent level. Nevertheless, Bruderer et al. \cite{Bruderer2008} suggested to describe the incoherent polaron current using the Esaki-Tsu relation, and indeed by fitting Eq.\eqref{eq:EsakiTsu} to their numerical data for a one-dimensional polaron model they found convincing agreement. 

In Ref.\cite{Grusdt2014BO} we pointed out two main short comings of the phenomenological Esaki-Tsu theory when applied to the polaron problem. First, in order to obtain reasonable agreement with the numerical data, the hopping $J$ in Eq.\eqref{eq:EsakiTsu} has to be replaced by an effective hopping $J_{\rm eff}$ which becomes an additional fitting parameter \cite{Bruderer2008}. This effective hopping is not related to the mass renormalization of the polaron, and not even the dependence of $J_{\rm eff}$ on the bare hopping $J$ (or on the interaction strength) is correctly predicted by the Esaki-Tsu theory \cite{Grusdt2014BO}. Thus, within the phenomenological model developed by Esaki and Tsu, there is no physical explanation for why the hopping should be used as a fitting parameter.

The second -- more severe -- shortcoming of the Esaki-Tsu model, when applied to the polaron problem, is that it predicts a generic Ohmic current-force relation $v_{\rm d}(F) \propto F$ in the weak-driving limit. This is in conflict with our analytical result in Eq.\eqref{eq:gmaPh}, which predicts a current-force relation 
\begin{equation}
v_{\rm d}(F) \propto F^d
\end{equation}
for weak driving $F \to 0$. This result depends sensitively on the dimensionality $d$. Only in the one-dimensional case $d=1$ studied in Ref.\cite{Bruderer2008} the Ohmic behavior is obtained that is characteristic for the Esaki-Tsu theory in arbitrary dimensions. For higher-dimensional systems, on the other hand, a \emph{sub-Ohmic} current-force relation is predicted.

In FIG.\ref{fig:vd_F} results from full time-dependent MF calculations are shown, where phonons in $d=3$ dimensions were simulated. In the double-logarithmic plot in part (b) sub-Ohmic current-force relations can be identified. For sufficiently small hoppings $J$ the analytical result in Eq.\eqref{eq:gmaPh} is valid and agrees well with the numerical data. Besides the sub-Ohmic behavior at small forces, the qualitative form of the curves $v_{\rm d}(F)$ agrees with the naive expectation from the Esaki-Tsu model, see Ref.\cite{Grusdt2014BO}. However from the point view of the analytical result \eqref{eq:gmaPh}, which describes in detail the dependence of the current on various system parameters like the scattering amplitude $\overline{V}_k$ or the phonon dispersion $\omega_k$, the qualitative agreement with the Esaki-Tsu curve seems rather accidental.

In the time-dependent MF theory discussed above, phonon correlations are discarded. This is certainly valid in the weak coupling regime, but becomes questionable for stronger couplings. Using a generalization of the RG procedure (see Chap.\ref{chap:RGapproach}) to real-time dynamics, we found in the continuum that time-dependent MF results provide a remarkably good description of the system up to rather large couplings \cite{Grusdt2015dRG}. To check whether this is true also for polarons in a lattice, quantum fluctuations have to be included. One possibility is to use time-dependent Gaussian variational wavefunctions (in generalization of the static case discussed in Sec.\ref{sec:varApproach}) which could be carried over to the lattice case.

\newpage
\chapter{Outlook}
\label{Chap:Outlook}

Before concluding these lecture notes we would like to review some of the open questions that need to be understood in the context of Bose polarons and related problems.

{\it Bose polarons beyond the Fr\"ohlich model}. 
The Fr\"ohlich model formulation of the Bose polaron focuses on processes in which an impurity atom scatters  atoms in and out of the condensate. Said differently, in any scattering process with the impurity atom one of the states of the BEC atoms is at $k=0$. The Fr\"ohlich Hamiltonian does not include processes in which uncondensed host atoms scatter between states at finite momenta. While scattering in and out of the $k=0$ state benefits from the Bose enhancement factor, scattering between finite momentum states is favored by the large available phase space in the case of a high UV cutoff. Scattering of atoms (phonons) at finite momentum is particularly important close to unitarity, where such processes play a crucial role in formation of the molecular bound state and defining the correct low energy limit of the scattering amplitude. At the two particle level these processes can be included using non-self consistent T-matrix approximation (see e.g. \cite{Rath2013}). However, this approximation does not allow for the creation of multiple phonon excitations. Thus it does not describe correctly either the full phonon dressing of the polaronic ground state or the dynamical spectral function. The broad spectral feature predicted by this approach (see Fig. 2 in Ref. \cite{Rath2013}) does not show the coherent polaron peak separated from the higher energy shake-off processes. Combining the physics of Feshbach resonances with the polaronic physics discussed in these lecture notes remains an open problem although first steps in this direction were taken by the use of self-consistent T-matrix approximation \cite{Rath2013} and variational calculations \cite{Li2014,Levinsen2015}.

{\it Bose polaron dynamics in traps}.  
 Most experiments with ultracold atoms include parabolic confining potential. Thus one of the readily available experimental tools is the analysis of impurity dynamics in the parabolic potential in the presence of the BEC. Measurements of the frequency and damping of oscillations can provide information about the polaron effective mass and mobility\cite{Catani2012}. Understanding polaron dynamics in such systems is more challenging than in homogeneous space. The total momentum of the system is no longer conserved and one can not do the Lee-Low-Pines transformation to integrate out the impurity atom. Reliable extensions of the formalism presented in this paper to systems without translational symmetry remains an open challenge (see Refs.\cite{Bonart2013,Bonart2012} for solutions of the simplified models of Caldeira-Leggett type).

{\it Dynamics of magnetic polarons}. 
We expect that many fruitful results can be obtained by extending techniques discussed in these lecture notes to other polaronic systems. One of the particularly promising directions is analysis of magnetic polarons in Mott insulators (see e.g. \cite{Auerbach1991,Berciu2009}) and other strongly correlated electron systems. Direct analogue of the RF spectroscopy of polarons that we discussed in these lecture notes is photoemission experiments \cite{Damascelli2003}. When an electron is ejected from the material, it leaves behind a hole, that can interact with spin waves, giving rise to characteristic spectral functions \cite{Kane1989,SchmittRink1988}. Another important class of experiments that can be analyzed using tools presented in this paper is resonant x-ray scattering\cite{Ament2011}, in which an electron photo-excited into the upper Hubbard band interacts with spin waves.

\newpage
\chapter{Appendices}
\label{Chap:Appendices}

\section{Lee-Low-Pines formalism in a lattice}
\label{sec:LLPlattice}

In this appendix we present a theoretical model of lattice polarons. It is based on the Lee-Low-Pines (LLP) formalism, which we reviewed for continuum polarons in Sec.\ref{sec:LLP}. Our starting point is an impurity confined to the lowest Bloch band of an optical lattice, interacting with a surrounding bath of Bogoliubov phonons describing elementary excitations of the BEC. Although we discuss the case of a continuous BEC here, our treatment can easily be generalized to a BEC in a lattice. A more detailed discussion can be found in Ref.\cite{Grusdt2014BO}. Here we will focus on the case of a three dimensional BEC ($d=3$) again, but analogous analysis can be done for other dimensions $d$. 

Our starting point is the following lattice polaron Hamiltonian, which is obtained from \eqref{eq:Hmicro} by introducing a deep species-selective optical lattice for the impurity. 
\begin{multline}
 \H = \int d^3 \vec{k} \Bigl\{ \omega_k \ad_{\vec{k}} \a_{\vec{k}} + \sum_j \cd_j \c_j e^{i k_x a j} \l \ad_{\vec{k}} + \a_{-\vec{k}} \r \overline{V}_{k} \Bigr\} + \\
 + g_\IB n_0 - J \sum_j \l \cd_{j+1} \c_j + \hc \r - F \sum_j j a ~ \cd_j \c_j.
  \label{eq:Hfund}
\end{multline}
Here $\cd_j$ creates an impurity at lattice site $j$, $\overline{V}_k$ is the scattering amplitude in the lattice (see Ref.\cite{Grusdt2014BO} for details), $J$ is the impurity hopping and $a$ the lattice constant. For concreteness we assume the lattice to be one-dimensional (pointing along $\vec{e}_x$), but the analysis can easily be carried over to arbitrary lattice dimensions. To study transport properties of the dressed impurity, we furthermore consider a constant force $F$ acting on the impurity. In experiments this force can e.g. be applied using a magnetic field gradient \cite{Anderson1998,Palzer2009,Atala2012}. 

The second term in the first line of Eq.\eqref{eq:Hfund} $\sim \cd_j \c_j$ describes scattering of phonons on an impurity localized at site $j$ (with amplitude $\overline{V}_k$). This term thus breaks the conservation of total phonon momentum (and number), and we stress that phonon momenta $\vec{k}$ can take arbitrary values $\in \mathbb{R}^3$, not restricted to the Brillouin zone (BZ) defined by the impurity lattice \footnote{Only if the bosons were also subject to an optical lattice potential, the phonon momenta $\vec{k}$ appearing in Eq.\eqref{eq:Hfund} would be restricted to the corresponding BZ. This is not the case for the model of a homogeneous BEC considered here.}.

\subsection{Coupling constant and relation to experiments}
Now we slightly adapt the definition of the dimensionless coupling constant for lattice polarons. The disadvantage of the definition given in Eq.\eqref{eq:defAlpha} is that there $\alpha$ depends on the (free-space) impurity-boson scattering length $a_\IB$. Here we use only $g_\IB$ to characterize the interaction strength, and define 
\begin{equation}
g_\eff^2 = \frac{n_0 g_{\text{IB}}^2}{\xi c^2} = \l \frac{E_\IB}{E_\text{ph}} \r^2.
\label{eq:defgeff}
\end{equation}
If we re-express $a_\IB$ in terms of $g_\IB$ using the free-space Lippmann-Schwinger equation, see Sec.\ref{Sec:LippmannSchwinger}, we find $\alpha= \frac{2}{\pi} m_\text{red}^{-2} n_0 g_\IB^2$. Thus our coupling constant \eqref{eq:defgeff} is related to $\alpha$ by
\begin{equation}
\alpha = \frac{1}{\pi} \left[1 + \frac{m_\text{B}}{M} \right]^{-2} g_\eff^2.
\end{equation}
In this expression the impurity mass $M$ enters as an additional parameter, which is not required to calculate $g_\eff$ however. Therefore we prefer to use $g_\eff$ instead of $\alpha$ in the case of lattice polarons.

\subsection{Time-dependent Lee-Low-Pines transformation in the lattice}
\label{sec:LLPlattice}
Now we simplify the Hamiltonian \eqref{eq:Hfund} by making use of the Lee-Low-Pines transformation (see Sec. \ref{sec:LLP}). This will make the conservation of polaron \emph{quasi}momentum explicit. We also include the effect of the constant force $F$ acting on the impurity. 

To do so, we start by applying a time-dependent unitary transformation,
\begin{equation}
\hat{U}_\text{B}(t) = \exp \l i \omega_{\text{B}} t \sum_j j \cd_j \c_j \r,
\label{eq:BOtransform}
\end{equation}
where $\omega_\text{B} = a F$ denotes the frequency of Bloch oscillations of the bare impurity. Next we apply the LLP transformation which is of the form
\begin{equation}
 \hat{U}_{\text{LLP}}=e^{i \hat{S}}, \qquad \hat{S}= \int d^3\vec{k} ~ k_x \ad_{\vec{k}} \a_{\vec{k}}  ~ \sum_{j}  a  j  \cd_j \c_j
 \label{eq:polaronTrofo}
\end{equation}
in the presence of a lattice, cf. Eq. \eqref{eq:LLPdef}. 

As described in Sec.\ref{sec:LLP}, the action of the Lee-Low-Pines transformation on an impurity can be understood as a displacement in quasimomentum space. Such a displacement  $q \rightarrow q + \delta q$ (modulo reciprocal lattice vectors $2 \pi /a$) is generated by the unitary transformation $e^{i \delta q \hat{X}}$, where the impurity position operator is defined by $\hat{X} = \sum_j a j \cd_j \c_j$. Comparing this to Eq.\eqref{eq:polaronTrofo} yields $\delta q = \int d^3 \vec{k}~ k_x \ad_{\vec{k}} \a_{\vec{k}}$, which is the total phonon momentum operator. Thus we obtain
\begin{equation}
 \hat{U}_{\text{LLP}}^\dagger \c_q  \hat{U}_{\text{LLP}} =  \c_{q+\delta q}.
 \label{eq:ULLP1}
\end{equation}
For phonon operators, on the other hand, transformation \eqref{eq:polaronTrofo} corresponds to translations in real space by the impurity position $\hat{X}$. One can easily see that
\begin{equation}
\hat{U}_{\text{LLP}}^\dagger \a_{\vec{k}} \hat{U}_{\text{LLP}} = e^{i \hat{X} k_x} \a_{\vec{k}}.
\label{eq:ULLP2}
\end{equation}

Now we would like to apply the time-dependent LLP transformation
\begin{equation}
 \hat{U}_{\text{tLLP}}(t)= \hat{U}_\text{LLP} \hat{U}_\text{B}(t) =e^{i \hat{S}(t)}, \qquad \hat{S}(t)= \l \int d^3\vec{k} ~ k_x \ad_{\vec{k}} \a_{\vec{k}} + F t \r ~ \sum_{j}  a  j  \cd_j \c_j
 \label{eq:tLLPdef}
\end{equation}
to the Hamiltonian Eq.\eqref{eq:Hfund}. In the new basis the (time-dependent) Hamiltonian reads
\begin{equation}
\H(t) = \hat{U}_\text{tLLP}^\dagger(t)  \H \hat{U}_\text{tLLP}(t) - i \hat{U}_\text{tLLP}^\dagger(t) \partial_t \hat{U}_\text{tLLP}(t).
\label{eq:Hefft}
\end{equation}
To simplify the result we first write the free impurity Hamiltonian in quasimomentum space, 
\begin{equation}
\H_\text{I} = -2 J \sum_{q \in \BZ} \cd_q \c_q \cos (a q),
\end{equation}
where we introduce the quasimomentum basis in the usual way,
\begin{equation}
 \c_q := \l L /a \r^{-1/2} \sum_j e^{i q a j} \c_j.
  \label{eq:BOoscillatingOps}
\end{equation}
Here $L$ denotes the total length of the impurity lattice and $q = -\pi/a,...,\pi/a$ is the impurity quasimomentum in the BZ. The transformation \eqref{eq:BOtransform} thus allows us to assume periodic boundary conditions for the Hamiltonian \eqref{eq:Hefft}, despite the presence of a linear potential $-F x$. 

Next we make use of the fact that only a single impurity is considered, i.e. $\sum_{q\in \BZ} \cd_q \c_q =1$ for all relevant states, allowing us to simplify
\begin{equation}
\cd_j \c_j e^{i k_x  \hat{X}  } = \cd_j \c_j e^{i k_x  a j} .
\label{eq:smplifyeikhatX}
\end{equation}
Note that although the operator $\hat{X}$ in Eq.\eqref{eq:smplifyeikhatX} consists of a summation over all sites of the lattice, in the case of a single impurity the prefactor $\cd_j \c_j$ selects the contribution from site $j$ only. 

We proceed by employing Eqs.\eqref{eq:ULLP1} - \eqref{eq:smplifyeikhatX} and arrive at the Hamiltonian
\begin{multline}
\H(t) = \sum_{q \in \BZ} \cd_{q} \c_q \left\{ \int d^3 \vec{k} \Bigl[ \omega_k \ad_{\vec{k}} \a_{\vec{k}} + \overline{V}_k \l \ad_{\vec{k}} + \a_{\vec{k}} \r \Bigr] -  \right. \\ \left.
 - 2 J \cos \l a q - \omega_\text{B} t - a \int d^3 \vec{k}' ~k_x' \ad_{\vec{k}'} \a_{\vec{k}'} \r + g_\IB n_0 \right\}.
 \label{eq:polaronHam}
\end{multline}
Let us stress again that this result is applicable only for a single impurity, i.e. when $\sum_{q \in \BZ} \cd_q \c_q=1$. We find it convenient to make use of this identity and pull out $\sum_{q \in \BZ} \cd_q \c_q$ everywhere to emphasize that the Hamiltonian factorizes into a part involving only impurity operators and a part involving only phonon operators. Notably the Hamiltonian \eqref{eq:polaronHam} is time-dependent and non-linear in the phonon operators. From the equation we can moreover see that, in the absence of a driving force $F=0$ (corresponding to $\omega_{B} = 0$), the total \emph{quasi}momentum $q$ in the BZ is a conserved quantity. We stress, however, that the total \emph{phonon}-momentum $\int d^3 \vec{k} ~ \vec{k} \ad_{\vec{k}} \a_{\vec{k}}$ of the system is not conserved.

Even in the presence of a non-vanishing force $F\neq 0$ the Hamiltonian is still block-diagonal for all times, 
\begin{equation}
\H(t) = \sum_{q \in \BZ} \cd_q \c_q \H_q(t),
\label{eq:qtConserved}
\end{equation}
and quasimomentum evolves in time according to 
\begin{equation}
q(t) = q - F t,
\label{eq:qt}
\end{equation}
i.e. $\H_q(t) = \H_{q(t)}(0)$. This relation has the following physical meaning: if we start with an initial state that has a well defined quasimomentum $q_0$, then the quasimomentum of the system remains a well defined quantity. The rate of change of the quasimomentum is given by $F$, i.e. $q(t) = q_0 - F t$. Thus states that correspond to different initial momenta do not mix in the time-evolution of the system.

\section{Renormalized impurity mass}
\label{apdx:renImpMass}
In this appendix we would like to show that the coupling constant $\mathcal{M}$ in the RG protocol can be interpreted as renormalized impurity mass. To this end we start from a Lee-Low-Pines type polaron model with UV cut-off $\Lambda_0$ and an impurity of mass $M$. Then we apply the RG to integrate out phonons at momenta larger than $\Lambda$, and show that the resulting low-energy model is equivalent to a Lee-Low-Pines type polaron model with a UV cut-off $\Lambda$ and for an impurity of mass $\mathcal{M}$.

For simplicity we restrict ourselves to the spherically symmetric case $q=0$. We start from the Fr\"ohlich Hamiltonian after the Lee-Low-Pines transformation,
\begin{equation}
\H = \int^{\Lambda_0} d^3\vec{k} ~ \left[ \omega_k \ad_{\vec{k}} \a_{\vec{k}} + V_k \l \ad_{\vec{k}} + \a_{\vec{k}} \r \right] 
 + \frac{1}{2M} \l \int^{\Lambda_0} d^3 \vec{k} ~ \vec{k}~  \ad_{\vec{k}} \a_{\vec{k}} \r^2
\end{equation}
Next, as in the main text, we apply the MF shift $\U_\MF$ and obtain
\begin{equation}
\tilde{\mathcal{H}}_{\Lambda_0} :=\Ud_\MF \H \U_\MF = \int^{\Lambda_0} d^3 \vec{k} ~ \Omega_{\vec{k}}^\MF \ad_{\vec{k}} \a_{\vec{k}} 
+  \int^{\Lambda_0} d^3 \vec{k} d^3\vec{k}' ~ \frac{\vec{k} \cdot \vec{k}'}{2 M} ~ : \G_{\vec{k}} \G_{\vec{k}'} : ~ ,
\end{equation}
where $\Omega_{\vec{k}}^\MF = \omega_k + k^2 / 2 M$ in this case. 

After the application of the RG from the initial UV cut-off $\Lambda_0$ down to $\Lambda$, we end up with the Hamiltonian
\begin{equation}
\tilde{\mathcal{H}}_{\Lambda} =  \int^{\Lambda} d^3 \vec{k} ~\left[ \Omega_{\vec{k}} \ad_{\vec{k}} \a_{\vec{k}} + W_{\vec{k}} \l \ad_{\vec{k}} + \a_{\vec{k}} \r \right] 
+  \int^{\Lambda} d^3 \vec{k} d^3\vec{k}' ~ \frac{\vec{k} \cdot \vec{k}'}{2 \mathcal{M}} ~ : \G_{\vec{k}} \G_{\vec{k}'} : ~ ,
\end{equation}
where the frequency is renormalized, $\Omega_{\vec{k}} = \omega_k + k^2 / 2 \mathcal{M}$, and $W_{\vec{k}} = \frac{k^2}{2} \l \mathcal{M}^{-1} - M^{-1} \r \alpha_{\vec{k}}$. 

Now, reversing the action of the MF shift, we end up with a Lee-Low-Pines type Hamiltonian again, but at the reduced cut-off $\Lambda$,
\begin{equation}
\U_\MF \tilde{\mathcal{H}}_{\Lambda} \Ud_\MF = \int^{\Lambda} d^3 \vec{k} ~ \left[  \omega_k \ad_{\vec{k}} \a_{\vec{k}} + V_k \l \ad_{\vec{k}} + \a_{\vec{k}} \r \right] 
 + \frac{1}{2 \mathcal{M}} \l \int^{\Lambda} d^3 \vec{k} ~ \vec{k}~  \ad_{\vec{k}} \a_{\vec{k}} \r^2 + \Delta E_0(\Lambda),
\end{equation}
with $\Delta E_0(\Lambda)$ describing a modified ground-state energy after integrating out phonon modes at larger momenta. This model is equivalent to our original model, but with an impurity of increased mass $\mathcal{M}(\Lambda)$ instead of $M$. 

That is, the effect of the RG is to introduce a renormalized \emph{impurity} mass. Note that this mass-enhancement goes beyond MF, and is different from the simple MF type enhancement of the \emph{polaron} mass. The latter originates from the fact that part of the polaron momentum $q$ is carried by the phonons, whereas the enhancement of $\mathcal{M}$ is due to quantum fluctuations of the impurity itself.

\section{Polaron properties from the RG -- derivations}
\label{apdx:MPfromRGqph}

In this appendix we show in detail how polaron properties can be calculated using the RG procedure introduced in the main text. In particular, we derive the RG flow equations of the polaron phonon number $N_\ph$, the phonon momentum $q_\ph$ and the quasiparticle weight $Z$.

\subsection{Polaron phonon number}
\label{subsecApdx:PhononNumber}
We start by the derivation of the RG flow equation \eqref{eq:RGflowNph} for the phonon number $N_\ph$. To this end we split the expression for $N_\ph$,
\begin{equation}
N_\ph = N_\ph^\MF + \int^{\Lambda_0} d^3 \vec{k} ~ \bra{\gs} \G_{\vec{k}} \ket{\gs},
\label{eq:RGflowNphStartAPDX}
\end{equation}
(see Eq.\eqref{eq:RGflowNphStart} in the main text) into contributions from slow and fast phonons,
\begin{equation}
N_\ph = N_\ph^\MF + \int_\s d^3 \vec{p} ~ \bra{\gs} \G_{\vec{p}}\ket{\gs} +  \int_\f d^3 \vec{k} ~ \bra{\gs} \G_{\vec{k}}\ket{\gs} .
\label{eq:NphRGstep0}
\end{equation}
Next we apply the RG step, i.e. the unitary transformation $\hat{U}_\Lambda$, to simplify both integrals. The ground state \emph{after applying the RG step} factorizes,
\begin{equation}
\ket{\gs} = \ket{0}_\f \otimes \ket{\gs}_\s.
\end{equation}
Thus using $\hat{U}_\Lambda^\dagger \G_{\vec{p}} \hat{U}_\Lambda =\G_{\vec{p}} + \mathcal{O}(\Omega_{\vec{k}}^{-2})$ for the slow phonons, we find
\begin{equation}
 \int_\s d^3 \vec{p} ~ \bra{\gs} \G_{\vec{p}}\ket{\gs} =  \int_\s d^3 \vec{p} ~_\s \bra{\gs} \G_{\vec{p}}\ket{\gs}_\s + \mathcal{O}(\Omega_{\vec{k}}^{-2}).
 \label{eq:GpRGforNph}
\end{equation}
In the subsequent RG step, we can treat the new term on the right hand side of Eq.\eqref{eq:GpRGforNph} in the same way as we treated our initial term in Eq.\eqref{eq:RGflowNphStart}.

The second term in Eq.\eqref{eq:NphRGstep0}, corresponding to fast phonons, reads
\begin{equation}
\int_\f d^3 \vec{k} ~\bra{\gs}\hat{U}^\dagger_\Lambda \G_{\vec{k}} \hat{U}_\Lambda \ket{\gs} = 
- 2 \int_\f d^3 \vec{k} ~ \alpha_{\vec{k}}  ~_\s \bra{\gs}  \F_{\vec{k}} \ket{\gs}_\s  + \mathcal{O}(\Omega_{\vec{k}}^{-2})
\label{eq:RGstepNphApdx}
\end{equation}
after the RG step. Here $\F_{\vec{k}}$ is defined in Eq.\eqref{eq:Fresult}, from which we directly obtain the first two terms in the square brackets of Eq.\eqref{eq:RGflowNph}, along with an additional renormalization term
\begin{equation}
N_\ph \rightarrow N_\ph - 2 \int_\f d^3 \vec{k} ~ \frac{|\alpha_{\vec{k}}|^2}{\Omega_{\vec{k}}} k_\mu \mathcal{M}_{\mu \nu}^{-1}  
 \int_\s d^3 \vec{p} ~ p_\nu ~_\s \bra{\gs} \G_{\vec{p}} \ket{\gs}_\s.
\label{eq:NphRGstepNonTrivial}
\end{equation}
We will see below (in \ref{subsecApdx:PolaronMomentum}) that the integral over slow degrees of freedom in the second line of Eq.\eqref{eq:NphRGstepNonTrivial} appears also in the expression for the phonon momentum, see Eq.\eqref{eq:qPhRGForm}. 

Furthermore, from symmetry considerations it follows that only $\nu=x$ can give a non-vanishing contribution, assuming that $\vec{q} = q \vec{e}_x$ points along $x$-direction. We can make use of Eq.\eqref{eq:qPhRGForm} by evaluating it not only at our current cut-off $\Lambda$, but also at $\Lambda=0$ where we obtain $q_\ph = P_\ph(0)$. From the expression for $q_\ph$ at $\Lambda$ (see Eq.\eqref{eq:qPhRGForm}) we thus obtain together with the result Eq.\eqref{eq:qFlowSolutions},
\begin{equation}
\int^\Lambda_\s d^3 \vec{p} ~ p_x \bra{\gs} \G_{\vec{p}} \ket{\gs} 
=  \frac{\mathcal{M}_\parallel(\Lambda)}{M} \left[ P_\ph(0) - P_\ph(\Lambda) \right] + \mathcal{O}(\Omega_{\vec{k}}^{-2}).
\end{equation}
Now the inclusion of the remaining terms from Eq.\eqref{eq:NphRGstepNonTrivial} in the RG flow of $N_\ph$ Eq.\eqref{eq:RGstepNphApdx} straightforwardly leads to our result, Eq.\eqref{eq:RGflowNph}.

\subsection{Polaron momentum}
\label{subsecApdx:PolaronMomentum}
To derive the result Eq.\eqref{eq:qPh} stated the main text, i.e. $q_\ph(\Lambda)=P_\ph(\Lambda)$, we start by writing the phonon momentum as
\begin{equation}
q_\ph = P_\ph^\MF + \int^{\Lambda_0} d^3\vec{k}~ k_x \bra{\gs} \G_{\vec{k}}  \ket{\gs},
\label{eq:qPhRGApdx}
\end{equation}
see Eq.\eqref{eq:qPhRG}. Similar to the first section \ref{subsecApdx:PhononNumber} of this appendix, we will decompose this expression into parts corresponding to slow and fast phonons respectively. However, anticipating the effect of the RG, we introduce a more general expression,
\begin{equation}
q_\ph = \overline{P}_\ph(\Lambda) + \chi(\Lambda) \int^\Lambda d^3 \vec{k} ~ k_x  \bra{\gs} \G_{\vec{k}} \ket{\gs},
\label{eq:qPhRGForm}
\end{equation}
where initially (i.e. before running the RG, $\Lambda=\Lambda_0$) we have
\begin{equation}
\chi(\Lambda_0) = 1, \qquad \overline{P}_\ph(\Lambda_0) = P_\ph^\MF.
\end{equation}
Here $\overline{P}_\ph(\Lambda)$ is some function of the running cut-off $\Lambda$ for which we will now derive the RG flow equation. We observe that in Eq.\eqref{eq:qPhRGForm} slow phonons contribute to this phonon momentum $\overline{P}_\ph$, but only with a reduced weight described by the additional factor $\chi(\Lambda)$.

Next, by applying the same steps as in Eqs.\eqref{eq:GpRGforNph} and\eqref{eq:RGstepNphApdx}, we can easily derive the following RG flow equations for $\overline{P}_\ph(\Lambda)$,
\begin{equation}
\frac{\partial \overline{P}_\ph}{\partial \Lambda} = - 2  \frac{\chi}{M} \int_\f d^2 \vec{k} ~  \frac{|\alpha_{\vec{k}}|^2}{\Omega_{\vec{k}}} k_x \Bigl[ - k_x \l P_\ph - P_\ph^\MF \r  + \frac{1}{2} k^2 - \frac{M}{2} \l \frac{k_x^2}{\mathcal{M}_\parallel} + \frac{k_y^2 + k_z^2}{\mathcal{M}_\perp} \r \Bigr],
\end{equation}
as well as for $\chi(\Lambda)$,
\begin{equation}
\frac{\partial \chi}{\partial \Lambda} = 2 \frac{\chi}{\mathcal{M}_\parallel} \int_\f d^2 \vec{k} ~ \frac{|\alpha_{\vec{k}}|^2}{\Omega_{\vec{k}}} k_x^2.
\end{equation}
By comparing these RG flow equations to those of $\mathcal{M}_\parallel^{-1}(\Lambda)$, see Eq.\eqref{eq:gsFlowM}, and of $P_\ph(\Lambda)$, see Eq.\eqref{eq:gsFlowQ}, a straightforward calculation shows that the solutions can be expressed in terms of the RG coupling constants,
\begin{equation}
\overline{P}_\ph(\Lambda) =P_\ph(\Lambda), \qquad \qquad \chi(\Lambda) = \frac{M}{\mathcal{M}_\parallel(\Lambda)},
\label{eq:qFlowSolutions}
\end{equation}
which apparently fulfills the required initial conditions at the initial cut-off $\Lambda=\Lambda_0$. Finally, when $\Lambda \to 0$, from Eq.\eqref{eq:qPhRGForm} we obtain a fully converged phonon momentum $q_\ph = \overline{P}_\ph(0) = P_\ph(0)$ as claimed in the main text.

\subsection{Quasiparticle weight}
\label{subsecApdx:ZRGflow}
Now we will derive the RG flow equation \eqref{eq:RGflowZp} of the logarithm of the quasiparticle weight, $\log Z$. To this end we introduce a unity $\hat{1} = \hat{U}_\Lambda \hat{U}_\Lambda^\dagger$ into the definition Eq.\eqref{eq:ZdefRG}, $Z = |\bra{- \alpha_{\vec{k}}}\bra{-\alpha_{\vec{p}}} \gs \rangle|^2$, and obtain
\begin{equation}
Z = \left| \bra{- \alpha_{\vec{k}}} \bra{- \alpha_{\vec{p}}}  \hat{U}_\Lambda \ket{0}_\f \otimes \ket{\gs}_\s \right|^2.
\label{eq:ZformulaAppendix}
\end{equation}
Here we used that $\hat{U}_\Lambda^\dagger \ket{\gs} = \ket{0}_\f \otimes \ket{\gs}_\s$ and introduced the short-hand notation 
\begin{equation}
\ket{- \alpha_{\vec{k}}} \ket{- \alpha_{\vec{p}}} \equiv \prod_{\vec{k} \in \f} \ket{-\alpha_{\vec{k}}} \otimes  \prod_{\vec{p} \in \s} \ket{-\alpha_{\vec{p}}}.
\end{equation}

To evaluate Eq.\eqref{eq:ZformulaAppendix}, we notice that $\ket{- \alpha_{\vec{p}}}$ is an eigenstate of $\hat{U}_\Lambda^\dagger$. To show why this is the case, let us first use that
\begin{flalign}
\G_{\vec{p}} \ket{- \alpha_{\vec{p}}} & = \G_{\vec{p}} \hat{U}^\dagger_\MF \ket{0}_\s \\
 & \stackrel{\eqref{eq:GkLLP}}{=} \hat{U}^\dagger_\MF \l \ad_{\vec{p}} \a_{\vec{p}} - |\alpha_{\vec{p}}|^2 \r \ket{0}_\s \\
 & = - |\alpha_{\vec{p}}|^2 ~ \ket{- \alpha_{\vec{p}}}.
 \label{eq:GammaEigenstates}
\end{flalign}
In the second line we used that
\begin{equation}
\hat{U}_\MF \G_{\vec{k}} \hat{U}^\dagger_\MF= \ad_{\vec{k}} \a_{\vec{k}} - |\alpha_{\vec{k}}|^2.
\label{eq:GkLLP}
\end{equation}
Thus, from the solution for $\hat{F}_{\vec{k}}$ Eq.\eqref{eq:Fresult} we obtain
\begin{equation}
\hat{F}_{\vec{k}} \ket{- \alpha_{\vec{p}}} = f_{\vec{k}} \ket{- \alpha_{\vec{p}}} + \mathcal{O}(\Omega_{\vec{k}}^{-2}),
\end{equation}
where
\begin{equation}
f_{\vec{k}} = \frac{1}{\Omega_{\vec{k}}} \left[  W_{\vec{k}} - \alpha_{\vec{k}} k_\mu \mathcal{M}_{\mu \nu}^{-1} \int_\s d^3 \vec{p} ~  p_\nu  |\alpha_{\vec{p}}|^2 \right].
\end{equation}
From the definition of $\hat{U}_\Lambda$ Eq.\eqref{eq:defU}, and noting that $f_{\vec{k}} \in \mathbb{R}$ is real, it now follows that
\begin{equation}
\hat{U}_\Lambda^\dagger \ket{- \alpha_{\vec{p}}} = e^{  - \int_\f d^3 \vec{k} ~ f_{\vec{k}} \left[  \a_{\vec{k}} - \ad_{\vec{k}} \right] + \mathcal{O}(\Omega_{\vec{k}}^{-2}) }  \ket{- \alpha_{\vec{p}}}.
\label{eq:UdLambdaCohsApdx}
\end{equation}
Using the last result, Eq.\eqref{eq:ZformulaAppendix} now factorizes, $Z=Z_\s Z_\f$, where
\begin{equation}
Z_\s =  |\bra{-\alpha_{\vec{p}}} \gs \rangle_\s |^2
\end{equation}
has the same form as $Z$ but includes only slow phonons. The contribution from fast phonons reads
\begin{equation}
Z_\f = | \bra{- \alpha_{\vec{k}}} e^{  \int_\f d^3 \vec{k} ~ f_{\vec{k}} \left[  \a_{\vec{k}} - \ad_{\vec{k}} \right] + \mathcal{O}(\Omega_{\vec{k}}^{-2}) } \ket{0}_\f  |^2,
\end{equation}
and can be simplified by recognizing the displacement operator by $- f_{\vec{k}}$,
\begin{equation}
Z_\f = | \bra{- \alpha_{\vec{k}}} - f_{\vec{k}} \rangle |^2 + \mathcal{O}(\Omega_{\vec{k}}^{-2} )
 = \exp \l - \int_\f d^3 \vec{k} ~ |\alpha_{\vec{k}} - f_{\vec{k}}|^2  + \mathcal{O}(\Omega_{\vec{k}}^{-2} ) \r.
\end{equation}

In summary, by applying a single RG step we can write
\begin{equation}
Z = |\bra{-\alpha_{\vec{p}}} \gs \rangle_\s |^2 \times e^{ - \int_\f d^3 \vec{k} ~ |\alpha_{\vec{k}} - f_{\vec{k}}|^2  + \mathcal{O}(\Omega_{\vec{k}}^{-2} ) }.
\end{equation}
I.e. in terms of the logarithm
\begin{equation}
\log Z = \log Z_\s  - \int_\f d^3 \vec{k} ~ |\alpha_{\vec{k}} - f_{\vec{k}}|^2  + \mathcal{O}(\Omega_{\vec{k}}^{-2} ),
\end{equation}
such that Eq.\eqref{eq:RGflowZp} immediately follows from 
\begin{equation}
\frac{\partial }{\partial \Lambda} \log Z = - \int_\f d^2 \vec{k} ~ |\alpha_{\vec{k}} - f_{\vec{k}}|^2  + \mathcal{O}(\Omega_{\vec{k}}^{-2} ).
\end{equation}

\newpage
\textbf{\Large Acknowledgements}\\
\addcontentsline{toc}{chapter}{Acknowledgements}

F.G. acknowledges invaluable support from Michael Fleischhauer. The authors are grateful to Wim Casteels for providing his results from variational Feynman path-integral calculations. They  would like to thank Dmitry Abanin, Michael Fleischhauer, Alexey Rubtsov, Yulia Shchadilova,  Richard Schmidt, and Aditya Shashi for invaluable discussions and for collaboration on problems discussed in these lecture notes. The authors are grateful to Immanuel Bloch, Thierry Giamarchi, Walter Hofstetter, Markus Oberthaler, Christoph Salomon,  Vladimir Stojanovich, Leticia Tarruell, Artur Widera,  Sebastian Will, and Martin Zwierlein for stimulating discussions. They would like to thank J. T. Devreese, S. N. Klimin, Hing Long and Tao Yin for useful comments on the manuscript. F.G. gratefully acknowledges financial support from the "Marion K\"oser Stiftung". He was supported by a fellowship through the Excellence Initiative (Grant No. DFG/GSC 266) and by the Moore Foundation. ED acknowledges support from the NSF grant  DMR-1308435, Harvard-MIT CUA, AFOSR New Quantum Phases of Matter MURI, the ARO-MURI on Atomtronics, ARO MURI Quism program, Dr.~Max R\"ossler, the Walter Haefner Foundation and the ETH Foundation.



\end{document}